\documentclass[a4paper,11pt]{article}
\pdfoutput=1 

\usepackage{jheppub} 

\usepackage[T1]{fontenc} 

\renewcommand{\thanks}[1]{\footnote{#1}}

\newcommand{\bea}{\begin{eqnarray}}
\newcommand{\eea}{\end{eqnarray}}

\def\a{\alpha}
\def\b{\beta}
\def\g{\gamma}

\def\l{\lambda}
\def\om{\omega}

\def\ep{\varepsilon}

\def\oab{\bar\omega_{\alpha}}

\def\oa{\omega_{\alpha}}
\def\ob{\omega_{\beta}}

\def\cC{{\cal C}}
\def\cD{{\cal D}}

\def\cG{{\cal G}}

\def\cN{{\cal N}}

\def\mE{\mathfrak{E}}

\def\mI{\mathfrak{I}}
\def\mJ{\mathfrak{J}}

\def\mM{\mathfrak{M}}

\def\half{ {1\over 2}}
\def\p{\partial}

\def\Gp{\cG}

\def\rS{{\rm S}}
\def\rAdS{{\rm AdS}}
\def\mbR{{\mathbb{R}}}

\def\no{\nonumber}
\def\sm{\smallskip}

\title{\boldmath M-theory Solutions Invariant under $D(2,1; \gamma) \! \oplus \! D(2,1;\gamma)$}
\preprint{Imperial/TP/2013/JE/01}

\author[a]{Constantin Bachas,}
\author[b]{Eric D'Hoker,}
\author[\,c]{John Estes}
\author[\,d]{and Darya Krym}

\affiliation[a]{Laboratoire de Physique Th\'eorique de l'\'Ecole Normale Sup\'erieure,\footnote{Unit\'e mixte
(UMR 8549) du CNRS
et de l'ENS, Paris.}
\\
24, rue Lhomond, 75231 Paris cedex, France}
\affiliation[b]{Department of Physics and Astronomy,
 University of California,\\  Los Angeles, CA 90095, USA}
\affiliation[c]{Blackett Laboratory, Imperial College,\\
 London, SW7 2AZ, UK}
\affiliation[d]{Physics Department,
New York City College of Technology, \\
The City University of New York, \\
Brooklyn, New York 11201, USA}

\emailAdd{bachas@lpt.ens.fr}
\emailAdd{dhoker@physics.ucla.edu}
\emailAdd{johnaldonestes@gmail.com}
\emailAdd{daryakrym@gmail.com}

\abstract{We simplify and extend the construction of half-BPS  solutions to 11-dimensional
supergravity,
with isometry superalgebra $D(2,1;\g) \oplus D(2,1;\g)$.
Their space-time  has the form  AdS$_3\times$S$^3\times$S$^3$ warped over a Riemann
surface $\Sigma$.
It describes  near-horizon geometries of  M2  branes  ending on, or intersecting
with, M5 branes along a common string.
The general solution to the BPS equations is specified by a reduced set of data $(\gamma,
h, G)$, where $\gamma$ is
the real parameter of the isometry superalgebra, and  $h$ and $G$
are functions on $\Sigma$ whose differential equations and regularity conditions
depend only on the sign of $\gamma$.
The magnitude of $\gamma$ enters only through the map of $h,G$ onto the supergravity
fields, thereby
promoting all solutions into families parametrized by $|\gamma|$.
By analyzing the regularity conditions for the supergravity fields,  we prove two
general theorems:
(i) that the only solution with a 2-dimensional CFT dual  is
AdS$_3\times$S$^3\times$S$^3\times \mbR^2$,
modulo discrete identifications of the flat $\mbR^2$, and
(ii) that solutions with $\g<0$ cannot have more than one asymptotic
higher-dimensional AdS region.
We  classify  the allowed singularities of $h$ and $G$ near the boundary of
$\Sigma$, and  identify four local solutions:
asymptotic AdS$_4/Z_2$ or AdS$_7^\prime$ regions; highly-curved M5-branes; and a
coordinate singularity called the "cap".
By putting these "Lego" pieces together we recover all known global regular
solutions with the above symmetry, including the self-dual strings
on M5 for $\gamma <0$, and the Janus solution for $\gamma >0$, but now promoted to
families parametrized by $|\gamma|$.
We also construct exactly new regular solutions which are asymptotic to AdS$_4/Z_2$
for $\gamma <0$,
and conjecture that they are a different superconformal limit of the self-dual string.
Finally, we construct exactly $\gamma >0$ solutions  with highly curved M5-brane
regions, which are the
formal continuation of the self-dual string solutions across the decompactification
point at $\gamma =0$.  }

\begin{document}
\maketitle
\flushbottom

\section{Introduction and summary}
\setcounter{equation}{0}
\label{sec:1}

The fundamental branes of M-theory are the M2-brane and the M5-brane.\footnote{For a  recent review  and more references
see   \cite{Berman:2007bv}.} There
has been  significant progress  \cite{Bagger:2006sk,Gustavsson:2007vu,Aharony:2008ug}  in recent years  towards
 elucidating
the gauge dynamics of multiple M2 branes,  which is dual to eleven-dimensional supergravity in AdS$_4\times$S$^7/Z_k$.
The dynamics of multiple M5 branes,
on the other hand,  and of intersections of M5 branes with M2 branes  remain  elusive. The two problems
are related, since it is believed that the M5-brane dynamics is described by self-dual strings,  which are
 the low-lying modes of
  M2 branes stretching between M5 branes.  At present, even counting
  the  degrees of freedom   on  M2/M5  intersections is
 an open question, see for instance \cite{Niarchos:2012cy}.

     In this paper we will analyze the supergravity solutions that  arise as  the near-horizon geometries of supersymmetric
     M2/M5-brane  intersections. These
     provide  a dual description
   of  the infrared dynamics of the field theories that live on  the  branes.
   All solutions have the form of  AdS$_3\times$S$^3\times$S$^3$ space-time warped  over a
   two-dimensional Riemann surface $\Sigma$.
       We  will build  upon the earlier works   in
\cite{D'Hoker:2008wc,D'Hoker:2008qm,D'Hoker:2009my,janus,Estes:2012vm},
  which we will   simplify and extend.

  The backgrounds of interest preserve one half of
  the maximal supersymmetry. They are left invariant by the  superconformal algebra
 $D(2,1; \gamma) \! \oplus \! D(2,1;\gamma)$ which   depends
  on a real parameter $\g$.   We will first rewrite the reduced equations and the   regularity conditions
  derived in \cite{Estes:2012vm}, so as to make it clear
    that any  solution
   can be continuously deformed by changing the magnitude  (but not the sign)  of $\g$.
   This generalizes the observation in  \cite{Estes:2012vm} that the maximally-symmetric  AdS$_4\times$S$^7$ and
   AdS$_7\times$S$^4$ solutions admit such  deformations.
  We   will see that  changing
   $\vert \g\vert$ actually   rescales the ratio of  the two kinds of
   M5 brane charge that are compatible with the half-BPS condition. When both types of
   charge are turned on,  $\vert\g\vert$ is thus a  rational rather than a continuous modulus
   of the solution.

\sm

  From the conditions of global regularity we will derive two other general results.
  The first is a "uniqueness theorem" for solutions dual to two-dimensional conformal theories,
  i.e.  solutions whose conformal boundary is two-dimensional. We will show that the
  only such solution  is AdS$_3\times$S$^3\times$S$^3\times$E$_2$, where E$_2$ is the Euclidean
  plane (or discrete identifications thereof). This
  is the near-horizon geometry of M2 branes suspended between  M5 branes, in the limit
  where the M5 branes  have been smeared  \cite{Boonstra:1998yu}.  Our   theorem
  implies that  the infrared dynamics on  the M2 branes always restores
  the translation symmetry, which is a priori broken  by the localized M5 branes.
  This should be contrasted with the  analogous situation of D3 branes
   suspended between NS5 branes and D5 branes  \cite{Hanany:1996ie}. There,
   the $3d$ field theory on the suspended D3 branes has
   a multitude  of strongly-coupled infrared fixed points \cite{Gaiotto:2008sa,Gaiotto:2008sd,Gaiotto:2008ak},
    in one-to-one correspondence with a rich set of  half-BPS solutions of the
   type-IIB supergravity equations \cite{Assel:2011xz,Assel:2012cj,Aharony:2011yc}.


\begin{figure}[htb]
\label{confBnrs}
\begin{center}
\includegraphics[width= 5.2in]{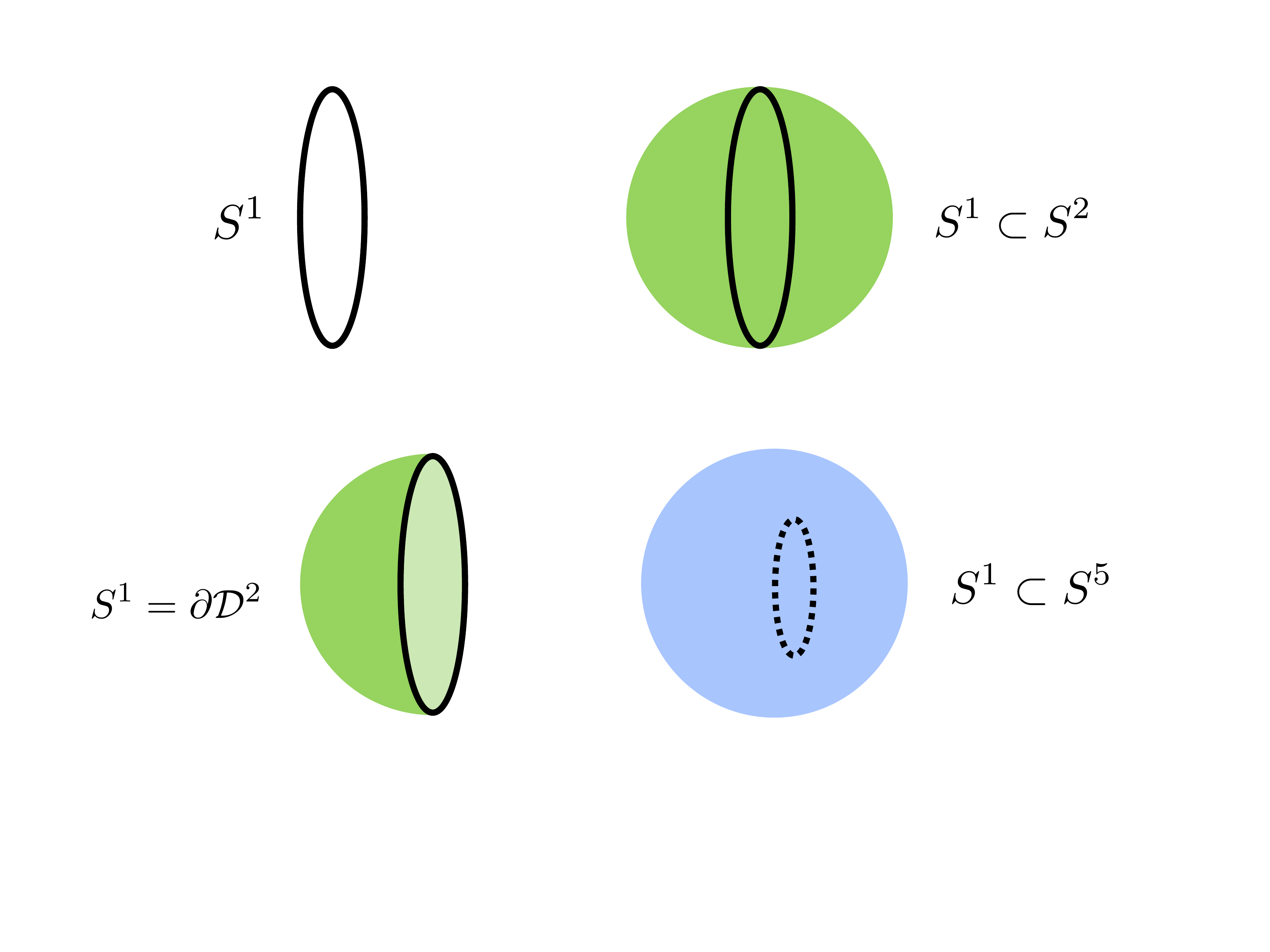}
\vskip -2.3 cm
\caption{\footnotesize The possible conformal   boundaries  of the near-horizon geometries
of intersecting M5/M2 branes.
 All boundaries share time and (in global coordinates)  a spatial circle which is :
  $(i)$  the entire space of the holographically dual CFT$_2$\,;
 $(ii)$ a  domain wall or $(iii)$ a boundary of the  CFT$_3$ that lives on M2 branes\,;
 or $(iv)$ a great-circle string defect of the CFT$_6$  that lives on M5 branes.
The boundary in the latter case need not be the round 5-sphere,  in general  it is a  deformed 5-sphere.
  }
\end{center}
\end{figure}


 Our second  general result can be stated more clearly by considering  the  conformal boundaries
 of   the Penrose compactifications of the supergravity  solutions.
  The boundaries that are   consistent with the symmetries of the problem are illustrated in Figure \ref{confBnrs}.
   The corresponding dual field theories are $(i)$  a two-dimensional CFT,  $(ii)$ a
   three-dimensional CFT
   with a domain wall,    $(iii)$  a three-dimensional CFT
   on a space with boundary, and $(iv)$ the N=(2,0)  six-dimensional CFT in the background of (self-dual)
   strings. Our theorem states that,  for $\g$ negative,  the second  possibility is not  allowed.
This comes about because  the boundaries
  $(iii)$ and $(iv)$   arise from  points on  $\Sigma$ where the radius of AdS$_3$ diverges, and we will show that
  the existence of more than one such points leads  unavoidably, when  $\g<0$,  to  conical singularities.
Since solutions that describe interfaces of the M2 brane theory require two (or more) boundaries with disk topology,  we
conclude that these are possible only when $\g$ is positive.  Our  argument also excludes  a
non-connected  conformal boundary when $\g<0$,    but this is anyway  ruled out under the much milder assumptions
 of ref.\,\cite{Witten:1999xp} (see also \cite{Galloway:1999br,Maldacena:2004rf}).
  \sm

 The above statements are,  of course,  compatible   with all  known regular   solutions of the M theory equations.
 In the last part of this paper we will discuss the known solutions in  a unified manner,
   and  calculate their
 M2-brane and M5-brane charges. Known exact solutions include\,:

 \begin{itemize}
 \item     the
  $\g$-deformed AdS$_7\times$S$^4$ and AdS$_4\times$S$^7$  found in \cite{Estes:2012vm}, which are  a special
  case of our general theorem stating that {\it all}  solutions can be $\g$-deformed;

 \item    the $\g=1$ Janus background discovered in  ref.\,\cite{janus}, which   admits  a $\g$ deformation like all    other
  solutions.
The ensuing two-parameter family of solutions is   identical to the one derived from
 $4d$ gauged supergravity
  in ref.\,\cite{Bobev:2013yra};

\item   the $\g=-{1\over 2}$ self-dual string solutions of ref.\,\cite{D'Hoker:2008qm}
which  are asymptotic to AdS$_7\times$S$^4$, and for which we will calculate the invariant charges;

\end{itemize}

\sm

\noindent  We will also present  some new global  solutions to the half-BPS equations, in particular\,:

 \begin{itemize}
 \item    regular solutions with  (AdS$_4/Z_2)\times$S$^7$
 asymptotics and $\g<0$, which should be dual to stacks of semi-infinite M2 branes;
 we  conjecture  that these solutions are different superconformal  limits of the same system also described by  the self-dual strings;

 \item  $\g>0$ solutions   with either (AdS$_4/Z_2)\times$S$^7$ or AdS$_7^\prime\times$S$^4$
 asymptotics,  which   contain highly-curved  regions  with  5-brane charge.
 These are the continuation to positive $\g$ of the regular,  self-dual string and semi-infinite M2-brane,  solutions
 at $\g<0$.

\end{itemize}


 \vskip 2mm
  All the solutions with the exception of  the last ones  are made out of three "Lego" pieces:    two throats asymptotic to
 AdS$_7^\prime \times$S$^4$ and  (AdS$_4/Z_2)\times$S$^7$, and a smooth region with a coordinate singularity that we
 call the "cap". Here AdS$_7^\prime$  indicates that the maximally-symmetric space-time  is, in general,   deformed.
 We will see that these  Lego pieces are local solutions of  the BPS equations for any value of   $\g$, both positive or negative.
 This has been noted for the AdS$_7^\prime \times$S$^4$ throat at $\g=1$  in ref.\,\cite{Berdichevsky:2013ija},
  and we generalize the observation to all three Lego pieces and all   $\g$.\footnote{Ref.\,\cite{Berdichevsky:2013ija} concluded that
   there is no obstruction, at the level of the
  symmetry algebra,    to the coexistence in the same solution of an  AdS$_7^\prime \times$S$^4$
  and a AdS$_4\times$S$^7$ throat.
  Nevertheless, we have not found any regular  solutions of this type. Indeed, since the
   conformal boundary of AdS$_7^\prime \times$S$^4$, item $(iv)$ in   Figure \ref{confBnrs}, is  a compact manifold,
   the presence of a second throat would lead to a disconnected boundary. With the help of  a
   physically-reasonable assumption,  this can be excluded  \cite{Witten:1999xp}.} The last solutions  also include
   a fourth Lego piece, with regular geometry but in which  the supergravity approximation breaks down. These regions
   are curved M5-brane sources that do not change the dimension of the conformal boundary;
 similar sources have been found in the type-IIB theory \cite{Assel:2011xz,Assel:2012cj,Aharony:2011yc}.
 \sm

The $\g$-deformed Janus solution, as well as our  new
 solutions with (AdS$_4/Z_2)\times$S$^7$ asymptotics,
 might seem to be in tension with the "rigidity" claim
 of ref.\,\cite{D'Hoker:2009my}. This  claim was based,   however,
 on certain regularity assumptions,  and on the assumption that the asymptotic symmetry
 of the solution is   $D(2,1;1) \oplus D(2,1;1)$.
 The regular new solutions   violate  the second requirement, while the  last solutions  in the above list
  do not obey the regularity assumptions made in   ref.\,\cite{D'Hoker:2009my}.
   \sm

 The   paper is organized as follows. In Section \ref{sec:1} we describe the M-brane configurations of interest
 and their symmetries, paying special attention to the supersymmetry algebra parameter $\g$. We also comment on
 similarities and differences between the problem at hand, and the analogous problem in type IIB string theory.
In Section \ref{sec:2} we review how  the Killing-spinor equations can be reduced to a mathematical
  problem
  formulated in terms of $\g$ and two functions, $h$ and $G$, on the base $\Sigma$.
  The main new point, compared to the analysis in \cite{D'Hoker:2008wc,Estes:2012vm}, is a redefinition
  of  $G$ that puts the two 3-spheres on equal footing, and shows that the reduced mathematical problem
  depends only  on the sign of $\g$. An immediate corollary is that all solutions come in families parametrized by
  the magnitude of $\g$.

  In Section \ref{sec:3} we compute the 3-form and 6-form gauge potentials, and the
  associated M2- and M5-brane charges,   in terms of the reduced data $(\g, h, G)$. We show,  in particular,
  that changing $\g$ rescales the M5 and M5$^\prime$ charges in opposite directions, while leaving the  product
  of these charges,  as well as
   the M2-brane charges,  invariant. In Section \ref{sec:4} we analyze the singularities of $h$, and deduce
   from them two general theorems: that the solution  with  CFT$_2$ dual is unique, and that there are no interface
   solutions with $\g<0$. In Section \ref{sec:5} we solve the equation and regularity conditions for $G$   locally,
   in the neighborhood of a boundary point. We exhibit the four local solutions that enter in the construction
   of globally regular solutions:   the two asymptotic throats and the "cap"
   for any value of $\g$, and the highly-curved M5 solution for $\g$ positive.

   In section \ref{sec:6} we put together the first three "Lego"  pieces to construct
   known solutions for $\g<0$, and calculate their charges.
   We also present the new solutions with (AdS$_4/Z_2)\times$S$^7$ asymptotics, which we conjecture
   to be different superconformal limits of the self-dual strings.
    New and old global solutions   for $\g>0$ are presented  in section  \ref{sec:7}.  The new solutions include
   the highly-curved M5 regions, and they describe either self-dual strings  or semi-infinite M2 branes.
   Details of many of the computations have been relegated to the appendices  \ref{sec:A} to  \ref{sec:E}.

%


\section{Brane configurations and symmetries}
\label{sec:1}

The fundamental branes of M-theory admit special arrangements in which  supersymmetry is only partially broken.
A~quarter-BPS intersection of a stack of coincident  M2-branes with   stacks of coincident  M5-branes may be
obtained by arranging the branes in $\mathbb{R}^{1,10}$ according to the following pattern (see for instance \cite{Gauntlett:1997cv,Smith:2002wn}):
\smallskip
\begin{table}[htdp]
\begin{center}
\begin{tabular}{|c||c|c|c|c|c|c|c|c|c|c|c|} \hline
 & 0 & 1 & 2 & 3 & 4 & 5 & 6 & 7 & 8 & 9 & 10
\\ \hline \hline
M2 & $\star$ & $\star$ & $\star$ & &&&&&&&
\\ \hline
M5 & $\star$ & $\star$ & & $\star$ & $\star$ & $\star$ & $\star$ &  &&&
\\ \hline
M5$^\prime$ & $\star$ & $\star$ & &  &&& & $\star$ & $\star$ & $\star$ & $\star$
\\ \hline
\end{tabular}
\label{table}
\end{center}
\caption{\footnotesize Quarter-BPS arrangement of M2-branes and M5-branes
 in flat  eleven-dimensional space-time. Stars indicate the dimensions along which the  brane worldvolumes extend.}
\end{table}

\noindent
As a reminder that all  branes share a  common string,  the above pattern is sometimes referred to
as   M2\,$\perp$\,M5\,$\perp$\,M5\,(1).
In general, the M2-branes may either intersect  the M5-branes,  or end on them.
All of these configurations exhibit  manifest $ISO(1,1) \oplus SO(4) \oplus SO(4)$ isometry, i.e.   Poincar\'e
invariance in the directions  \{0,1\}, and  symmetry  under rotation of  the dimensions \{3,4,5,6\} and
of the dimensions \{7,8,9,10\}. \footnote{The exact supergravity solutions with fully-localized M-brane sources  remain  elusive,
despite several interesting attempts, see for example
\cite{Smith:2002wn,Youm:1999zs,Hosomichi:2000iz,Rajaraman:2000ws,Niarchos:2012pn,Lunin:2007mj}.
 To simplify the equations, one usually smears the  M5-branes along  their  (common) transverse direction 2,
   thereby  providing  an extra  $U(1)$ isometry  if $x^2$ is compact,  or $\mathbb{R}$ if $x^2$
   is a non-compact coordinate. }


\sm
In the near-horizon limit, one expects  $ISO(1,1)$ to be  promoted to the conformal group
 $SO(2,2)$. This  can be realized  geometrically as the isometry of  a warped ${\rm AdS}_3$ factor.
 The near-horizon
geometry must thus have the fibered  form $({\rm AdS}_3\times {\rm S}^3\times {\rm S}^3) \ltimes \Sigma$, where $\Sigma$ is a two-dimensional
base over which the three (peudo)spheres are  fibered.
The precise nature of the near-horizon limit
depends, of course, on the  details of the  M2-M5 system, and possibly also on the  manner in which the
 horizon is being approached.


\subsection{The Lie superalgebra $D(2,1;\gamma) \oplus D(2,1;\gamma)$ and the parameter $\gamma$}

The symmetries of the supergravity solutions in this paper are governed by the Lie superalgebra
$D(2,1;\gamma) \oplus D(2,1;\gamma)$. More specifically, it is the real form $D(2,1;\gamma, 0)$,
whose bosonic subalgebra is $SO(2,1) \oplus SO(3) \oplus SO(3)$, which enters here
\cite{D'Hoker:2008ix,Gunaydin:1986fe,Sevrin:1988ew}.
We designate the generators of the bosonic subalgebra by $T^{(a)} _i$ with $a=1$ corresponding to
$SO(2,1)$ and $a=2,3$  to the remaining two $SO(3)$ subalgebras. The index $i=1,2,3$
labels the three linearly independent generators within each subalgebra, and we have
\bea
[T_i ^{(a)} , T^{(b)} _j ] = i \delta ^{ab} \ep _{ijk} \eta_a ^{k\ell} T_\ell ^{(a)}\ .
\eea
Here, $\ep_{ijk}$ is the completely antisymmetric tensor, and $\eta _a^{k \ell}$ is the canonical metric on each
simple Lie algebra factor. The fermionic generators of $D(2,1;\gamma,0)$ form an 8-dimensional spinor
$F_{\alpha_1  \alpha_2 \alpha _3}$ whose index $\alpha _a$ transforms under the 2-dimensional spinor representation
of $T^{(a)}_i$ for $a=1,2,3$. The anti-commutator of two fermionic generators is given as follows,
\bea
\{ F_{\alpha _1 \alpha _2 \alpha _3} , F_{\beta _1 \beta _2 \beta _3} \}
& = &
c_1 C_{\alpha _2 \beta _2} C_{\alpha _3 \beta _3 } \left ( C \sigma ^i \right ) _{\alpha _1 \beta _1} T^{(1)} _i
\no \\ &&
+c_2 C_{\alpha _1 \beta _1} C_{\alpha _3 \beta _3 } \left ( C \sigma ^i \right ) _{\alpha _2 \beta _2} T^{(2)} _i
\no \\ &&
+c_3 C_{\alpha _1 \beta _1} C_{\alpha _2 \beta _2 } \left ( C \sigma ^i \right ) _{\alpha _3 \beta _3} T^{(3)} _i\ .
\eea
Here, $\sigma ^i$ are the Pauli matrices and $C=i \sigma ^2$. The real parameters $c_1, c_2, c_3$ satisfy
$c_1+c_2+c_3=0$. A rescaling $c_a \to \lambda c_a$ by any real non-vanishing $\lambda$
can always be absorbed into the normalization of the spinor, and is immaterial. As a result, only a single
ratio of the $c_a$ matters,  and is chosen to be $\gamma = c_2/c_3$, the parameter of the Lie superalgebra
$D(2,1;\gamma,0)$.

\sm

The bosonic subalgebra of $D(2,1;\gamma,0) \oplus D(2,1;\gamma,0)$ is $SO(2,2) \oplus SO(4) \oplus SO(4)$
and is the maximal isometry algebra of the space $AdS_3 \times \rS^3 \times \rS^3$. The transformation
$\gamma \to 1/\gamma$ corresponds to the interchange of the two $SO(3)$ subalgebras in $D(2,1;\gamma,0)$.
Accompanied by a suitable transformation on the supergravity fields, the transformation $\gamma \to 1/\gamma$
simply interchanges the two 3-spheres.   This  is physically irrelevant,  so inequivalent supergravity solutions may
be parametrized by $\g$ in the restricted  range
\bea
\label{1a2}
-1 \leq \g \leq 1\ .
\eea
Within this range, certain special values of $\gamma$ may be distinguished. For  $\g=-1$, the ${\rm AdS}_3$
factor reduces to  Minkowski space-time $\mathbb{R}^{1,2}$, and its $ SO(2,2)$ isometry  undergoes a
Wigner-Inonu contraction to $ISO(1,2)$. For  $\g =0,\infty$ on the other hand, one or the other of the two ${\rm S}^3$
decompactifies to  Euclidean $\mathbb{R}^3$,  and the corresponding $SO(4)$ factor  undergoes a contraction to $ISO(3)$.
Besides the decompactification points, two other special values in the above range are  $\g=-1/2$ and $\g=1$,
where the exceptional superalgebra $D(2,1;\g,0)$ reduces to a classical Lie superalgebra,
\bea
\label{1a3}
D(2,1;\g,0) &=& OSp(4^*|2) \hskip 0.96in {\rm at} \ \ \ \g=-1/2\  ,
\no \\
D(2,1;\g,0) &=& OSp(4|2,\mathbb{R}) \hskip 0.85in {\rm at} \ \ \  \g = 1\ .
\eea
All these facts are summarized in  the figure below.

\begin{figure}[htb]
\begin{center}
\includegraphics[width=4.5in]{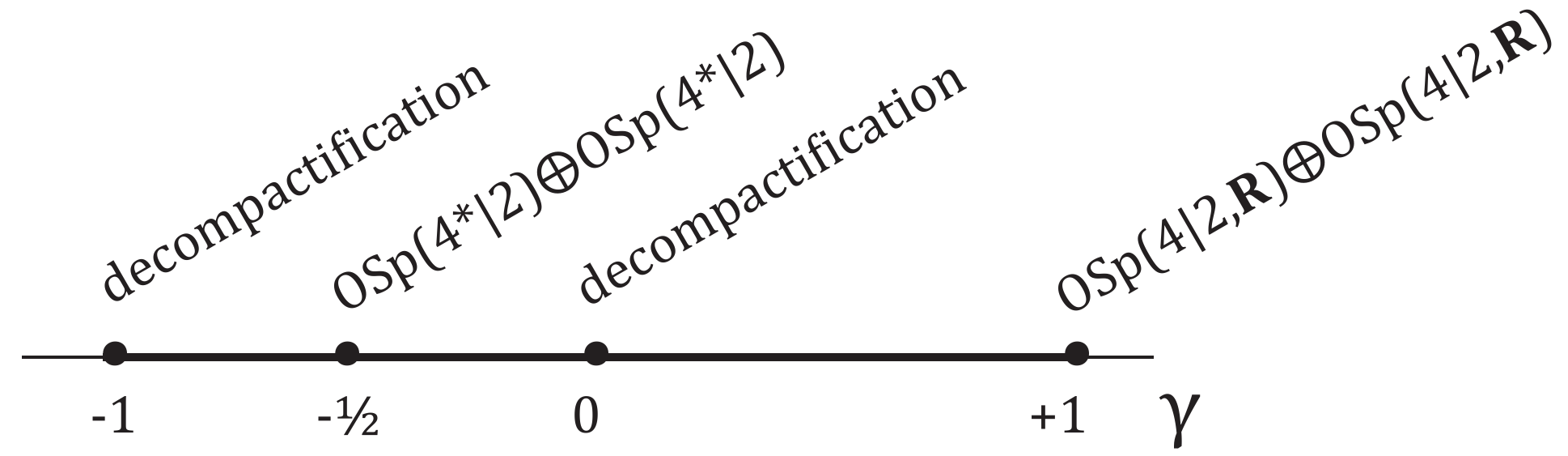}
\caption{\footnotesize The irreducible parameter space for $\g$.
Away from the  points $\g = -1, 0$, the symmetry algebra is $D(2,1;\g,0) \oplus D(2,1;\g,0)$.
At the decompactification point $\g=-1$ the ${\rm AdS}_3$ factor becomes Minkowski $\mathbb{R}^{1,2}$, while at
the point $\g=0$ one of the ${\rm S}^3$ factors becomes Euclidean $\mathbb{R}^3$. The two regions,
$\g>0$ and $\g<0$,  are thus disjoint regions of parameter space.}
\end{center}
\label{gammaparam}
\end{figure}


The solutions to M-theory, to be constructed here, will be invariant under the Lie superalgebra
$D(2,1;\gamma,0) \oplus D(2,1;\gamma,0)$, for any given value of $\gamma$.
The bosonic subalgebra $SO(2,2) \oplus SO(4) \oplus SO(4)$, which is independent of $\gamma$,
singles out the space-time of the form $\rAdS_3 \times \rS^3 \times \rS^3$, warped over a Riemann surface $\Sigma$.
The number of fermionic generators in $D(2,1;\gamma,0) \oplus D(2,1;\gamma,0)$ is 16, independently of $\gamma$,
so that these solutions are  half-BPS.

\sm

To make contact between the superalgebra structure of half-BPS solutions of M-theory  and their brane structure,
we present a brief discussion of the various known solutions. For pure M2-branes  or pure M5-branes
the near-horizon  geometries are smooth and respectively given by  ${\rm AdS}_4 \times {\rm S}^7$ with isometry
algebra $OSp(8|4,\mbR)$, and $\rAdS_7 \times \rS^4$ with isometry algebra $OSp(8^*|4)$. The respective bosonic
subalgebras,  $SO(8) \oplus SO(2,3) $ and $SO(2,6) \oplus SO(5)$, contain the bosonic subalgebra
of $D(2,1;\gamma,0) \oplus D(2,1;\gamma,0)$ for all values of $\gamma$. However, inclusion does not extend
to the full Lie superalgebras, and one has (for $\gamma \in [-1,1]$),
\bea
D(2,1;\gamma,0) \oplus D(2,1;\gamma,0) \subset OSp(8|4,\mbR) & \quad \Longleftrightarrow \quad & \gamma = 1\ ,
\no \\
D(2,1;\gamma,0) \oplus D(2,1;\gamma,0) \subset OSp(8^*|4) ~~ \, & \quad \Longleftrightarrow \quad & \gamma = -1/2\ .
\eea
Sufficiency of these conditions is evident from (\ref{1a3}), along with the canonical inclusions
$OSp(4|2,\mbR)\oplus OSp(4|2,\mbR) \subset OSp(8|4, \mathbb{R})$ and
$OSp(4^*|2)\oplus OSp(4^*|2) \subset OSp(8^*|4)$. Necessity of the conditions is non-trivial, and
was proven in \cite{D'Hoker:2008ix}. We stress that, for generic values of $\g$,  $D(2,1;\gamma,0) \oplus D(2,1;\gamma,0) $
is  a subalgebra of neither $OSp(8^*|4)$ nor of $OSp(8|4, \mathbb{R})$.

\sm

The   regular   solutions  of \cite{D'Hoker:2008qm} and \cite{janus} were obtained for  these special values of the
parameter $\g$. Each solution contain at least one region asymptotic to either ${\rm AdS}_4 \times {\rm S}^{7}$ or
to ${\rm AdS}_7 \times {\rm S}^{4}$, in which  the superconformal  symmetry is maximally enhanced. Conversely,
the simultaneous presence of M2-brane and M5-brane charges reduces the asymptotic maximal superconformal
symmetry algebras, characteristic of each pure brane, to the subalgebras $OSp(4|2,\mbR)\oplus OSp(4|2,\mbR)$ or
$OSp(4^*|2)\oplus OSp(4^*|2)$. By the holographic correspondence, the maximal symmetry algebra of the ground state
of either the pure M2- or the pure M5-brane is reduced to a subalgebra, following the customary patterns of symmetry breaking.

\sm

For generic values of $\gamma$ a crucial step  was taken in ref.\,\cite{Estes:2012vm} where it was shown that the
two maximally symmetric $\rAdS_4 \times \rS^7$ and $\rAdS_7 \times \rS^4$ solutions can be deformed to arbitrary
values of the modulus $\gamma$ (but not the sign) of $\g$. A complementary observation was made in \cite{Berdichevsky:2013ija},
where it was noted that   $\g=1$ is compatible with solutions of the supergravity equations that  are asymptotic to
${\rm AdS}_7^\prime\times {\rm S}^{4}$.  In the present paper, we shall extend and unify these results, by showing that
all known solutions associated with special values of $\gamma$ may be continuously deformed to solutions with the
same sign of $\gamma$, but different modulus $|\gamma|$. Given that, for generic values of $\gamma$,
$D(2,1;\gamma,0) \oplus D(2,1;\gamma,0) $ is  a subalgebra of neither $OSp(8^*|4)$ nor of $OSp(8|4, \mathbb{R})$,
the existence of these solutions raises a challenge to the holographic interpretation of the symmetry breaking
patterns of the corresponding superconformal field theories.


\subsection{M  theory versus type IIB }

A given M2\,$\perp$\,M5\,$\perp$\,M5(1) configuration  of Table 1  depends a priori  on a large number of parameters.
A general setup may  contain  several different
 M5-brane and ${\rm M5}^\prime$-brane stacks, located at different points $\delta$ and $\hat\delta$ in the $x^2$ direction.
 Each stack  is  characterized
 by its M5-brane charge,   and by  the net number of oriented M2-branes that end  on it.
We can  denote this data for the $i$th M5-brane stack by  $(N_5^{(i)}, N_2^{(i)},   \delta^{(i)})$, and
likewise for the $j$th  ${\rm M5}^\prime$-brane stack by $(\hat N_5^{(j)}, \hat N_2^{(j)},   \hat \delta^{(j)})$.
The microscopic world-volume theory depends on  all these discrete and continuous parameters.
Not  all of them, however,  need survive in the infrared limit and thus characterize the dual supergravity geometry.
\sm

A similar situation is  encountered in type-IIB string theory,
 for  the   configuration
of D3-branes, D5-branes and NS5-branes shown in Table 2.
  A given  configuration is characterized by  the data $(N_5^{(i)}, N_3^{(i)},   \delta^{(i)})$
   that specifies the 5-brane and 3-brane charges and the position
along $x^3$
of the $i$th NS5-brane stack, and by the analogous data
$(\hat N_5^{(j)}, \hat N_3^{(j)},   \hat \delta^{(j)})$
  for the $j$th D5-brane stack.
 There may be  in addition $n$ D3-branes that  intersect
all of the 5-branes without ending on any one of them, in which case
 one must  also  specify   the asymptotic
values of the dilaton  field.
\vskip 1mm

\smallskip
\begin{table}[htdp]
\begin{center}
\begin{tabular}{|c||c|c|c|c|c|c|c|c|c|c| } \hline
 & 0 & 1 & 2 & 3 & 4 & 5 & 6 & 7 & 8 & 9
\\ \hline \hline
D3 & $\star$ & $\star$ & $\star$ &$\star$ &&&&&&
\\ \hline
D5 & $\star$ & $\star$ &  $\star$ && $\star$ & $\star$ & $\star$ &  &&
\\ \hline
NS5 & $\star$ & $\star$ & $\star$&  &&& & $\star$ & $\star$ & $\star$
\\ \hline
\end{tabular}
\end{center}
\label{table2}
\caption{\footnotesize  The quarter-BPS D3$\perp$D5$\perp$NS5(2)  configurations of type-IIB string theory discussed
in the text. They  can be  related to
 those  of Table 1 by T-duality of  the coordinate $x^2$ and lift to eleven dimensions.  The near-horizon
 geometries of such dual configurations are not, however,  related in any simple way.}
\end{table}

 The low-energy  theory on the D3-branes
 is  a defect field theory  consisting of  an ${\cal N}$=4 supersymmetric   gauge theory in three dimensions,
coupled, when $n\not=0$,  to  four-dimensional   ${\cal N}$=4   super-Yang Mills
 theory \cite{DeWolfe:2001pq,Erdmenger:2002ex}
\cite{Gaiotto:2008sa,Gaiotto:2008sd}.
    It has been  conjectured by Gaiotto and Witten
  \cite{Gaiotto:2008ak}
that under suitable conditions on the brane charges, such field theories
 flow to non-trivial strongly-coupled fixed points in the infrared. By the holographic principle,
  these  should have in turn
 dual  solutions of the  form $({\rm AdS}_4\times {\rm S}^2\times {\rm S}^2)\ltimes \Sigma$, which
 realize geometrically the   ${\cal N}$=4  superconformal symmetry
 $OSp(4,4)\supset SO(2,3)\times SO(3)\times SO(3)$.\footnote{This setup
has been actually  proposed  as  a possible realization  of  (locally) localized gravity in string theory \cite{KR2}.  For
more recent analyses of  this idea  see   \cite{Aharony:2003qf,Bachas:2011xa}.}
  The  precise correspondence   has been derived in
\cite{Aharony:2011yc,Assel:2011xz,Assel:2012cj}.  It was also shown in these references
 that the  positions of the 5-brane stacks are not free parameters in the infrared, but rather prescribed functions of the brane
charges.
When there are no D3-branes extending
  to  infinity in the $x^3$ direction,
 these solutions provide a
 rich  class of holographic duals to  strongly-coupled three-dimensional   ${\cal N}$=4  superconformal theories
 \cite{Assel:2011xz,Assel:2012cj}.
 \sm

It is   natural to ask  whether similar conclusions can be drawn    for  the M2\,$\perp$\,M5\,$\perp$\,M5(1)  configurations
considered here. One would like, for instance,  to know whether the field theory on the M2-branes
 admits  supersymmetric interfaces, other than Janus, which flow
 to non-trivial fixed points in the infrared.\footnote{For an analysis of supersymmetry-preserving
 boundary conditions on semi-infinite M2-branes, see for instance \cite{Berman:2009kj,Berman:2009xd}.} 
 Does the existence of such fixed points impose
 conditions  on the brane charges  analogous to those of
  ref.\,\cite{Gaiotto:2008ak}? And are there  new
 AdS$_3$ solutions   dual  to strongly-coupled  two-dimensional  CFTs ?

We will see in this paper that the two problems have close similarities  but also some important differences.\footnote{In
   flat Minkowski space-time  the type-IIB and   M-theory setups are related by
  T-duality.
     The T duality requires, however,  that the  D3 branes  be wrapped around a circle, and this
changes completely the infrared behavior of their world-volume theory.
  At the level of the supergravity solutions, this corresponds to  T-dualizing  an  AdS$_4$
  coordinate, which gives a singular solution in M theory.
  }
  The construction of the type-IIB
   solutions can be reduced to the choice of two real harmonic functions $(h_1,h_2)$,  which must be positive in the interior
of $\Sigma$ and obey  (Dirichlet, Neumann) or (Neumann, Dirichlet) conditions at generic points of the boundary $\partial\Sigma$
 \cite{D'Hoker:2007xy,D'Hoker:2007xz}.
  Likewise, the construction of the M-theory solutions can be reduced to the choice of two functions $(h, G)$,
  where $h$ is real harmonic, and $G$ is a complex function that obeys a simple linear equation
   \cite{D'Hoker:2008wc,Estes:2012vm}. Regularity requires  Dirichlet conditions for both functions in $\partial\Sigma$  and,
   as we will show in the following section, two positivity conditions,  $h>0$ and $ \g( G\bar G -1)>0$,  in the interior.

   The local analysis of these problems around a point in $\Sigma$  gives  "Lego" pieces
   which must be put together  to construct global solutions.
    In the the type-IIB problem there are asymptotic  near-horizon regions  of flat D3, D5 and NS5 branes, but also
    local solutions corresponding to  five-branes with (highly-curved) AdS$_4\times$S$^2$
   world-volumes. It is furthermore  possible to cap-off smoothly an asymptotic region
   by taking the corresponding brane charge to zero,   leaving
    behind a coordinate singularity \cite{Aharony:2011yc,Assel:2011xz}.  We will find similar Lego pieces in the
    M theory problem, but the rules  for putting these ingredients together are different.
   We will see, for example,  that the only solution
   with a two-dimensional conformal boundary
    is AdS$_3\times$S$^3\times$S$^3\times$E$_2$, in contrast with the rich set of type-IIB solutions
   that are dual to 3-dimensional superconformal theories.

   \sm

   Another  close parallel, once again with   noteworthy differences, can be drawn
   between some self-dual string  solutions  in the present work,
  and the  half-BPS solutions  in type-IIB theory that are dual to Wilson
 line operators in $\cN=4$ super-Yang-Mills \cite{D'Hoker:2007fq}.  The space-time of these type-IIB
solutions  consists of
 $\rAdS_2 \times \rS^2 \times \rS^4$ warped over a Riemann surface $\Sigma$ with
boundary, and they have a single asymptotic $\rAdS_5 \times \rS^5$ region.
 The solutions exhibit alternating
 $\rS^3$ and $\rS^5$ cycles, whose bases are  open curves on $\Sigma$  that start and end
on segments of the boundary at which
  the $\rS^2$ or the $\rS^4$ spheres, respectively, shrink to a point.
 A pictorial representation
 of the topology of these solutions was given in Figure 1 of \cite{D'Hoker:2007fq}.
 \sm

 There is a completely analogous story in the self-dual string solutions of
 M-theory for $\gamma <0$. The space-time is now $\rAdS_3 \times
\rS_2^3 \times \rS_3^3$
 warped over a Riemann surface $\Sigma$, and there is  a single
 asymptotic $\rAdS_7  \times \rS^4$ region.
 If one replaces   the $\rS^2$ of
type IIB by  $\rS_2^3$,
 and the $\rS^4$ of type IIB by $\rS_3^3$, then the  alternating type-IIB cycles $\rS^3$ and
$\rS^5$  map to two alternating kinds of $\rS^4$ spheres, carrying M5-brane and  M5$^\prime$-brane
charge.
 With these replacements, Figure 1 of \cite{D'Hoker:2007fq} provides the correct
pictorial representation of the topology of our solutions in M-theory.
 In both cases, the supergravity data determines a Young diagram, and an irreducible representation
 of $SU(N)$ along the lines explained
  in refs.\,\cite{Yamaguchi:2006te,Gomis:2006sb,Okuda:2008px}. We will review  the details of this argument
  in section \ref{stringsols} of the present work.

\section{Reduction of the BPS Equations and Regularity}
\setcounter{equation}{0}
\label{sec:2}


In order to realize the bosonic symmetry $SO(2,2) \times SO(4) \times SO(4)$
the  space-time manifold must be  AdS$_3 \times$S$^3  \times $S$^3$ warped over a two-dimensional base
space $\Sigma$. Since $\Sigma$ inherits an orientation and a metric from supergravity,
it  is automatically a Riemann surface, endowed with a complex structure.
The reduction of the Killing spinor equations on such space-time manifolds has been carried out
in refs.\,\cite{D'Hoker:2008wc} and \cite{Estes:2012vm}  (see also \cite{Yamaguchi:2006te,Lunin:2007ab} for
earlier work).
 In this section we will review and simplify the results of these references.

After all the dust has settled, the background bosonic fields can be
expressed in terms of the following data: the parameter $\g$, and two functions on $\Sigma$, namely
 a real harmonic function  $h$,  and  a complex function  $G$ which
satisfies a first-order partial differential equation. In  a  system
of local complex  coordinates $(w, \bar w)$   the two basic equations read
\bea
\label{1a1}
\p_{\bar w}\p_w h=0\ , \qquad  2 h\, \p_w G =  (G + \bar G)\, \p_w h\ .
\eea
Note that the $G$  equation is also linear, so a superposition of solutions
with real coefficients is also a  solution.
Furthermore, both equations  are preserved by conformal reparametrizations
of the Riemann surface $\Sigma$.

  The reduction of the half-BPS solutions to the data ($h,G$),
as well as the derivation of suitable regularity conditions on this data, was carried out in \cite{D'Hoker:2008wc}
for the special values $\g =-1/2$ and $\g= 1$. In  \cite{Estes:2012vm}  it was shown that
the data ($h$, $G$)
still provides the complete reduction of the half-BPS equations,  for general   values of   $\g$.
 The regularity conditions on $h$,  namely $h >0$ in the interior of $\Sigma$ and $h=0$ on $\p \Sigma$ for surfaces with boundary,
 continue to be those of  reference \cite{D'Hoker:2008wc}.
But the regularity conditions on $G$ proposed in \cite{Estes:2012vm} depend  non-trivially  on
the parameter $\g$. Furthermore, the expressions for the flux fields in terms of $h$ and $G$,  
given in either \cite{D'Hoker:2008wc} or \cite{Estes:2012vm},
are sufficiently complicated so  that   the M2-brane and M5-brane charges of the solutions could be evaluated
only in  the simplest cases.

\sm
The new result in this section, compared to the above references, is a simple redefinition of  the function $G$
that leaves (\ref{1a1}) invariant,  while
 greatly simplifying   the conditions needed for regularity of the supergravity solution.
In terms of the redefined $G$ these conditions read
 \bea
\label{1a4}
h >0\quad {\rm and}  \quad  \g (\, |G|^2-1) >0 \hskip 0.5in \hbox{  in  the {\it interior}  of} \ \   \Sigma \ ,
\eea
 \bea
\label{1a5}
h =0 \quad {\rm and}\quad  G \in \{ +i, -i \} \hskip 0.6in \hbox{ on the {\it boundary}} \ \  \p \Sigma\ .
\eea
Note that  the   reduced problem only depends on the sign of $\g$, but not on its modulus.
An immediate consequence of this result   is the following general
\medskip

\underline{Theorem}:  All solutions to the half-BPS equations come in   families
                                                  parametrized by $\vert\g\vert$.

 \medskip

In the remainder of this section we  review the reduction of the supergravity equations to the data $(\g, h, G)$,
and we derive the above regularity conditions. Our redefinition of  $G$ also leads to simple, calculable
expressions  for the brane charges, but we   postpone these calculations  to section \ref{sec:3}.


\subsection{Invariant ansatz for supergravity fields}
\label{sec:2b}

11-dimensional supergravity \cite{Cremmer:1978km}
 contains the metric $g_{MN}$ and the real-valued field strength $F_{PQRS}$,
which is often recast in terms of a 4-form $F= F_{PQRS} dx^P\wedge dx^Q \wedge dx^R \wedge dx^S/24$.
Here, indices range over the 11 dimensions of space-time, $M,N,P,Q,R,S=0,1, \cdots , 9, 10$.
The field strength $F$ satisfies the Bianchi identity $dF=0$, or equivalently derives from a real 3-form $C$
by the relation $F=dC$.  The bosonic part of the action is given by
\bea
\label{2a1}
S={1\over 2 \kappa_{11}^2} \int d^{11}x \sqrt{-g} \Big ( R-{1\over 48} F_{MNPQ}F^{MNPQ} \Big )
-{1\over 12 \kappa_{11}^2} \int C \wedge F \wedge F\ ,
\eea
where $\kappa_{11}^2$ is the 11-dimensional Newton constant.
The Einstein equations read
\bea
\label{2a2}
R_{MN}-{1\over 12} F_{MPQR}F_N^{\;\; PQR} +{1\over 144} g_{MN} F_{PQRS}F^{PQRS}=0\ ,
\eea
while the  field equation for $F$ takes the form
\bea
\label{2a3}
d*F + {1\over 2} F \wedge F=0\ .
\eea
Finally, the  BPS equations can be written as
\bea
\label{2a4}
\nabla_M \ep +{1\over 288} \Big(\Gamma_M^{\;\; NPQR}
- 8 \delta_M{}^N \Gamma^{PQR} \Big) F_{NPQR} \, \ep=0\ ,
\eea
where $\ep$ is an eleven dimensional Majorana spinor, and $\nabla _M$ is
the covariant derivative with respect to the Levi-Civita spin connection associated with the metric $g_{MN}$.

The $SO(2,2) \times SO(4) \times SO(4)$ invariant reduction of the supergravity
fields on AdS$_3 \times$S$^3  \times $S$^3$ warped over $\Sigma$ is given by
the following expressions:
\bea
\label{2b1}
ds^2 & = & f_1^2 \;ds_{AdS_3}^2+ f_2^2 \;ds_{S^3_2}^2 + f_3^2 \; ds_{S^3_3}^2+ \rho ^2 |dw|^2\ ,
\no \\
C & = & b_1 \hat e^{012} +  b_2 \hat e^{345} + b_3 \hat e^{678}\ ,
\no \\
F & = & d b_1 \wedge \hat e^{012} + d b_2 \wedge \hat e^{345} + d b_3 \wedge \hat e^{678}\ .
\eea
Here  $ds_{AdS_3}^2$,  $ds_{S^3_2}^2$, and $ds_{S^3_3}^2$ are the metrics of unit-radius
(pseudo)spheres,
invariant under the action of $SO(2,2)$, $SO(4)$ and $SO(4)$, and
$\hat e^{012}$, $\hat e^{345}$ and $\hat e^{678}$ are the respective volume forms.
The metric on $\Sigma$ is expressed in terms of a system of local complex coordinates $w,\bar w$.
The metric factors $f_1, f_2, f_3, \rho$ are real-valued positive functions on $\Sigma$, and
 $db_1, db_2, db_3$ are real one-forms expressed in terms of locally-defined  potentials $b_1,b_2,b_3$.
\sm

  The invariant ansatz for the Killing spinor $\ep$ reads
  \bea
  \ep = \sum_{\eta_j=\pm}  \chi^{\eta_1, \eta_2, \eta_3} \otimes \zeta_{\eta_1, \eta_2, \eta_3}
  \eea
where $\chi^{\eta_1, \eta_2, \eta_3}$  is the
tensor product  of the three 2-component Killing spinors on AdS$_3$, S$_2^3$ and S$_3^3$, while
the eight corresponding $\zeta_{\eta_1, \eta_2, \eta_3}$ are four-component spinors.
These are subject to a reality condition that is inherited from the Majorana condition in eleven dimensions.
Furthermore, the symmetries of the BPS equation relate all the eight spinors to each other, leaving one
independent spinor which can be parametrized  as follows:
 \bea\label{zeta}
\zeta_{+++} \ =\  \left(
  \begin{matrix}  \bar\alpha \\  -\bar\beta \\\alpha  \\  \beta  \end{matrix}
   \right)\ ,
 \eea
where
$\alpha$ and $\beta$ are complex-valued functions on the base-space $\Sigma$.
The other $\zeta_{\eta_1, \eta_2, \eta_3}$ can be obtained by rotations or boosts of the
(pseudo-)spheres AdS$_3$, S$_2^3$ and S$_3^3$.


\subsection{The reduced data ($\gamma, h, G)$ }
\label{sec:2c}

 Projecting the BPS equation in the directions of the symmetric spaces gives six algebraic conditions
for $\alpha$ and $\beta$ that  involve the metric and 4-form fields.  Three of these can be solved   for the
(pseudo)sphere radii with the   result
 \bea\label{metricKill}
f_1 = {1\over c_1} (\vert\alpha\vert^2 + \vert\beta\vert^2)\ , \quad
f_2  = {1\over c_2} (\vert\beta\vert^2 - \vert\alpha\vert^2)\ , \quad
f_3  = {i\over c_3} ( \alpha\bar\beta  - \bar\alpha  \beta )\ ,
\eea
where the $c_j$ are integration constants subject to the constraint $c_1+c_2+c_3 =0$.
Since an overall multiplicative factor can be reabsorbed in $\alpha$ and $\beta$, only one of these
constants has physical significance. It can be chosen to be
  the ratio  $c_2/c_3 \equiv \g$. Note that $c_1$ must be positive, but $c_2, c_3$ and $\g$ can
  in principle have either sign.
\sm


Much of the time we will work with the parameters $c_i$. Sometimes it will be convenient
to fix by a rescaling $\vert c_2c_3\vert = 1$, in which case these parameters can be written as
\bea
c_1 = \g^{1/ 2} + \g^{-1/ 2}\ , \quad c_2 =  -\g^{1/ 2}, \quad c_3 = -\g^{-1/ 2}\qquad (\g>0)\, ; \no
\eea
\bea
c_1 = \vert\g\vert^{-1/ 2} -  \vert\g\vert^{1/ 2}\ , \quad c_2 =  \vert\g\vert^{1/ 2}, \quad c_3 = -\vert\g\vert^{-1/ 2}\qquad (0> \g>-1)\, ; \no
\eea
\bea\label{fixedci}
c_1 = \vert\g\vert^{1/ 2} -  \vert\g\vert^{-1/ 2}\ , \quad c_2 =  -\vert\g\vert^{1/ 2}, \quad c_3 = \vert\g\vert^{-1/ 2}\qquad (-1> \g )\ .
\eea
This makes in particular manifest the fact that $\g=0, -1$ correspond to decompactification points, as anticipated
in Figure 2.


\sm

 The remaining BPS equations reduce to  conditions on the flux field, and to
  two first-order non-linear differential equations for $\alpha$ and $\beta$.
  We refer the reader to   \cite{D'Hoker:2008wc,Estes:2012vm} for details.
  The upshot of the analysis in these references  is that the  non-linear BPS equations
  for $\alpha$ and $\beta$ can be further
  reduced to linear equations for two auxiliary functions on $\Sigma$, $h$ and $G$. The function $h$ is
  real-valued and harmonic, whereas $G$ is a complex-valued  function
  which obeys a single remaining
equation,\
\bea
\label{2c1}
\p _w G = \half (G + \bar G) \p _w \ln h\ .
\eea
The equation is expressed in local complex coordinates $w,\bar w$ on $\Sigma$. A remarkable property of this equation,
discovered in \cite{Estes:2012vm}, is its independence of the parameter $\gamma$. Another important property is
its invariance under conformal reparametrizations of $\Sigma$.
\sm

 All supergravity background fields can be expressed in terms of the reduced data $(\g, h, G)$, and conversely
 any half-BPS solution in the above ansatz corresponds to a unique choice of this data.
Note that, for given $h$,  since  (\ref{2c1})  is a linear equation, any   linear combination  of  solutions with real  coefficients
is also a solution.  Global regularity imposes  however, as we will see, stringent  constraints on   $G$ which are
not satisfied by general linear combinations of admissible solutions.
Enforcing these regularity conditions  is, in fact,  the
main technical obstacle to finding global  solutions of the BPS equations.


\subsection{Metric factors}
\label{sec:2d}

 The conditions of global regularity given in  reference \cite{Estes:2012vm} have  a complicated  dependence on
 the parameter $\g$. We have found that one can simplify these conditions by exploiting
the invariance of (\ref{2c1})  under the linear transformations  $G \to i a  + b  G$ and $h\to \lambda h$,
where $a,b$ and $\lambda$ are real constants.   In appendix \ref{sec:A} we use this freedom   to  define
a  more convenient pair of functions,  $h$ and $G$, which   still obey   the same linear equations.
In terms of these redefined functions the scale factors in the metric (\ref{2b1})   read
\bea
\label{2d2}
f_1 ^6 =  { h^2 W_+ W_- \over c_1 ^6 \, ( G \bar G -1)^2}\ ,
& \hskip 1cm &
\rho^6 =  { |\p_w h |^6  \over   c_2^3 c_3^3\,  h^4 }  ( G \bar G -1) W_+ W_-\ ,
\no \\ && \no \\
f_2 ^6 =  { h^2 ( G \bar G -1) W_- \over c_2 ^3 c_3^3\,  W_+^2}\ ,
&&
f_3 ^6 =  { h^2 ( G \bar G -1) W_+ \over c_2 ^3 c_3^3\, W_-^2}\ ,
\eea
where the auxiliary functions  $ W_\pm$  are given  by
\bea
\label{2d1}
W_+ =  | G+  i  |^2 + \g  \, ( G \bar G -1)\ , \quad
{\rm and}\quad
W_- =  | G-  i  |^2 + \g ^{-1}  ( G \bar G -1)\ .
\eea

\sm
From these expressions  one easily derives a  useful identity
  relating  the product of the three pseudo-sphere radii   to the harmonic function $ h$,
\bea
\label{6a5}
  c_1 c_2 c_3 \, f_1 f_2 f_3  \, = \, \sigma    h\  ,
\eea
where $\sigma = \pm 1$, chosen so as to allow $h \geq 0$. In addition,  from the parametrization
 (\ref{metricKill}) of the radii in terms of components of the Killing spinor
  one finds  the following   inequality,
 \bea\label{bAdSs}
  (c_1 f_1)^2 \geq  (c_2 f_2)^2 +  (c_3 f_3)^2\ .
 \eea
Both  \eqref{6a5} and  \eqref{bAdSs}
  will be very  useful in the analysis of  the singularities of the data $(\g, h, G)$,  which  determine
  the topological characteristics  of the solutions.
\sm
\setcounter{footnote}{0}

An alternative useful form of the eleven-dimensional
 supergravity metric is  obtained by the following   Weyl rescaling:
$$
ds^2 =
 e^{2A} ( \hat f_1^2 \;ds_{AdS_3}^2+ \hat f_2^2 \;ds_{S^3_2}^2 + \hat f_3^2 \; ds_{S^3_3}^2+ \hat \rho ^2 |dw|^2 )
$$
\vskip -6mm
\bea\label{weyl1}
{\rm with}  \qquad e^{6A} = {h^2\, (G\bar G -1)W_+W_-\over  c_2^3 c_3^3}\ .
\eea
The  rescaled metric factors are given by
\bea\label{weyl2}
 \hat f_1^{-2} =  (\g +{1\over \g}+ 2) (G\bar G -1)\ , \quad \hat f_2^{-2} = W_+ \ , \quad
 \hat f_3^{-2} = W_-\ , \quad \hat \rho^2 = {\p_w h \p_{\bar w}  h\over h^2}\ .
\eea
Using as local  coordinate  $2 z =  -\tilde h + i    h$, where $\tilde h$ is the
  the dual harmonic function,\footnote{Defined so that $z$ is a holomorphic function of $w$}
   shows that  $\hat\rho^2 dwd\bar w$  is the constant-negative-curvature metric,
    \bea
  d\hat s^2_\Sigma =   {d\tilde h^2 + dh^2\over 4 h^2}\ .
  \eea
Furthermore, the three rescaled  radii  satisfy, at all points on $\Sigma$,  the condition
\bea
\hat f_2^{-2} +  \hat f_3^{-2}  - \hat f_1^{-2}  =  4\ .
\eea
Thus, solutions of the BPS equations   induce  a map from (a domain of) the hyperbolic plane to the
above  SO(1,2) invariant  hyperboloid.\,\footnote{Note also that
as  $\vert \g\vert$ ranges from zero to infinity, the   inverse rescaled radii trace a trajectory on the
hyperboloid. It is unclear to us whether there is some  mathematical
 reason behind this observation.}


\subsection{Regularity conditions}
 \label{sec:25}

The regularity conditions on $(\g, h,G)$  may be
read off from the dependence of the metric factors on these data. In brief, these
conditions require that the supergravity fields of the solution be smooth everywhere.
 This does not of course exclude singularities that are  artifacts of the parametrization of the surface,
 implicit in our choice of ansatz.

In particular, if   $\Sigma $ has a boundary,
points on the boundary  do not map  generically to points on any
boundary of the eleven-dimensional space-time. Rather, generic points on $\p \Sigma$
map to interior points of the supergravity space-time where one or the other of the
$S^3$ spheres is pinched to zero radius. This pinching should be viewed as arising from
 the radial slicing of a regular manifold which is locally diffeomorphic
 to $\mathbb{R}^4  \simeq \mathbb{R}^+\times$S$^3$. Put differently, the coordinate normal to
 $\p \Sigma$ must be the radial  coordinate in a local polar
parametrization of  some open region in $\mathbb{R}^4$.
\sm

Now the identity (\ref{6a5}) says that   $h$ is proportional to the sphere radii, so
 we must have $h=0$  on  the boundary of $\Sigma$. Furthermore, for the other radii to stay
finite, we must   require that   either $W_+$ or $W_-$  also vanishes, and that $\vert G \vert = 1$.
This implies that $G= \pm i$.
Therefore, the  boundary conditions on the reduced data read, for all $\g$,
 \bea\label{bnryC}
 h = 0 \ ; \qquad G= \pm i\qquad {\rm on\ the \ boundary\ of }\ \Sigma\ .
 \eea

These conditions are not only necessary, they are also locally sufficient. To see why,  choose again as coordinate
$z= x+ iy$ such that  $h= -iz + c.c. = 2y$,  and $\partial\Sigma$ is the real-$z$ axis.
Assume furthermore  that
$G=-i$ on the boundary, the case $G=i$  can be treated similarly. As a result, we may set  $W_- \simeq 4$
in  the expressions (\ref{2d2}) for   $f_3$ and $\rho$, which
 leads to
 the following metric,
\bea\label{bnryds}
ds^2 \simeq f_1^2 ds^2_{AdS_3} + f_2^2 ds^2_{S^3} +
\left[ {4 (\bar G G - 1)  W_+ \over  c_2^3 c_3^3 h^4} \right]^{1/3} (dx^2 + dy^2 + y^2 ds^2_{S^3})\ .
 \eea
Thus  the metrics on $\Sigma$ and on the second sphere combine
 nicely into a polar parametrization of $\mathbb{R}\times\mathbb{R}^4$.
To avoid singularities we only need that the scale factor of this $\mathbb{R}^5$ patch, as well as
the factors $f_1$ and $f_2$, approach   finite values when $y\to 0$.

This is indeed generically the case. Setting
  $G = \Delta_1 + i( \Delta_2 -  1)$,   where the $\Delta_i$ vanish at $y=0$, and keeping only the
leading-order terms gives
\bea
(\bar G G-1) \simeq \Delta_1^2 - 2\Delta_2\ ,\quad
\quad W_+ \simeq   (1+\g) \Delta_1^2  -2 \g\Delta_2\   .
\eea
Furthermore, from the real part of equation \eqref{2c1}   we find $\partial_x \Delta_1 + \partial_y \Delta_2 = 0$.
Assuming $G$ to be  real-analytic near the boundary implies
 then the following leading behavior for small $y$,   $\Delta_2 \sim  \Delta_1^2 \sim  y^2$.
With this leading behavior all scale factors in \eqref{bnryds} approach finite values at $y=0$, as required.
\footnote{This analysis is valid at generic points on the boundary of $\Sigma$. A finer analysis
 is needed at special points where
  the leading corrections to $G+i$   vanish, or if $h$ and/or $G$ are singular.}
\sm

Note that the harmonic function $h$ cannot vanish at isolated points in the interior of
 $\Sigma$. It may vanish on a one-dimensional locus that  divides  the original $\Sigma$  into several
  disconnected pieces, in each of which $h$ has a definite sign. Each piece corresponds to disjoint
  space-time manifold  in eleven dimensions.
We  can therefore  restrict attention to a single  connected piece, and choose $h>0$ (if not
we trade $h$ for $-h$).

\sm

  Inspection of  the expressions (\ref{2d2}) for the metric factors shows, furthermore, that $W_+$, $W_-$
  and $\g (G\bar G -1)$ must all be positive in the interior of $\Sigma$. This is required for positivity of these
  metric factors, if one recalls that $c_2 = \g c_3$.  Actually, the last of the above three conditions implies
  immediately the other two, so the regularity conditions in the interior of $\Sigma$ can be summarized as follows:
   \bea\label{BulkC}
h>0\ ; \qquad   \g(\vert G\vert^2- 1) > 0   \qquad {\rm in\ the\ interior\ of} \ \Sigma  .
\eea
This is precisely the condition we anticipated in \eqref{1a4}.
 Let us stress once more
  that,  in contrast to the regularity conditions in  \cite{Estes:2012vm}, the above condition  only involves  the sign of
the parameter $\g$.
  \sm

  One final question concerns potential singularities of $G$ and $h$.
  This question will be discussed in section \ref{sec:4},
   but  we here note by anticipation that singularities on   $\p \Sigma$
 do indeed arise in the  vacuum space-time solutions AdS$_4 \times$S$^7$ and AdS$_7 \times$S$^4$.
 These singularities are coordinate artifacts,
and mathematically provide the extra non-compact direction  in these higher-simensional  AdS  spaces.
\sm

 {In conclusion}, the
  problem of finding  $D(2,1; \gamma) \! \oplus \! D(2,1;\gamma)$
  invariant   solutions of M theory   can be  reduced to the simpler mathematical
problem of finding solutions to
  equation   (\ref{2c1}),  subject to the regularity conditions   (\ref{bnryC}) and (\ref{BulkC}).
 For the reader's convenience, we have already summarized this reduced problem in
 a single page, c.f.  the (in)equalities  \eqref{1a1},  \eqref{1a4} and  \eqref{1a5} of the introduction to this section.



\section{Gauge Potentials and Charges}
\setcounter{equation}{0}
\label{sec:3}

Generally, our supergravity solutions  can have three different types of  charge:
electric charge carried by M2 branes, and magnetic charges carried by the M5 and M5$^\prime$ branes
of Table 1.\footnote{We use   "electric" for the  M2-branes and the 7-form flux that they source,
 and  "magnetic" for the  M5-brane sources and their  corresponding 4-form fluxes.}
In this section we will evaluate the gauge potentials and associated charges   in terms of  the
reduced data $(\g, h, G)$ of the solutions. As we will see, the final expressions are considerably simpler than the
expressions for the flux fields in \cite{Estes:2012vm}, and they will allow us  to calculate the charges for
all exact solutions.

The  cohomologically non-trivial  pieces of the potentials,  which
 are responsible for the brane charges, have a particularly simple dependence on $\g$.
From this dependence we  will see that $\vert \g\vert$ controls the ratio of the two types of M5-brane charge.
When both types of charges are turned on, $\gamma$ is   a rational parameter rather than a continuous modulus
of the solutions.  We also  point out, in passing,  the  $\mathbb{Z}_2$ symmetry that  exchanges the roles of
the two S$^3$, and inverts  $\g\to  1/\g$.   By virtue of this symmetry,
the   inequivalent supergravity solutions are parametrized by   $\g$  in the
interval $[-1, 1]$.


\subsection{The "magnetic"  three-form potential }
\label{sec:2e}

We first consider  the 3-form potential $C$, whose reduction   is given by the ansatz  (\ref{2b1}).
The components $b_i$ in this ansatz can be expressed  readily
in terms of the  harmonic function $\tilde h$ dual to $h$, which obeys
\bea\label{dualharm}
\p_w \tilde h = - i \, \p _w h\ ,
\eea
and a  real auxiliary  function $\Phi$ which is  defined in terms of $h$ and $G$   by the relation
\bea
\label{2e3}
 \p_w \Phi \, =\, \bar G\,  \p _w h \ .
\eea
The fact that such a $\Phi$ indeed exists  locally can be seen by  choosing as local coordinate
$w= - \tilde h + ih \equiv x+iy$, so that the basic equation (\ref{2c1}) takes the form
\bea
 y  (\p_x -  i\p_y)( {\rm Re}G + i\, {\rm Im}G) = - i\, {\rm Re}G\ .
\eea
The real  part  of this equation can be integrated  in terms of the function $\Phi$,
\bea
\p_x {\rm Re}G +  \p_y {\rm Im}G = 0\ \Longleftrightarrow\ G =  {1\over 2} (\p_y\Phi -  i \p_x \Phi)\,  =\, - i \p_{\bar w} \Phi\ .
\eea
Inserting this  in the imaginary part of the equation leads to a second-order partial-differential equation
for $\Phi$, or more conveniently for the function $\Psi = y^{-1} \Phi$, \cite{D'Hoker:2008wc}
\bea\label{eqfor Psi}
\left ( \p_x^2  + \p_y^2  +{1 \over y} \p_y  -{ 1 \over y^2} \right ) \Psi  =0\ .
\eea
 In order not to stop the flow of the text, we give the details of  the
 computation of $b_i$  in terms of $\tilde h$ and $\Phi$ in  Appendix \ref{sec:B}. The final result  can be written as
\bea
\label{2e1}
b _i = { \nu _i \over c_i ^3}  ( b_i ^s + b_i ^c ) \ , \
\eea
where the $\nu_i$ are simple signs that satisfy
$\nu_1\nu_2\nu_3=-\sigma$,   and $b_i^s$, $b_i^c$ are given by the following expressions:
\bea
\label{2e2}
b_1 ^s =   - { h (G + \bar G) \over 1-G \bar G}  \ , ~~ & \hskip 1cm & b^c _1 = b_1 ^0 + (2 + \gamma + \gamma^{-1}) \Phi - (\gamma - \gamma^{-1}) \tilde h\ ,
\no \\
b_2 ^s =   - \g \, { h (G + \bar G) \over W_+} \ , && b^c _2 = b_2 ^0 +\gamma (\Phi - \tilde h)\ ,
\no \\
b_3 ^s =   + { 1 \over \g}  \, { h (G + \bar G) \over W_-} \ , && b^c _3 = b_3 ^0 - \frac{1}{\gamma} (\Phi + \tilde h)\ .
\eea
  Here $b_i^0$ are arbitrary integration constants  which may be viewed   as a  residual   freedom of gauge.
The  superscripts $s,c$ denote respectively the single-valued and the
cohomological contributions to the gauge potential.
\sm

  The logic behind this nomenclature is as follows. In a non-singular gauge, the potentials  $b_2$  and $b_3$
  should vanish on those parts of $\p\Sigma$  where  the  corresponding sphere,   $S_2^3$
  or $S_3^3$,   shrinks to a point. More generally, however, the potentials   take constant values on
  these segments of $\p \Sigma$,  and the differences between these values measure non-trivial M5-brane charges.
  This is a higher-dimensional generalization of the way in which one  computes  the magnetic charge
  as the difference of  the  monopole field, $A = g_m (1 \pm \cos\theta)d\phi/4\pi$,  between the north and south poles
  of the 2-sphere. The logic behind the break-up \eqref{2e1} is that
   the  $b_j^s$ vanish on the $f_j=0$ segments, so they  make no contribution to the brane charges.
These
    come entirely from the
  cohomologically non-trivial terms,  $b_i^c$.


An important remark is that  the auxiliary functions $\Phi$ and $\tilde h$ are independent of $\g$,
so that  $b_2^c$  and $b_3^c$ are proportional, respectively, to $\g$ and $\g^{-1}$.
Since the 5-brane charges are given by integrals of  $b_i^c/c_i^3$,
the  effect of  rescaling
 $\vert\g\vert$ is   to rescale the ratios of M5-brane to  M5$^\prime$-brane  charges by the same factor,
 while keeping their products fixed.
 This  can be seen  with the help of the standardized expressions \eqref{fixedci} for the $c_i$.
 Note however that  $b_2^s$  and $b_3^s$ have a more involved dependence
on the parameter $\g$, so the full 4-form magnetic  fluxes  are not simply rescaled in the same way as the
5-brane charges.


\subsection{The  "electric" six-form potential}
\label{sec:2f}

In order to calculate M2-brane charges, we also need
  the dual 6-form potential  whose exterior derivative is a 7-form field strength.
Here one encounters a well-known problem (see for instance \cite{Marolf:2000cb}),
namely that  the naive Poincar\'e dual   $*F$  is not closed,
because of the   Chern-Simons interaction. One can construct a closed 7-form $d \Omega$ at the expense
of gauge invariance as follows:
\bea
\label{2f1}
d \Omega = * F + \half C \wedge F\  ,
\eea
where $\Omega$ is the   6-form potential.
Note that the field equation (\ref{2a3}) implies  that
  $d\wedge d \Omega=0$, so that $\Omega$ can be indeed  defined, at least in local patches.

 The effect  of a  gauge
transformation of $C$ on $\Omega$  is  easy to  compute. Transforming   $C \to C + d \varpi$, where  $\varpi$ is a 2-form,
sends  $\Omega \to \Omega + \varpi \wedge F /2$.
Since the  M2-brane charge is given by  the integral of
$d\Omega$ over a compact 7-cycle, this charge will transform under large gauge transformations of $C$,
that is  under gauge transformations such that the integral of  $ d\varpi$ over a 3-cycle does not vanish.

\sm

\setcounter{footnote}{0}

Actually, there is an even greater ambiguity in  the definition  \eqref{2f1} of the 7-form.
Not only can we shift it  by a gauge transformation, but we can add to it an arbitrary   $d\eta$,
since both  $\Omega$ and $\Omega + \eta$  integrate  the same
Maxwell-Chern-Simons equation in some local patch.
  We are ultimately interested in integrating the 7-form  over compact
7-cycles,  so   $\eta$ should be chosen judiciously such that  $d \Omega$ is everywhere well-defined on the cycle
in question.
We will see a concrete example of   this   in a minute.

\sm

The reduced form of the fields $C$ and $F$, equation (\ref{2b1}), leads to
 the following natural decomposition of  $d \Omega$  onto products of the volume forms of unit radius,
\bea
\label{2f2}
d \Omega = - d \Omega _1 \wedge \hat e^{345678} + d\Omega _2 \wedge \hat e^{678012} +
d \Omega _3 \wedge \hat e^{012345}\ .
\eea
The components $d\Omega_i$  are one-forms on $\Sigma$,   given
by the following expressions\,\footnote{The completely anti-symmetric tensor $\ep_{ijk}$  on $i,j,k=1,2,3$ is normalized by
$\ep _{123}=+1$; the indices $j,k$ are raised using the Minkowski metric with signature $(-++)$;
and the Poincar\'e duality ${}*_\Sigma$ on $\Sigma$ is given in local complex coordinates $w,\bar w$ by,
${}*_\Sigma db_i = - i dw \p_w b_i + i d\bar w \p_{\bar w} b_i$.}
\bea
\label{2f3}
 d\Omega _i  =    - {f_1^3 f_2^3 f_3^3 \over f_i^6} \left ( {}* _\Sigma db_i \right ) + \half \ep_i{}^{jk}  b_j db_k + d\eta_i\ ,
\eea
where $d\eta_i$ represents  the ambiguity mentioned above.
Conservation of $d \Omega$ is equivalent to the closure of each component $d\Omega _i$
as a one-form on $\Sigma$, a relation which is in turn  equivalent to one of the components
of the field equations for $F$.

\sm

Consider now the M2-brane charge  obtained by integrating $d\Omega$ over a compact seven-cycle
in the space-time manifold of the solution. In view of the general decomposition
  (\ref{2f2}), the contributions $d\Omega_2$  and $d\Omega_3$
always correspond to non-compact cycles, partly  subtended  by the AdS$_3$  factor.
Thus, the only part of interest for the calculation of M2-brane charges is the one  proportional to $d \Omega _1$.
Furthermore, in the cycles over which we will be integrating $d\Omega _1$,  either $b_2$ or $b_3$
(but not necessarily both) will be well-defined. By  choosing
\bea\label{ambg}
\eta_1  = - {\epsilon \over 2} \, b_2 b_3 \ , \qquad {\rm with} \ \ \epsilon= 1,0, \ {\rm or}\ -1\ ,
\eea
we can thus ensure that the ill-defined component appears through its  derivative, which is well-defined.
Of course, when both $b_2$ and $b_3$ can be defined over the entire cycle, all three choices for
$\epsilon$ lead to the same M2-brane  charge.
\sm

We may  integrate $d\Omega_1$ to obtain $\Omega_1$ along the same lines as we integrated $db_j$ to obtain
  the   magnetic  potentials  $b_j$.
  One first defines an auxiliary real function $\Lambda$, which satisfies
\bea\label{lambda3.12}
\p_w \Lambda =  ih\p_w\Phi  - 2i \Phi\p_w h\ .
\eea
The  existence of such a function  is established
   in Appendix \ref{sec:C}. The electric potential can then  be broken up into a cohomologically trivial and
 a  non-trivial  piece,
  \bea
\label{2f5}
\Omega _1 = { \nu_1 \sigma \over c_2^3 c_3^3}\, ( \Omega _1 ^s + \Omega _1 ^c )   \ ,
\eea
where, as shown in  appendix \ref{sec:C},  in the $\eta_1 = b_j^0=0$ gauge
 \bea
 \Omega _1 ^s  =
 { h \over 2 W_+} \left[  \g h ( \vert G\vert^2-1) +  ( \Phi  +\tilde h)  (G + \bar G) \right]
- { h \over 2 W_-} \left [ { h \over \g}  ( \vert G\vert^2-1) +  ( \Phi - \tilde h) (G + \bar G)   \right] \no
\eea
\bea
{\rm and}\qquad \label{2f7}
\Omega _1 ^c =  \Omega _1 ^0 -  \tilde h \Phi + \Lambda\ .
\eea
Here  $\Omega _1^0$ is an arbitrary  integration constant, $\Phi$ was defined in (\ref{2e2}),
$\sigma=- \nu_1 \nu_2 \nu_3$,  and the signs factors $\nu_i$ have  been
defined earlier in equation  (\ref{2e1}).

In order  to  avoid cumbersome formulae, we have  assumed in  \eqref{2f7}  that  $\eta_1= b_j^0= 0$.
The constant magnetic potentials can be added back  to the above expressions with no sweat.
In what concerns
 $\eta_1$, recall that it is  proportional to
$b_2b_3 \hat e^{345}\wedge  \hat e^{678}$,  and it  is needed precisely
  when this 6-form  is not globally defined  on the integration cycle.\footnote{If it were, the addition of
   $\eta_1$ would only affect the cohomologically trivial part of the electric potential.}
   In such cases  the $\Omega _1 ^s$ given  in \eqref{2f7} will
  contain   a piece that does not vanish on the relevant parts of $\p\Sigma$. This is
    cancelled precisely by  the addition of $\eta_1$, which  changes  however also the expression
    for the cohomological
    part $\Omega _1 ^c$.

 To be more specific, when $b_2$ is   well-defined on the integration cycle, we must choose
  $\epsilon = +1$ and subtract from  the   potential  $b_2 b_3/2$ which includes the
  non-trivial piece $\sim b_2 b_3^c $. If the well-defined cycle
  is associated with $b_3$, we must choose $\epsilon = -1$ and   the non-trivial piece is $\sim   b_2^c b_3 $.
  These extra terms can be computed easily from eqs.\,\eqref{2e2}.
After all the dust has settled, the general expression for $\Omega _1 ^c$ reads
 \bea\label{finalomegac}
\Omega _1 ^c = \Omega _1 ^0 -  \tilde h \Phi + \Lambda +  {\epsilon\over 2} ( \Phi^2 - \tilde h^2 )\ .
\eea
 To keep the expressions simple, we have set  here again
  $b_2^0=b_3^0=0$. We
 do not give the general formula  for $\Omega _1 ^s$ because it won't be needed.

 Like $\Phi$ and $\tilde h$ defined in the previous subsection, also the auxiliary function $\Lambda$
does not depend on  the parameter $\g$. This follows from the  definition \eqref{lambda3.12}.
 Thus the cohomological piece $\Omega _1 ^c$, and hence also the M2-brane charges of a
 solution,  are $\g$-independent.


\subsection{M5-brane charges}
\label{sec:2g}

The M5-brane charges of a solution  are obtained by integrating   $F$ over compact four-cycles
in the space-time manifold. Given  our metric ansatz, such four-cycles must
consist of one of the  spheres,  $S_2^3 $ or $S_3^3$,  fibered over a  non-contractible curve
$\cC$ in $\Sigma$.

\begin{figure}[]
\begin{center}
{\includegraphics[width=0.9\textwidth]{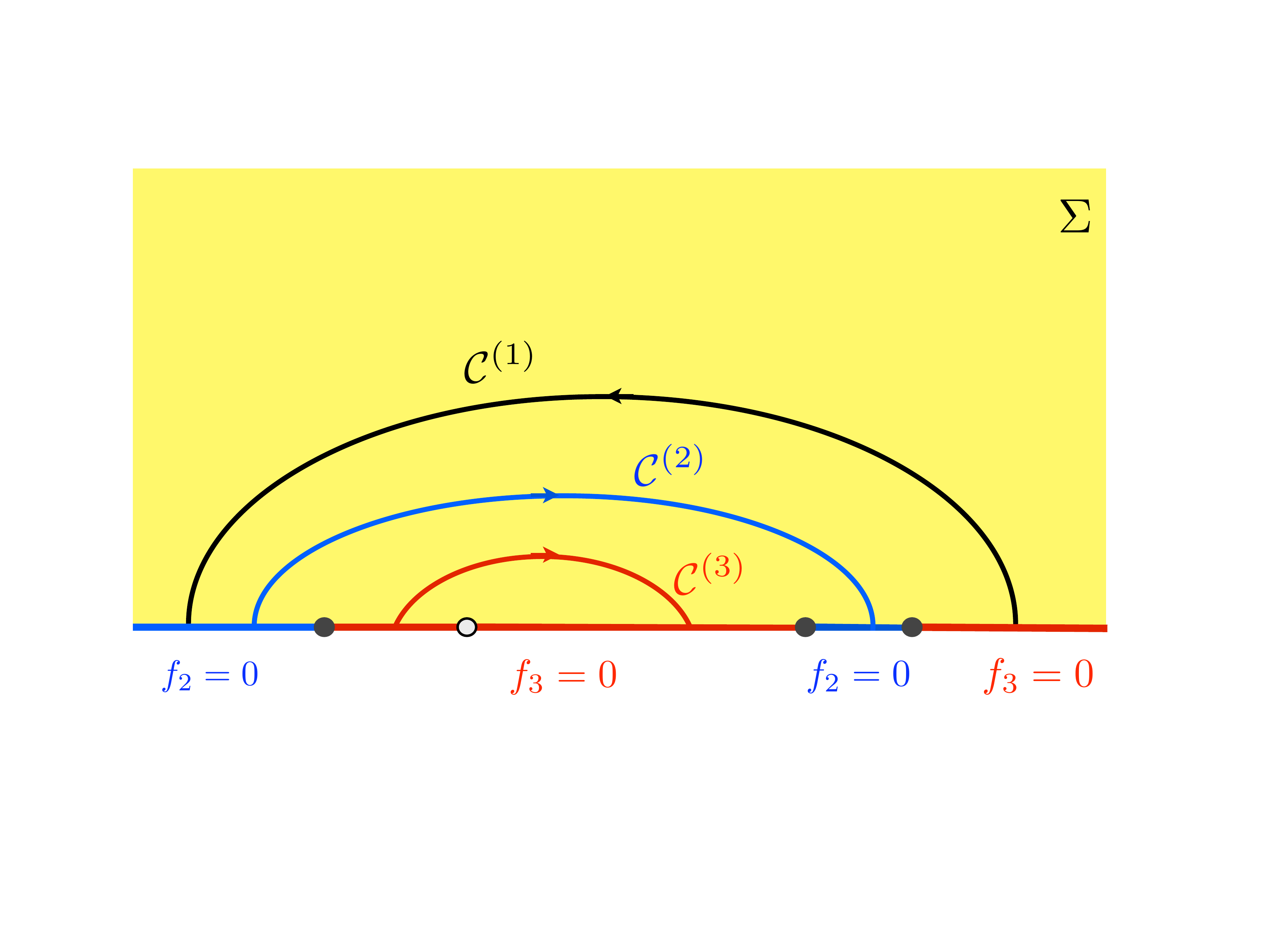}}
\end{center}
\vskip -2cm
\caption{\footnotesize The three types of curves,  $\cC^{(1)}, \cC^{(2)}$ and $\cC^{(3)}$, discussed in the text.
The surface $\Sigma$ is parametrized by the upper-half complex plane, and its boundary segments are marked in red or blue
according to whether it is $S_2^2$ or $S_3^3$ that shrinks there to a point.
 The curves $\cC^{(i=2,3)}$ have the same sphere $S_i^3$  shrinking to zero size at the two endpoints, while
 the curves  $\cC^{(1)}$ have different spheres  collapsing at its two endpoints.
 Thus $S_i^3\times_w \cC^{(i)}$ ($i=2,3$) are topological 4-spheres, while  $(S_2^3\times S_3^3)\times_w \cC^{(1)}$ is a
 topological 7-sphere. These cycles  are non-contractible when the endpoints are separated by one or more singularities,
 as will be explained later in section \ref{sec:5}.
 }
\label{fig:2}
\end{figure}

On  generic Riemann surfaces,   $\cC$ can be a  non-contractible  closed curve. We will,
however, mainly  consider the case where $\Sigma$ is  a disk, so a
homology basis of compact four-cycles may be obtained by restricting to curves $\cC$ that are open and
connected, and   end on the boundary $\p \Sigma$. To form a compact four-cycle, the
associated $S_i^3$ must shrink to zero radius at both ends of the open curve,  so that this latter has the form
\bea
\label{2g2}
\cC^{(i)} = \{ z  (t) \in \Sigma, ~ t \in [0,1], ~ f_i(z (1)) = f_i (z (0))=0\}\qquad i=2,3\ .
\eea
A schematic representation of the curves $\cC^{(i)}$ is given in Figure \ref{fig:2}.
These curves are  non-contractible whenever
  their endpoints are separated by one or more singular points  on the boundary of $\Sigma$, as will become
  clear   later.
\sm


The magnetic  M5-brane charges  $\mM^{(i)} $
associated to the four-cycles $\cC^{(i)}\times_w S_i^3$  read
\bea
\label{2g3}
\mM^{(i)}  \equiv  { 1 \over 2 \pi^2} \int _{ S_i ^3 \times_w \cC^{(i)}} F\ .
\eea
The normalization factor $2 \pi^2$ is the volume of the three-sphere  with unit radius.
In terms of the reduced fields $b_i$, and their cohomological decomposition in  (\ref{2e1}),
these charges can be simplified as follows,
\bea
\label{2g4}
\mM^{(i)} =   \int _{ \cC^{(i)} } db_i =   { \nu _i \over c_i^3} \int _{ \cC^{(i)} } db_i^c = { \nu _i \over c_i^3}\ .
\, b_i ^c \, \bigg | ^{z(1)} _{z(0)}
 \eea
In the right-most expression, the flux potentials are  evaluated at the end points
of the curves $\cC^{(i)}$, and we only need to keep  the non-constant piece  of
the cohomological part  $b_i^c$.
This is
  because  $b_i^s=0$ on the boundary,  and the constant piece
$b_i^0$ makes no contribution.
  The M5-brane charges are gauge invariant by construction.


\subsection{M2-brane charges}
\label{sec:2h}

The M2-brane charges of a solution are obtained by integrating $d\Omega$ over compact seven-cycles.
 Given the reduced form of   $d \Omega$  in (\ref{2f2}),
it is clear that such cycles consist of the product  $S_2^3 \times S_3^3$ fibered
over a curve  $\cC$ in $\Sigma$. When the Riemann surface $\Sigma$  has the topology of  the disk,
we have three different types of compact seven-cycles given as follows:
\bea
\label{2h2}
\cD ^{(1)} & = & \left ( S_2 ^3 \times S_3^3 \right ) \times _w \cC^{(1)}\ ,
\no \\
\cD ^{(2)} & = & \left ( S_2 ^3 \times_w \cC^{(2)}  \right ) \times S_3^3\ ,
\no \\
\cD ^{(3)} & = & S_2 ^3 \times \left ( S_3 ^3 \times_w \cC^{(3)}  \right )\ .
\eea
The second and third cycles have topology $S^4 \times S^3$,
where the $S^4$ is one of the compact four-cycles that carry  M5-charge.
The first type of 7-cycle has topology $S^7$, and it is obtained
by fibering the product $S_2^3 \times S_3^3$ over a curve $\cC \subset \Sigma$ such that one endpoint
of $\cC$ corresponds to vanishing $S_2^3$, while the other endpoint corresponds to vanishing $S_3^3$.
We  denote such curves by $\cC ^{(1)}$,
\bea
\label{2h1}
\cC^{(1)} = \{ z  (t) \in \Sigma, ~ t \in [0,1], ~ f_3(z (1)) = f_2 (z (0))=0\}\ .
\eea
  Of course, if  $\Sigma$ has non-contractible one-cycles, as when it has the topology of the annulus,
 there will be  also seven-cycles which are  topologically  equivalent to $S^3\times S^3\times S^1$.

 The charges corresponding to the cycles $\cD ^{(j)}$ for $j=2,3$  give the net number of M2 branes  ending
respectively on M5 and  M5$^\prime$ branes. The charge corresponding to the cycle  $\cD ^{(1)}$, on the
other hand, gives the number of semi-infinite M2 branes.
 The actual values of these "electric"   charges  read
\bea
\label{2h3}
\mE^{(i)} =  { 1 \over (2 \pi^2)^2} \int ( *F + {1\over 2} C\wedge F + d\eta) =
  \int _{\cD^{(i)} } d \Omega_1 =   \   { \nu_1 \sigma \over c_2^3 c_3^3}
\, \Omega_1 ^c    \bigg | ^{z  (1)} _{ z (0)}\ .
\eea
The normalization factor $(2 \pi^2)^2$ is the volume of the product of the unit-radius spheres
$S_2^3 \times S_3^2$,  which is common to all M2-brane charges. In  the last step we have used
the cohomological decomposition in  (\ref{2f5}), where $\Omega _1^c$ is given by \eqref{finalomegac}.
The choice of $\epsilon$ is dictated by the fact that $b_2$ can be  globally defined on
${\cD^{(3)} }$ cycles, and $b_3$ can be defined  on
${\cD^{(2)} }$ cycles. We therefore choose
 $\epsilon^{(2)} = -1$ and  $\epsilon^{(3)} = 1$. The choice for $\cD ^{(1)}$ is irrelevant,
 provided we work in  a gauge where $b_2(z(0))= 0$ and $b_3(z(1))= 0$, which can be always
 achieved by adjusting the constant pieces $b_j^0$.

We stress once more  that, contrary to   M5-brane charges, the M2-brane charges are  not gauge invariant, as can
be seen  by shifting  the potentials  $b_i$  by a (quantized)  constant.

\sm

 In general, the above three types of M2-brane charges are not mutually   independent.   If a solution contains
at least one boundary component on which $f_2=0$, and at least one boundary component on which
$f_3=0$, then the cycles $\cD^{(1)} $ will produce all charges by linear combination, and one
may retain only those charges built from the cycles $\cC^{(1)}$.
An exception occurs  when the entire boundary $\p \Sigma$ is characterized by either $f_2=0$ or $f_3=0$,
so that there are no cycles of type $\cC^{(1)}$.
This is for instance the case for  the purely $AdS_7 \times S^4$ solution. We will return to the choice of a basis for
7-cycles in later sections.

 \if
 One last important remark is here in order.
Recall   that the auxiliary functions $\Phi ,  \Lambda$ and $\tilde h$ were  defined in terms of
  $h$ and $G$, and do not depend on  the parameter $\g$.  From the expressions  \eqref{2e2} and
  \eqref{finalomegac} for the cohomological pieces  $b_j^c$ and $\Omega _1 ^c$, we see that the dependence
  of  the charges on $\g$ enters  through  a multiplicative factor,  common for all charges of same type.
  Thus the $\g$ deformation  rescales all
  existing charges, but does not generate  new ones. Furthermore, from eqs.\,\eqref{2e1} and \eqref{2e2}
  one finds that  the ratio of   M5 to M5$^\prime$ charges is rescaled precisely by    $  \g /\g_0 $,
 where $\g_0$ is some reference value of the parameter. This gives
  the physical interpretation of   $\vert \g\vert $  as the ratio of M5 to M5$^\prime$ charges.
 \fi


\subsection{The transformation $\mI$}
\label{sec:2i}

Inspection of the expressions (\ref{2d2}) and (\ref{2d1}) for the space-time metric shows that the  following
  transformation,
 \bea
 \mI = \left \{   \begin{matrix}   \g \to 1/\g \\  G \to -G    \end{matrix}   \right . \ ,
\eea
leaves the AdS$_3$ metric function $f_1$   invariant, while permuting the three-sphere factors, $f_2$ and $f_3$,   into
one another. The transformation also exchanges the  gauge potentials $b_2$ and $b_3$, given by eqs.\,\eqref{2e2}.
Since the  exchange of the two spheres
is a reparametrization,  we   identify any solution
with its image under $\mI$,  and restrict   $\g$ to the  interval
\bea
-1 \leq \g \leq +1\ .
\eea
The point $\g=0$, which is equivalent under $\mI$ to $\g =\infty$,  does not correspond to a regular solution because
 one of the three-spheres and the base $\Sigma$ decompactify.\footnote{But one can try to make sense of these solutions
 as a limit.}
 One may consider this
point to be removed from the range  $[-1,+1]$ of  $\g$. It will be sometimes convenient to allow
$\g$ to range over all real numbers and  fix  the sign of $G$ on some  boundary segment.

\sm

The  net effect of
the transformation  $\mI$ is to exchange the
 M5 branes  and M5$^\prime$ branes of Table 1.
  This is a trivial symmetry in M theory, but it has non-trivial consequences
  after compactification on a two-torus. To see why,
 note first that compactifying
  $x^{10}$ on a small circle gives   a type-IIA
 configuration   with D2-branes intersecting stacks of NS5-branes and D4-branes. Further compactifying and T-dualizing
 the coordinate $x^6$ leads (after a relabeling of coordinates) to  the type-IIB configurations of Table 2.
 The  M2, M5  and   M5$^\prime$ branes
 of the original setup
 have been converted by these operations
 respectively to D3 branes,  NS5 branes  and D5 branes.
 The exchange of M5  and M5$^\prime$-branes descends in this way to the
   S-duality  transformation that exchanges the NS5  and D5 branes of the type-IIB setup.
   For  the gauge  theory  on the D3-branes,
this S-duality  acts as mirror symmetry
exchanging  electric and magnetic quivers
 \cite{Intriligator:1996ex,deBoer:1996mp,deBoer:1996ck}.

\section{The Function $h$  and Two Simple  Corollaries}
\setcounter{equation}{0}
\label{sec:4}

     As has been shown in  sections \ref{sec:2} and \ref{sec:3},  any solution
     of eleven-dimensional  supergravity  with $D(2,1;\g,0) \oplus D(2,1;\g,0)$ symmetry can be expressed in terms of
 $(\g, h, G)$.  The search for such backgrounds is thus  reduced to the mathematical problem of finding solutions of the linear
equations  \eqref{1a1},  subject to the regularity conditions \eqref{1a4} and  \eqref{1a5}.
  In this section we will focus on the harmonic function $h$.  As we will see, the
   conditions that $h$ should be  positive  in the interior of $\Sigma$
  and vanishing  on its boundary  determine almost completely the allowed form of the harmonic function.

 These considerations have two  general implications,   even before one attempts  to solve for
 the   function  $G$. Note indeed that   the relation (\ref{6a5}) together with  the inequalities (\ref{bAdSs})
 imply that the AdS$_3$ scale factor diverges at singularities of  $h$. Such singularities  change  the
 conformal boundary of the solution from $S^1\times \mathbb{R}$,  the conformal boundary of AdS$_3$,
  to a conformal boundary of higher dimension.
  Furthermore, at points   where $\p_w h=0$,
 the metric on $\Sigma$ develops a conical singularity except, possibly,  if $G$ diverges at this same point.
 That this is so can be seen from the expression  for   $\rho$ given in  \eqref{2d2}. Our general "theorems"   follow
  from these two simple observations,
and  from the study of the  zeroes and singularities of the meromorphic function $\p_wh$.


\subsection{Admissible singularities of   $h$}
\label{generalh}

 A  smooth harmonic function that  vanishes on the boundary of a compact Riemann surface  is identically zero.
To find  non-trivial solutions,  we must  either allow $h$ to be singular, or assume  that $\Sigma$ is a surface
with  no boundary,
  in which case $h$ can be a non-zero constant. We will discuss the second possibility in the following  subsection.
 Here we focus on possible singularities of   $h$.

In the interior of   $\Sigma$, the only allowed singularities of a positive harmonic function are logarithmic
  (this is sometimes referred to as B\^ocher's theorem).\footnote{We only need to
   consider isolated singularities. Extended singularities make $h$ discontinuous  and,  since $h$ is
     the product of
   the (pseudo-)sphere radii,  they  give non-regular metrics.  }
A simple example is the function $h = - \alpha \, {\rm log} w + c.c.$ with $\alpha > 0$, which   is harmonic,
   positive in the interior of the unit disk,  and vanishes  on its boundary.  A function with many
   logarithmic singularities
   is
   \bea\label{bulksings}
   h^{[1]} (w)  = -  \sum_{j=1}^N    \alpha_j  \, \log \left( { w- \beta_j\over w-  \bar \beta_j} \right)  + c.c.
   \eea
  where  $w$ is  defined in the upper-half complex plane, the $\alpha_j$ are  real and positive, and
    Im$(\beta_j) >0$   for all $j$.
   With  these assumptions,  $h^{[1]}=0$ on the real-$w$ axis, and $h^{[1]}>0$ in the interior
   where  Im$(w) >0$. Note that \eqref{bulksings} is the electrostatic potential for a collection of pointlike  charges
   $\alpha_j$ located at $\beta_j$, in the presence of a conducting boundary.
   \sm

 Regularity of the supergravity metric does not  exclude such
 logarithmic singularities in  $h$.
 For instance, if  at the
position $\beta_j$ of a singularity   $\vert G\vert$= constant$\,\not= 1$,  then  the  Weyl-rescaled
  form of the space-time metric, equations \eqref{weyl1} and \eqref{weyl2}, reads
  \bea
  ds^2 \,   \simeq \,  C^2  x^{2\over 3}  \left[ R_1^2\,  ds^2_{AdS_3} +
  R_2^2\, ds^2_{S_2^3} + R_3^2\, ds^2_{S_3^3} + {  dx^2 + d\theta^2 \over x^2}\right] \ ,
  \eea
  where $-\log(w- \beta_j) \simeq x + i \theta$, and $C$, $R_j$ are finite constants.
  As $x\to\infty$  the geometry approaches
  AdS$_3\,\times\,$S$_2^3\,\times\,$S$_3^3\,\times\,$H$_2$ times  a Weyl factor that diverges like  $x^{1/3}$.
   If $G$  also blows up at the singularity, then $C$   diverges and  the $R_j$   vanish   in such a way that
   $CR_j$ stays  finite. In either case
   the geometry asymptotes to flat  space-time, with one of the dimensions being  a  circle of shrinking radius.

  Solutions with such asymptotically-flat  regions may  be regular,
   but  have  no holographic interpretation.
  We will encounter an example  in section \ref{sec:71}.
  A holographic interpretation might  be only possible   if $\vert G\vert \to 1$ at all  singular points, but we have not found
  any exact solutions of such kind. In any case,  we will allow  $h$ to have
   logarithmic singularities  when deriving our  general "theorems" in the following subsections.

\sm

 Singularities on the boundary $\p\Sigma$ arise when  Im$(\beta_j)= \delta \to 0$, so that the  charge and its image
 pinch the real axis. This creates a pole in the analytic part of $h$ with residue $2\delta\alpha_j$,
  which can be held fixed in the limit.
  To see that this is the only   type of singularity that is allowed on $\p\Sigma$,
   choose local  coordinates $(w,\bar w)$
such that  the putative singularity   is  at  $w=0$,   the  boundary
is a segment of the real axis,  and the $\Sigma$  interior is Im$(w)>0$.
Consider an arbitrary power-law dependence of $h$
  as $w\to 0$,
 \bea
 h \sim   i \, h_n w^n + c.c. =   i  r^n (h_n e^{in\theta}  -   \bar h_n  e^{-in\theta})\ ,
 \eea
where $w = r e^{i\theta}$.
For  $h$ to vanish  at $\theta =0$,   the coefficients  $h_n$ must be real.
But $h$ must also vanish at $\theta=\pi$, so $n$ must be integer.
In addition,   $h \sim  - 2 h_n\,  r^n  {\rm sin}(n\theta)$   must be positive   for all $0<\theta <\pi$,
which  is only possible if $n=-1$ and
$h_{-1}>0$, or   if $n=1$ and $h_1<0$.
   Thus  the leading behavior of  $h$ near a point on  (a smooth piece of) the  boundary of $\Sigma$ is either a simple zero,
   or a simple pole.\footnote{Strictly speaking, when talking here and later about a zero or a pole
   of $h$, we   mean of  the holomorphic  part of $h$. And when we say that the holomorphic part of $h$ has a zero, we
     assume a local coordinate in which  an irrelevant  imaginary constant has been absorbed.
     We hope the reader will not be confused by this loose use of
   language.}

   In the electrostatic analogy, the simple pole of $h$ on $\p\Sigma$ comes from the pinching of an electric dipole
   made out   of a (positive)  charge and of its (negative)  image. A double pole of  $h$ would, in this language,
   correspond to a collapsing quadrupole. This would require,  however,  opposite charges on the same side
   of the boundary, which  would destroy the positivity of the electrostatic  potential $h$.

  \sm

   \setcounter{footnote}{0}

This intuitive argument makes it, on the other hand,  clear
that we are free to superpose any number of boundary poles, by bringing several   charges of the same sign,
and their images,  to the real axis. Consider  indeed the function
\bea\label{hDisk}
h^{[2]}(w) =  \sum_{j=1}^n    {i  a_j\over w - d_j} + c.c. \ ,
\eea
where  the parameters $\{   a_j, d_j\}$ are real and the $  a_j$ are positive.
 Clearly,  $h^{[2]}=0$ on the real
axis, and $h^{[2]}>0$ when Im$(w) >0$, so   the regularity conditions are satisfied.
  The most general  harmonic function on $\mathbb{C}^+$  that obeys our regularity conditions
is found by  combining
    \eqref{bulksings} and \eqref{hDisk},
 \bea
h(w) = h^{[1]}(w) + h^{[2]}(w)\,.
\eea
This is the most general  $h$ when  $\Sigma$ has  the topology of a disk  parametrized  by the
upper-half  complex plane.  On higher-genus surfaces,  the admissible   bulk and boundary singularities are the same,
but the explicit expressions for $h$ are more involved.

 \if
 \footnote{An  alternative  parametrization of  $h^{[2]}$ is in terms of
 coordinates $(z, \bar z)$ that
live on the infinite strip  $\Sigma= \{z=x+iy \in \mathbb{R}+ i  [0, \pi/2]\, \}$.
In  these  coordinates
$$
i h^{[2]}(z)   =   a \,  {\rm sinh} (2z)  + \sum_{i=1}^{N} \gamma_i\, {\rm tanh}(z-\delta_i) -    \sum_{j=1}^{\tilde N}
 \tilde \gamma_j\,  {\rm coth}(z-\tilde \delta_j)  -  c.c.\ ,
$$
where all   parameters are real and,
in addition,   $\{a, \gamma_i, \tilde\gamma_j\}$  must be  positive.
One can  check  that the  holomorphic part of  $h^{[2]}$
is purely imaginary on  $\p\Sigma$, so the function  vanishes  there as it should, and
that in the interior of the strip  the imaginary part of each  summand  is
positive,
so  this is also true   for $h^{[2]}$  as a whole. The above function has
$n= N+\tilde N +2$ singularities, located
at $z= \pm \infty$,  $z= \tilde \delta_j$ and
 $z = \delta_i + i\pi/2$, and  $2(N+ \tilde N) -1$ free parameters,  the same as the number of free
 parameters of  (\ref{hDisk}).}
\fi

\sm

In practice, in all  known solutions,  $h$ has at most two singularities on the closure
of the Riemann surface, $\bar\Sigma$. For
 backgrounds with holographic duals
 we will   argue later that there can  be no more.  Nevertheless, the above
 explicit    form of   $h$   will be useful  in   the proof of our second theorem,  in section \ref{sec:4.4}.


\subsection{Uniqueness of solution dual to  $2d$   CFTs}

 Solutions that
     are dual to two-dimensional conformal theories should have a conformal boundary $S^1\times \mathbb{R}$,
     which is the conformal boundary of AdS$_3$.
    This means that  the radius of the AdS$_3$ fiber must  be
      finite everywhere, both in the interior and on the boundary  of $\Sigma$, since
   the  space-time metric
  must only   diverge on  a sequence of  two-dimensional  submanifolds.   From
  the relation (\ref{6a5}) and the inequalities (\ref{bAdSs}) we know that this is only possible   if
  $h$ is everywhere smooth. The only   option is therefore
   $$
      h= {\rm constant} \not=0\qquad  {\rm on\  a }\  \Sigma \ \ {\rm without  \ boundary. }
      $$
 Now when $h$ is constant  the expression (\ref{2d2}) for the $\Sigma$
   metric   degenerates, and
      the  reduction  of the BPS  equations  must be
      carried out  from scratch. This is done      in appendix \ref{sec:D}. The final result is that all
    four scale factors ($f_1, f_2, f_3, \rho$) are in this case constants,  subject to the two  relations
 \bea\label{CFT2relns}
  f_1^2 = {f_2^2\, f_3^2\over f_2^2 + f_3^2}\qquad {\rm and}\qquad   {f_3^2  \over f_2^2 }   = \g\ .
 \eea
   Thus,  the only   space-time manifold dual to a 2-dimensional CFT
    is   AdS$_3\times $S$^3\times $S$^3\times $E$_2$,   where  E$_2$ denotes the Euclidean plane,
   possibly with  discrete identifications.
\sm

  The  AdS$_3\times $S$^3\times $S$^3\times $E$_2$  background is known to   describe
     the near-horizon region of
  the
  M2\,$\perp$\,M5\,$\perp$\,M5\,(1) configuration of Table 1, when  all  M5-branes
  are  smeared\footnote{The solutions of M-theory corresponding to smeared intersecting branes
  were constructed in \cite{Tseytlin:1997cs,Gauntlett:1997cv}.  Smearing creates an additional
  Killing isometry, which is doubled  in the near-horizon limit.
  }
   in the common transverse dimension (parametrized by $x^2$)  \cite{Boonstra:1998yu,de Boer:1999rh}.
  In this background all of the components of the 4-form  flux  point in  a common direction of  E$_2$.
   Compactifying this  direction on  an "invisibly-small"  circle,
 gives
  a solution of  type IIA string theory which is exact at all orders in the
  sigma-model ($\alpha^\prime$)  expansion \cite{Antoniadis:1989mn}.
   One may further  compactify E$_2$ on a two-torus, though  this  solution cannot be anymore connected
   to the asymptotically-flat region of the M2\,$\perp$\,M5\,$\perp$\,M5\,(1) configuration.\footnote{The reason is
   that one of the E$_2$ coordinates is a combination of   two radial coordinates in the $\{3,4,5,6\}$
    and $\{7,8,9,10\}$
   subspaces of Table 1, so it cannot be compact. More generally, however,
  we  can  consider solutions with a   non-diagonal torus, and  an arbitrary orientation for the flux fields.}

  \sm

It is interesting to compare the
 uniqueness of half-BPS  M-theory solutions which have   two-dimensional CFT duals,   to
    the  analogous situation in one higher dimension in type-IIB theory.
  As   shown in \cite{Assel:2011xz,Assel:2012cj}, there exists a rich set
    of   half-BPS solutions of type-IIB string theory
 that are dual to   three-dimensional  CFTs.
  These latter  are the conjectured infrared limits of quiver gauge theories  that  describe  the D3-brane dynamics
  in the configurations of Table 2  \cite{Gaiotto:2008sa,Gaiotto:2008sd,Gaiotto:2008ak}.
   The D3-branes have finite extent in the $x^3$ direction, since they
  are suspended between the NS5-branes and D5-branes of the configuration.
    The geometry of the  type-IIB solutions
 captures  precisely the quiver data, i.e. the partition of the D3-branes among the five-branes \cite{Assel:2011xz,Assel:2012cj}.

   Performing a T-duality of  the transverse coordinate $x^3$, and
     lifting to eleven dimensions,  transforms these configurations to M2-branes
    in the background of two types of Kaluza-Klein monopoles \cite{Assel:2012cj}. Interestingly, on the M-theory side  the
      near-horizon geometry  is   (almost) unique -- it is  the  orbifold of AdS$_4\times$(S$^7/(\mathbb{Z}_k
      \times \mathbb{Z}_{\hat k}) $, where $k, \hat k$ are the numbers of KK monopoles. The  missing  data of the quiver CFT
must thus be  captured either by   discrete torsion
 of the 3-form potential in the M-theory orbifold \cite{Imamura:2008ji}, or by non-geometric backgrounds
 of the wrapped M2-brane field.
 \sm

   Can a similar story hold for the case at hand? In the flat probe-brane limit, one can suspend a collection  of $N$ M2-branes
   between sets of M5- and $M5^\prime$-branes in many different ways, as described by  pairs
   of partitions of $N$.  Our result suggests that in the
   infrared limit this information is lost since,  in contrast to  AdS$_4\times$(S$^7/(\mathbb{Z}_k
      \times \mathbb{Z}_{\hat k}) $, the
   AdS$_3\times $S$^3\times $S$^3\times $E$_2$ geometry   does not   have the required  homology to encode
   the data. This  uniqueness of the infrared SCFT$_2$
    could  be  a consequence of  the Coleman-Mermin-Wagner theorem in two dimensions:
       the M5-branes are effectively  smeared, restoring dynamically   the broken invariance under
   translations in the $x^2$ direction.


\subsection{Why  $\g$  is not a continuous parameter}

The AdS$_3\times $S$^3\times $S$^3\times $E$_2$ background  has two free parameters,
which can be chosen to be the radii of the two S$^3$, or alternatively the AdS$_3$ radius and the parameter $\g$.
The meaning of this latter parameter becomes clear if we  make  the dimension of
 E$_2$  in  which the 4-form fluxes are aligned into
   a circle $C_x$,
of circumference  $\ell_x$. The geometry now has two 4-cycles that support M5-brane charge: ${\cal C}_x \times S^3_2$
and  ${\cal C}_x \times S^3_3$.
The corresponding charges,  $\mM^{(2)}$ and $\mM^{(3)}$,  are calculated
 in appendix \ref{sec:D}, where it is shown that
 \bea
  \mM^{(3)} =  \pm \g\,    \mM^{(2)}\ .
 \eea
 Thus  $\g$ is  the ratio of the two M5-brane charges, up to a sign that can be chosen freely.
 This agrees with the general conclusion
  drawn from the  form of the cohomological potentials,   at the end of section \ref{sec:2h}.
Since M5-brane charges are quantized, $\g$ must take rational values. As a result,  even though $\g$
is a continuous parameter  of the
supergravity  solutions, it  is not  necessarily a continuous modulus in M theory.

\sm

  Compactifying  E$_2$ on a cylinder allows us also to view the solution as a configuration in  type-IIA string theory.
Since all 4-form backgrounds have an index in the direction  $C_x$, the string theory background only involves the
Neveu-Schwarz 2-form field. It has therefore an exact world-sheet description in terms of Wess-Zumino-Witten
models for the product group manifold
  $SL(2, \mathbb{R})_{k_1}\times SU(2)_{k_2}\times SU(2)_{k_3}$ \cite{Antoniadis:1989mn}.
The relations
\eqref{CFT2relns} become relations between the levels $k_j$ of the WZW models, in particular $\g=k_3/k_2$.

Note that the circumference $\ell_x$ translates into a constant dilaton field. Since the M5-branes of M theory descend
to orthogonal NS5-branes in type-IIA, one might have expected the dilaton to run  away  to strong coupling
in the near-horizon limit. The fact that this does not happen is probably because of  background fundamental strings, even if
their charge cannot be defined unambiguously, as discussed in appendix \ref{sec:D}.


\subsection{All interface CFTs  have  $\g>0$ }
\label{sec:4.4}

The  structure of  admissible harmonic functions $h$ leads to another general conclusion,
which can be stated as follows,
\medskip

\underline{Theorem}: Regular solutions with  $\g<0$ cannot have more than one $h$-singularity.

\medskip

  \noindent The idea behind the proof   is that  when $h$ has more than one singular points,
$\p_w h$ develops  zeroes in the interior of $\Sigma$. These give rise  to conical singularities unless $G$
is allowed to   diverge.
But this is forbidden by  the regularity condition $\vert G\vert <1$ when $\g<0$, so we must
conclude that  $h$ cannot have more than  one singularity in the closure of $\Sigma$.

\sm
More concretely, consider
 first the case of  $h$ with only boundary poles, as  given by the expression
\eqref{hDisk}. The equation $\p_w h^{[2]}=0$  is equivalent to a polynomial  equation
\bea
(\sum a_j) \, w^{2n-2} + \cdots = 0\ ,
\eea
where $n$ is the number of boundary poles. Since the $a_j$ are positive,
the leading term is non-vanishing, and  there are  $2n-2$ roots.
 Furthermore,
on the real axis
\bea
i\p_w h^{[2]} = \sum_j {a_j\over (w - d_j)^2} >0 \ ,
\eea
so all the zeroes are complex. Finally, since  all
coefficients of the polynomial equation are real, the zeroes come in complex-conjugate pairs,
 and $n-1$ of them  lie in the interior of $\Sigma$.

 The argument can be extended easily to  the general function $h= h^{[1]} + h^{[2]}$,
      with both interior and boundary singularities. From  the
  expression   \eqref{bulksings}  we find
        \bea\label{dhpoles2}
 \p_w h  =    - \sum_i  { 2 \alpha_i {\rm Im}(\beta_i)  \over  (w-\beta_i)(w- \bar\beta_i)}
 -i \sum_j {a_j\over (w - d_j)^2}\ .
\eea
 The zeroes of this meromorphic function come again  in complex-conjugate pairs, and they do not lie on the real axis
since both the real and the imaginary parts  are
 manifestly positive when $w$ is real [this follows from the positivity of  $\alpha_i$,   ${\rm Im}(\beta_i)$ and $a_j$].
 Thus half of the zeroes of \eqref{dhpoles2} lie in the interior of $\Sigma$.
 Now for any meromorphic function, the number of zeroes equals  the number
of poles,  counted with multiplicities  and   including  the point at infinity.
The function \eqref{dhpoles2} has a double zero at infinity
and
$2n$ poles, where $n$
  is the number of terms in the sum. Thus the number of  zeroes inside $\Sigma$ is $n-1$.

  \sm

  Now consider the geometry in the vicinity of one of the zeroes of $\p_wh$, at $w=w_0$.
  From the expression  \eqref{2d2} for the metric factors we have
\bea
\rho ^2dw d\bar w  =  \left( {(G\bar G-1)W_+W_-\over c_2^3 c_3^3 h^4}\right)^{1\over 3}\, \vert \p_w h\vert^2 dw d\bar w \
\sim  \vert (w-w_0)\vert^2 dw d\bar w\ ,
\eea
where in the second relation we have assumed that the factor inside the parenthesis  stays  finite at $w_0$.
This metric has a conical singularity, which could be "cured" by an appropriately diverging $G$, but this
is not allowed when $\g<0$.  To avoid such singularities we are thus forced
to  request that $n=0$ or $1$, which is the statement in the theorem.

 \sm

   We have actually assumed in this argument that $\Sigma$ is  a disk
   with marked points, and one may wonder whether this assumption is too restrictive.
   In this regard, note that $\Sigma$ actually  inherits  a metric from  supergravity, which is encoded in the data
   $(\g, h, G)$.
 Non-trivial  topology can thus only arise   in  two ways:  $(a)$  by modding out $\Sigma$ by a discrete group, $\Gamma$,  of symmetries of   the  data; or $(b)$  if the metric develops  branch-point singularities which are resolved by higher topology.
   It can be easily seen that branch-point singularities require divergent $G$ and are excluded when $\g<0$,
   whereas modding out
   can only reduce the number of singularities,
     so our theorem still holds.

\sm

 A corollary of this theorem  is that superconformal interfaces of the  ABJM membrane theory
 must necessarily have $\g>0$. Indeed, as
we  will see in the following section, space-time throats with either   (AdS$_4/Z_2)\times$S$^4$ or AdS$_7\times$S$^4$
geometry arise at
  singularities of $h$ on $\p\Sigma$.\,\footnote{The reason
why one   finds  AdS$_4/Z_2$  can be understood as follows: all  anti-de Sitter
space-times can be always written as   warped products  AdS$_{d+1} \simeq  (AdS_{d-n}\times S^n)\times_w \mathbb{R}^+$
for any $0\leq n< d$. The case $n=0$ is however special, because the 0-dimensional sphere consists of two points
and it is not connected. A diverging AdS$_3$ fiber near a singularity of $h$
covers therefore only half of the boundary of AdS$_4$.
Note that this would continue to be the case, even if the conformal boundary of our solutions were to arise  from the (not fully
chartered) logarithmic singularities of $h$ in the interior of $\Sigma$, see section \ref{generalh}.
}
The conformal boundaries of these throats, illustrated by  the entries $(iii)$ and $(iv)$ in  Figure \ref{confBnrs}, are
the 3-dimensional hemisphere and the 5-dimensional sphere. Since
the conformal boundary of solutions dual to interfaces in the ABJM theory
must have  at least two such hemispheres, there can be  no such solutions  with  $\g<0$.

\setcounter{footnote}{0}

\section{ Expansions near $\p\Sigma$ and Lego Pieces}
\setcounter{equation}{0}
\label{sec:5}

 Having characterized completely the allowed functions $h$, we turn now to the second function
 of the reduced problem, the complex function $G$. This obeys  the second equation
 in  \eqref{1a1},  subject to the regularity conditions \eqref{1a4} and  \eqref{1a5}.
 We will  now analyze this mathematical  problem locally, in a neighborhood of a   boundary point
 of $\Sigma$.

 \sm
 We  will identify four local solutions that are the "Lego pieces"  in  the construction of
 global solutions in the following sections. These  are:
   $(a)$  an asymptotic AdS$_7\times$S$^4$ throat;         $(b)$ an asymptotic  (AdS$_4/Z_2)\times$S$^7$ throat;
   $(c)$ a coordinate
   singularity, called the "flip",  which  changes  the boundary value of $G$ from $+i$ to $-i$; and
  $(d)$ a highly-curved asymptotic region whose M-theory interpretation we will discuss.
The first three Lego pieces are compatible with
any value  of $\g$, though   for  $\g\notin \{-1/2, -2\}$  the AdS$_7$ throat is  deformed.  The last local solution
 requires  $\g$ to be positive.

   This list of local solutions,  in which one or both of the functions $h$, $G$ have singular behavior,
   may not be exhaustive. We did not classify all singularities in the interior of $\Sigma$, nor all possible boundary
   solutions when $G$ is allowed to diverge. The above four Lego pieces will however suffice for
      all the exact solutions in sections \ref{sec:6} and \ref{sec:7}.


\if
Generic, non-singular point:
 \bea
 h = h_0 + w + \bar w\ .
 \eea
 \bea
 h_0 (\p_r - {i\over r} \p_\theta) (G_1+iG_2) = 2G_1 e^{i\theta}
 \eea

Setting  $G(r, \theta) = r^{\nu} G(\theta)$ gives
 \bea
 h_0 (\nu G_1 + \p_\theta G_2) = 2r\cos\theta \, G_1\ , \quad
 h_0 (\nu G_2 -  \p_\theta G_1) = 2r\sin\theta \, G_1\ .
 \eea
 \fi


\subsection{Solving for  $G$ with  Legendre polynomials}

As   shown in section \ref{generalh}, the holomorphic part of $h$ has either a simple zero or a simple pole
at any given point of  (a linear segment of) $\p\Sigma$.  By an analytic   change of the coordinate $w$,
which leaves  invariant the origin  and
 a piece of  the  real axis, one can   bring $h$  locally  to one of the following two canonical  forms\,\footnote{The
 generic $h$ admits a Laurent expansion with imaginary coefficients,
 $h =  i\,  {h_{-1} \over w}   +  i \sum_{n=1}^\infty  h_n  w^n  + c.c. \ ,$ where $h_{\pm 1}$ cannot both be zero.
The change of coordinate that brings this to the canonical forms \eqref{2forms} leaves invariant the piece of the
real axis inside a certain  radius of convergence.}
\bea\label{2forms}
 h =  {i \over w} + c.c. \ , \ \qquad {\rm or}\  \ \ \ h =  -i w + c.c.  \ .
\eea
Generic   points  of  $\p\Sigma$  correspond to  interior points of the eleven-dimensional
 geometry, and at such points    $h$ has   a simple zero.
  At points where $h$  has a pole   the radius of the AdS$_3$  fiber diverges, as we have argued in the previous section.
\sm

Inserting   (\ref{2forms})  for  $h$ in the  equation for  $G$,
and using polar coordinates ($w= re^{i\theta}$)  gives
 \bea\label{eqG1}
  (  \p_\theta + i r \p_r  ) G =  {\rm Re}(G) (  \cot\theta \pm i ) \ .
\eea
  The sign `$+$'  corresponds to the case when $h$ has a zero,   and the sign `$-$'  when it has a pole.
 We recall that  on the boundary     $G = \pm i$. This sign determines which of the two 3-spheres shrinks to a point,
 and it should not be confused with   the
 sign in \eqref{eqG1}.

Exploiting the linearity of the equation, we  break up $G$  into a $r$-independent solution
that obeys the required boundary conditions, and
  a solution that vanishes on the boundary,
  \bea
   G(r, \theta) = G_0(\theta) + \delta G(r, \theta)\  \qquad {\rm with}\ \ \ \delta G\Bigl\vert_{z=\bar z} = 0\ .
  \eea
 There are two possibilities for the $r$-independent solution $G_0$: it  either
 changes  sign as $\theta$ goes from $0$ to $\pi$,  or it does not. The corresponding solutions read
   \bea\label{twocases}
\underline{\rm no\  flip}: \ \  G_0 =  \,  i \ , \qquad
\underline{\rm   flip}: \ \ G_0 =  \, i\,   e^{ \pm i   \theta} = \, i \cos\theta \mp \sin\theta   \ .
\eea
To simplify the expressions  we have  assumed that $G=i$ at $\theta=0$.
If $G=-i$ on this half axis, one should simply replace  $G_0$ by $-G_0$.
\sm

  Next we turn  to   the homogeneous piece, $\delta G$.
  Since equation \eqref{eqG1} is invariant under dilatations $r\to\lambda r$,
we may  assume  a power-law  dependence on the radial coordinate.
We also  decompose the function into its real and imaginary parts,
\bea
 \delta G (r, \theta) = r^\nu G_\nu (\theta) \equiv r^\nu\, [G_{1, \nu} (\theta)  + i G_{2, \nu} (\theta) ]\ .
\eea
Inserting this ansatz  in (\ref{eqG1})
 leads to the following   coupled ordinary differential equations for the two real functions $G_{1, \nu}$ and  $ G_{2, \nu}$:
    \bea\label{F12}
( {d\over d\theta}  - { \cot   \theta }) G_{1, \nu} =   \nu  G_{2, \nu} \,
 \hskip 6mm {\rm and} \ \ \  \
{d G_{2, \nu} \over d\theta}  =  (-\nu \pm 1) G_{1, \nu}   \ .
\eea
 We eliminate  $G_{2, \nu}$ from  the first equation,
 define  $G_{1, \nu} := \sin\theta\, f_{1, \nu}$,  and  change variable to $x := \cos\theta$,
to finally get
 \bea
 (1-x^2)\, {d^2f_{1, \nu}\over dx^2} -2x\, {df_{1, \nu} \over dx} + \nu(\nu\mp 1)\, f_{1, \nu} = 0\ .
 \eea
 This  we recognize as the Legendre equation with $\mu=0$ \cite{grad}.
   The two independent solutions of this equation are the Legendre functions of the first and second kind,
   \bea
   f_{1, \nu}(\theta) =   a_\nu\, P_\kappa(\cos\theta)  + \tilde a_\nu\,   Q_\kappa(\cos\theta)\ ,
   \eea
 where $\kappa \geq -1/2$ solves  the quadratic equation
 \bea\label{kappa}
 \kappa (\kappa +1) = \nu (\nu \mp 1)\,\  \Longrightarrow\, \ \kappa = {\rm max} ( \mp\nu\,  , \, \pm\nu -1 )\ .
 \eea
 To complete the calculation, we need to find the  imaginary part of $G_\nu$  from the first  of the
 differential equations   (\ref{F12}),
 \bea
  \nu G_{2, \nu} =  \sin\theta\, {df_{1, \nu}\over d\theta} =  (x^2 -1)  {df_{1, \nu}\over d\theta}\ .
 \eea
The derivative of the Legendre functions can be re-expressed using the  identity
$$
(x^2-1) P_{\kappa}^\prime (x) = (\kappa +1) \left[  P_{\kappa+1}  (x) - x P_\kappa (x) \right]\ ,
$$
 which  holds also for the functions $Q_\kappa$. Putting everything together we  arrive at
 \bea\label{Gnu}
 G_\nu(\theta)  =  &a_\nu \Bigl[ \sin\theta\,  P_\kappa(\cos\theta) + i {\kappa+1\over \nu}
   \left[  P_{\kappa+1}  (\cos\theta) - \cos\theta\, P_\kappa (\cos\theta) \right] \Bigr]  \nonumber \\
  &+ \tilde a_\nu [ {\rm \ same\  with} \ \ P_\kappa \to Q_\kappa]\ .
 \eea

\smallskip

 The general solution of (\ref{eqG1})  can be expanded as   a   linear superposition $  \sum   r^\nu G_\nu(\theta)$.
From the properties of the Legendre functions one can check that  $G_\nu(0) =  G_\nu(\pi) = 0$ for all $\nu\not= 0$, so the
 Dirichlet boundary condition  puts no  restrictions on  the coefficients $a_\nu, \tilde a_\nu$.
 A non-trivial condition comes  however from the $\theta$-derivatives at $0$ and $\pi$,  which diverge logarithmically  for generic $\nu$.
This would lead to curvature singularities of the geometry, and   should be excluded. From
 \bea
  \partial_\theta G_\nu\Bigl\vert_{\theta= \epsilon }\    \sim \      \tilde a_\nu  \,  Q_\kappa(1- \epsilon^2/2)  \sim \   -   \tilde a_\nu \log \epsilon\qquad {\rm as}\ \epsilon\to 0
 \eea
we conclude  that all  coefficients $ \tilde a_\nu$ must vanish. Furthermore, from
 \bea
  \partial_\theta G_\nu\Bigl\vert_{\theta= \pi -\epsilon}\  \sim \   -  a_\nu  \,  P_\kappa(-1+\epsilon^2/2)  \sim
  - {2\over \pi}\,  a_\nu \, \sin(\pi\kappa)\, \log\epsilon \ \qquad {\rm as}\ \epsilon\to 0
 \eea
we find that the $a_\nu$ also  vanish  unless  $\kappa =$integer.
 The relation (\ref{kappa}) between $\kappa$ and $\nu$ shows
that,  after all the dust has settled,  one
is left with  an expansion involving  integer powers of $r$ multiplied by  Legendre polynomials.

 \smallskip

  The upshot of this analysis is that  the  solution of the $G$-equation  in the neighborhood of any
   boundary point can be expanded as follows:
  \bea\label{nusum}
 G = G_0(\theta) +  \sum_{0\not= \nu\in \mathbb{Z}}    a_\nu\,  r^\nu
 \Bigl[ \sin\theta\,  P_\kappa(\cos\theta) + i {\kappa+1\over \nu}
   \left[  P_{\kappa+1}  (\cos\theta) - \cos\theta\, P_\kappa (\cos\theta) \right] \Bigr] , \ \ \ \,
  \eea
 where $G_0$ is given by eq.\,(\ref{twocases}),   and
 $\kappa$ is determined  by eq.\,(\ref{kappa}).
 The $a_\nu$ are at this stage arbitrary real coefficients,  restricted only by the conditions  of global regularity
 in the interior of $\Sigma$.
 Recall  that this expression    depends,
 via $G_0$ and $\kappa$,  on whether
 $h$ has a simple zero or a simple pole at $r=0$.


 \subsection{Two  AdS throats  and a  cap }
 \label{sec:52}

    The general  solution \eqref{nusum} has terms with    positive and negative powers of $r$.
 We consider first  the case where $G$ is bounded at $r=0$, so that
 only  positive integers $\nu$ can enter in the sum (\ref{nusum}).
For  $\g<0$, this condition is required to keep $G$ finite.
   From
 eq.\,(\ref{kappa}) we deduce  that $\kappa = \nu -1$ when $h$ has a zero,  and $\kappa = \nu$ when $h$ has a pole.
The expansion (\ref{nusum}) in these two cases therefore reads
  \bea\label{expansionZ}
\underline{ {\rm zero }  }: \ \ \
&&G(r, \theta) =  \, G_0(\theta)  + i  \sum_{n=1}^\infty a_n\,  r^n \bigl[ P_{n}  (\cos\theta)   - e^{i\theta}
  P_{n-1}(\cos\theta)
   \bigr]
  \  \nonumber \\
 &&\simeq  G_0(\theta)   + a_1 r \sin\theta + a_2 r^2 (\sin\theta \cos\theta  - \frac{i}{2} \sin^2\theta) + \cdots \ ,
\eea
 \bea\label{expansionP}
\underline{ {\rm pole }  }: \ \ \
G(r, \theta) &&\hskip -2mm  =  G_0(\theta)  + \sum_{n=1}^\infty    a_n\,  r^n  \Bigl[ \sin\theta\,  P_n(\cos\theta)
  + i {n +1\over n}
   \left[  P_{n+1}  (\cos\theta) - \cos\theta\, P_n (\cos\theta) \right] \Bigr]   \nonumber \\
   && \hskip -12mm \simeq  G_0(\theta)  + a_1 r  \sin\theta\, e^{-i\theta}  +
  \frac{a_2}{2} r^2 \bigl[ \sin\theta (3\cos^2\theta -1) - 3i
 \cos\theta \sin^2\theta
 \bigr]
  + \cdots \ .  \nonumber  \\
  &\, &
\eea
 The coefficients $a_n$ are all  real,  and $G_0(\theta)$ is given by (\ref{twocases}).
\sm

These solutions satisfy  the boundary conditions $G=\pm i$, so the last thing we need to impose
is  the  regularity condition
 $\g(G\bar G -1)>0$ in the interior of  $\Sigma$.
 Near $r=0$,  the  sign of     $\g(G\bar G -1)$  depends on   the linear and quadratic terms of the
expansions \eqref{expansionZ} and  \eqref{expansionP}.
A little calculation gives in the case of a $h$-zero
  \bea
  \vert G\vert^2  -1\ \simeq      \left\{   \begin{matrix}   & (a_1^2 - a_2) r^2 \sin^2 \theta  \qquad \mbox{(zero, no  flip)} , \\
 & - 2 a_1 r \sin^2 \theta \qquad  \mbox{(zero, flip)}  ,    \end{matrix}  \right .
\eea
whereas in the case of a $h$-pole
\bea
 \vert G\vert^2  -1\ \simeq    \left\{   \begin{matrix}   &-2 a_1 r  \sin^2 \theta  \qquad\quad  \mbox{(pole, no  flip)} ,  \\
&(a_1^2 - a_2) r^2 \sin^2 \theta \qquad  \mbox{(pole, flip)}  .   \end{matrix} \right .
\eea
Note that in all cases   the right-hand side  has a definite sign, at all angles $\theta$.
 By choosing therefore  expansion coefficients so that $\g(a_1^2 - a_2) >0 $   or   $\g a_1<0 $,
 according to the case at hand,
 we can always  satisfy the regularity condition locally,   for any  $\g$,
 whether $h$ has a zero or a pole,  and whether $G$ flips or does not
 flip sign as $\theta$ goes from $0$ to $\pi$.
  \vskip 2mm

  Let us take a closer look at the above solutions.  Inserting the
  expansions  (\ref{expansionZ}) and (\ref{expansionP}) in   (\ref{2d2}) leads, after some
  straightforward algebra, to the following  expressions for the  space-time metric,
  \bea\label{asymototicCases}
  &&\underline{ {\rm zero,   no\,  flip } }:  \ \ \  ds^2 \simeq  B_1\, ds^2_{AdS_3} + B_2\, ds^2_{S^3_3} +
  B_3 (dx^2 + dy^2 + y^2  ds^2_{S^3_2}) \ ; \nonumber \\
 &&\underline{ {\rm zero, flip } }:  \ \ \  ds^2 \simeq  B_1 \, ds^2_{AdS_3} + B_2\,  r\,  \left[  {1\over 4r^2} (dr^2 + r^2 d\theta^2) +
  \sin^2({\theta\over 2}) ds^2_{S^3_2} +  \cos^2({\theta\over 2})\,  ds^2_{S^3_3}
 \right] \ ; \nonumber   \\
 &&\underline{ {\rm pole, no\,  flip } }:  \ \ \  ds^2 \simeq  {1\over r} (B_1\, ds^2_{AdS_3} + B_2\, ds^2_{S^3_3})
 + B_3 \left({dr^2\over r^2}  +  d\theta^2 + \sin^2\theta\,  ds^2_{S^3_2}\right)\ ;
 \nonumber \\
 &&\underline{ {\rm pole, flip } }:  \ \ \  ds^2 \simeq {B_1 \over r^2}  \, ds^2_{AdS_3} + B_2\,
  \left[ {1\over 4r^2} (dr^2 + r^2 d\theta^2)  +
 \sin^2({\theta\over 2})\, ds^2_{S^3_2} +  \cos^2({\theta\over 2}) \, ds^2_{S^3_3}
 \right] \, .
 \nonumber \\
 &&\,
 \eea
In the "zero, no flip"  case, we have used the Cartesian rather than the polar  parametrization
of  $\Sigma$, $w=   x+iy $.
  The constants $B_1, B_2, B_3$ in these expressions are combinations of the $c_i$ and of the expansion coefficients,
 $a_j$,  of the function $G$.
 They are different combinations in each case, but we  use  the same symbols for economy
 of notation.  In the "pole, no flip" case for example they read
\bea\label{b1b2b3}
B_1 =  {2 \over c_1^2} \left( { -\g a_1 }\right)^{-1/3} \ ,\quad
B_2 =  {2\g \over c_2c_3} \left( { -\g a_1 }\right)^{-1/3} \ ,\quad
B_3 = {4 \over c_2c_3}\left( { a_1^2\over \g }\right)^{1/3} \ .
\eea
There are  similar  fomulae  in the  other three cases.

\if
  In the "zero, no flip" case for example they read
    $$
 B_1 =   {2\over c_1^2} \left( {  a_1 \over 2}\right)^{-2/3} \ ,
 \qquad  B_2 = - {8\over c_2 c_3} \left( {a_1\over 2}\right)^{1/3}\ ,
 $$
 \fi

 \sm

  The first expression in (\ref{asymototicCases}) is the metric of AdS$_3\times$S$^3\times\mathbb{R}^5$.
  This is the metric near a generic   point of $\p\Sigma$, where both $h$ and $G$ are continuous,
  as was already noted in the discussion of regularity conditions,
  in section \ref{sec:25}. The second expression asymptotes to the metric of AdS$_3\times \mathbb{R}^8$.
This can be  seen by  changing  coordinates to $r= \tilde r^2$,
so that  $\tilde r$ is a local polar coordinate for
$\mathbb{R}^8$, while $\theta$ parametrizes  the line-segment coordinate of  a  seven-sphere written as a
fibration  (S$^3_2\times$S$^3_3)\ltimes I$ (with $I$ the line segment).
Both  $h$-zero cases correspond   therefore to regular interior regions of the supergravity geometry.
The singularity of $G$ in the second case is a coordinate singularity, analogous to  the "smoothly-capped"
D3-brane throats found in the type-IIB context, in  \cite{Aharony:2011yc,Assel:2011xz}.\,\footnote{For general
$h \simeq ih_{-1}/w +  i h_1 w + c.c.$, taking the residue, $h_{-1}$, of the pole to zero amounts
to sending to zero  the number of   (semi-)infinite  M2-  or M5-branes that create the throat. The  limit is not
strictly-speaking smooth, since the pole changes the topology and the boundary of space-time.
  } We will refer to it as a "cap". Note that there are no non-trivial cycles of any dimension at a cap.

\sm

The two other solutions near a $h$-pole are higher-dimensional Anti-de Sitter  throats.
 The third  metric in (\ref{asymototicCases}) is AdS$_7^\prime \times$S$^4$, where the prime here  signifies  that
 in general   the maximally-symmetric AdS$_7$
space-time is deformed. Maximal symmetry requires that
 $B_1=B_2$,  which   from    \eqref{b1b2b3}   can be seen to imply $\g=-2$.
 This is indeed the value that corresponds to the superconformal symmetry of  pure AdS$_7  \times$S$^4$
 space-time \cite{D'Hoker:2008wc}.\,\footnote{The reason why we don't also find
 $\g=-1/2$ is because  we broke  the symmetry $\mJ$ by choosing  on the boundary
 $G_0=+i$ rather than $-i$, see eq.\,(\ref{twocases}).
  }
   Finally, the fourth metric in (\ref{asymototicCases}) is (AdS$_4/Z_2) \times$S$^7$, which is the
   near-horizon geometry of semi-infinite M2 branes. Interestingly,  the  (AdS$_4/Z_2) \times$S$^7$ asymptotics
   are  compatible with
   any value of $\g$, even though the maximally-symmetric  AdS$_4  \times$S$^7$  background
   has $\g= 1$.


\subsection{ M5-brane singularities}
\label{sec:53}

 For $\g$ positive,  $G$ is   allowed to diverge, so one
  should also consider  negative powers in the expansion \eqref{nusum}.
  Inspection of the expressions  \eqref{2d2} for the metric factors shows that,  if  $G$ diverges as
  $ \sim r^{-n}$, then   $f_j \sim  h^{1/3}$ and
 $\rho \sim  r^{-n}  \vert \p_w h\vert\, h^{-2/3}$. Since the holomorphic part of  $h$ has either
  a simple zero or a simple pole,  we  find
 \bea\label{verycurvedasym}
 \underline{\rm zero}:\ \   f_j \sim  r^{1/3}, \ \ \rho\sim r^{-(n+2/3)}\ ;\qquad
  \underline{\rm pole}:\ \  f_j \sim  r^{-1/3}, \ \ \rho\sim r^{-(n+4/3)}\ .
 \eea
  In both cases, the origin at $r=0$ is  infinitely far  in the eleven-dimensional metric.
 In   the case of a $h$-pole all scale factors diverge, and the geometry is
 asymptotically flat. This behavior is legitimate, and we will see an example in section \ref{sec:71}.
However,  such solutions  have no holographic interpretation and will not really concern us here.

 \sm
 The case of an $h$-zero is different, because the (pseudo-)sphere radii vanish   as $r\to 0$.
 Such regions do not therefore change the dimension of the conformal boundary of space-time,
  but  the geometry
 is highly curved and  the supergravity approximation breaks down.
 For $n>1$, these local solutions do not carry any M5-brane charge,  and they have no obvious
 interpretation in M theory. To see why  there is no 5-brane charge, note that this
 is given by the discontinuity
 of the auxiliary function $\Phi$, defined in \eqref{2e3}, across the singular point on the real axis.
 But $\Phi$   scales near the singularity  as $\sim r^{-n+1}$,
 so the only consistent value for the M5-brane charge, when $n>1$,  is zero.

\sm

  Let us then focus  on
   the case $n=1$,  which corresponds to the  $\nu=-1$ term in the general expansion
  \eqref{nusum}.  Since we are near a zero of $h$, eq.\,\eqref{kappa}  implies that
 $\kappa=1$. Using the standard Legendre polynomials we find
 \bea\label{singM5}
G = G_0 + \delta G\ , \qquad  {\rm with}\quad
 \delta G =  {a  \over r } ( \sin\theta \cos\theta +  i  \sin^2\theta ) =   a  {  w \,{\rm Im}(w)  \over   \vert  w\vert^3}\ ,
 \eea
where  $a$ is a shorthand notation for the arbitrary coefficient $a_{-1}$ in  \eqref{nusum}.
 From   \eqref{2e3} we find the following contribution of $\delta G$ to $\Phi$,
\bea\label{dPhi}
\delta\Phi = -2a \cos\theta = - a\, (\sqrt{w\over \bar w} + \sqrt{\bar w\over  w})\,  .
\eea
 Inserting this in the formulae  \eqref{2e2} and  \eqref{2g4} for the 3-form potential  and M5-brane charges,
 and noting that $\tilde h$ is continuous,   yields
 \bea\label{624}
 \mM^{(j)} \,=\,  {\nu_j\over c_j^3}\, b_j^c\Biggl\vert_{z(0)}^{z(1)}
\  =\  \left \{   \begin{matrix}    4 a  {\nu_2 \g / c_2^3}\ ,    \ \ \ j=2\,,     \\   \\ - {4 a }  {\nu_3/ (\g c_3^3) } \ ,
\ \ \ j=3\,.   \end{matrix}   \right .
 \eea
The upper result  ($j=2$)  is relevant  if  $G_0=i$,  so that S$_2^2$ shrinks to a point  on $\p\Sigma$, while the lower
result ($j=3$)  applies when it is  S$_3^2$ that collapses  to a point on $\p\Sigma$, which happens if $G_0=-i$.
The case where $G_0$ is a flip need not be treated separately, since it is always possible to separate it from
the singularity of $\delta G$ on  $\p\Sigma$.
\sm

The upshot of this analysis is that the (local) solution \eqref{singM5} describes an M5-brane
or M5$^\prime$-brane stack with a
 AdS$_3\times$S$^3$ world-volume. Unlike the AdS$_7^\prime\times$S$^4$ throat, this solution  does not change
 the dimension of the conformal  boundary of space-time.  Strongly-curved five-brane regions of a similar kind
 entered also  in the type-IIB solutions of refs.\,\cite{Aharony:2011yc,  Assel:2011xz}.
Finally,  let us   calculate the M2-brane charge of this  local solution.
The contribution from $\delta G$ to the auxiliary
 function $\Lambda$ defined by
  eq.\,\eqref{lambda3.12} reads
  \bea
  \delta \Lambda = a \left( {w^2\over \vert w\vert} + {{\bar w}^2\over \vert w\vert }  + 6 \vert w\vert \right)\ .
  \eea
This is continuous  at the position $w=0$ of the singularity, so it does not contribute to
the M2-brane charge. The only terms in the expression \eqref{finalomegac} for
 the cohomological
piece,   $\Omega_1^c$, which have a discontinuity on the real axis  are  the terms  $-\tilde h \delta\Phi + \epsilon \Phi^2/2$.
We will use this fact in section \ref{sec:73}, where we discuss global solutions with  $G$ singularities of this type.

\if
Inserting $\delta\Lambda$ and $\delta\Phi$ in \eqref{finalomegac} shows that the cohomological
piece  $\Omega_1^c$  has no  discontinuity, so it does not
contribute an  M2-brane charge.
Recall however that \eqref{finalomegac}  was obtained by setting the
constant pieces of the magnetic potentials, $b_j^0$, to zero.
Putting these back gives a gauge-variant
M2-brane  charge  proportional to  $b^0_2\, \mM^{(3)}$ if $G_0= -i$,
or  $- b^0_3\, \mM^{(2)}$   if $G_0=i$.
We will use these expressions   in section \ref{sec:73}
\fi

\section{Global  solutions for $\g <0$}
\setcounter{equation}{0}
\label{sec:6}

 We turn now to  global solutions of the supergravity equations,  considering  first the case of negative $\g$.
 The theorem of   section \ref{sec:4.4} restricts $h$ to be either constant, in which case the only solution is
 AdS$_3\times$S$^3\times$S$^3\times$E$_2$ with $\g>0$, or to
  have a single singularity. Without loss of generality, a singular $h$ takes one of two forms,
  according to whether the singularity is on the boundary or in the interior of $\Sigma$,
  \bea
  \label{section6h}
h = -i   (w - \bar w) + c.c. \ , \qquad {\rm or} \quad\ \  h = a \log\left( {w+ i\over w- i}\right) + c.c. \ .
\eea
 Here $\Sigma$ is the upper-half complex plane.
 Note that  the singularities of $h$ are at  $w= \infty$ or at $w=i$, but they could  be placed anywhere else
 on $\p\Sigma$ or $\Sigma$ by a fractional linear transformation of the  complex half-plane.

\sm
     In this section we will only consider the case of the boundary singularity.  We will review the exact
     solutions found in refs.\,\cite{D'Hoker:2008qm, Estes:2012vm}
      from a unified perspective, and  calculate their M2-brane
      and M5-brane charges. This will establish a correspondence between the solutions in
        \cite{D'Hoker:2008qm},  and self-dual strings in
          arbitrary representations  of $SU(N)$ labelled  by   Young tableaux.
          We  also exhibit  a dual description of these configurations, as  boundary conditions for semi-infinite
          M2 branes.


\subsection{$\g$-Deformed AdS$_ 7\times$S$ ^4$}

The simplest solution   is the AdS$_ 7\times$S$ ^4$ background, which
corresponds to the near-horizon geometry of a stack of M5-branes.
The data $(\g, h, G)$ for this background reads
\bea\label{ads7S4hG}
  \underline{{ AdS}_7 \times { S}^4} :\quad    \g = -{1\over 2}\, , \quad     h = -   i   (w - \bar w), \quad
G =   i \left(-1 + \frac{w + \xi }{|w + \xi |} - \frac{w-\xi }{|w-\xi |} \right)\,.
\eea
There is one (positive) free parameter $\xi  = 2 L^3 c_1^3$\,,\,  where  $L$ is   the radius of  S$^4$.
This is a solution with one pole in $h$, and two cap singularities at $w = \pm \xi$.
There exists, as we explained in  section \ref{sec:2i}, an equivalent solution
 with $\g=-2$ and $G$ equal to minus the above expression. This follows from the symmetry $\mI$ of the problem.

\sm

 To check that  \eqref{ads7S4hG}  is indeed the AdS$_ 7\times$S$ ^4$ solution, it is convenient to
  change  coordinate to  $ w = \xi \cosh(2z)$, thereby mapping  the upper-half complex plane
 to the semi-infinite strip $z = x + iy \in [0, \infty) + i  [0, \pi/2]$, as illustrated in  Figure 3.   In the new coordinate system
\bea\label{AdS7S4}
 h =  - i\xi  \cosh(2z) + c.c. \,, \qquad
 {\rm and}\quad G = - i \,\left[  1 + 2 \,   {\sinh( z- \bar z)\over \sinh(2\bar z)}  \right]  \ .
\eea
The two flip singularities are now located at the corners of the semi-infinite strip, and  the metric reduces to the
simple form
\bea\label{AdS7S4metric}
ds^2 =  4L^2 \left[ \cosh^2 \hskip -0.8mm x  \,
ds^3_{AdS_3} + \sinh^2 \hskip -0.8mm x  \, ds^2_{S^3} + dx^2\right] +   L^2   \, ds^2_{S^4}  \ ,
\eea
with $2y$ playing the role of  latitude on  the four-sphere. This is precisely AdS$_ 7\times$S$ ^4$
written as a AdS$_ 3\times$S$ ^3\times$S$^3$ fibration over $\Sigma$.
Some of the intermediate calculations necessary to derive the above metric can be found   in
appendix \ref{sec:E}.


\begin{figure}[htb]
\label{fig:3}
\begin{center}
\includegraphics[width= 5.1in]{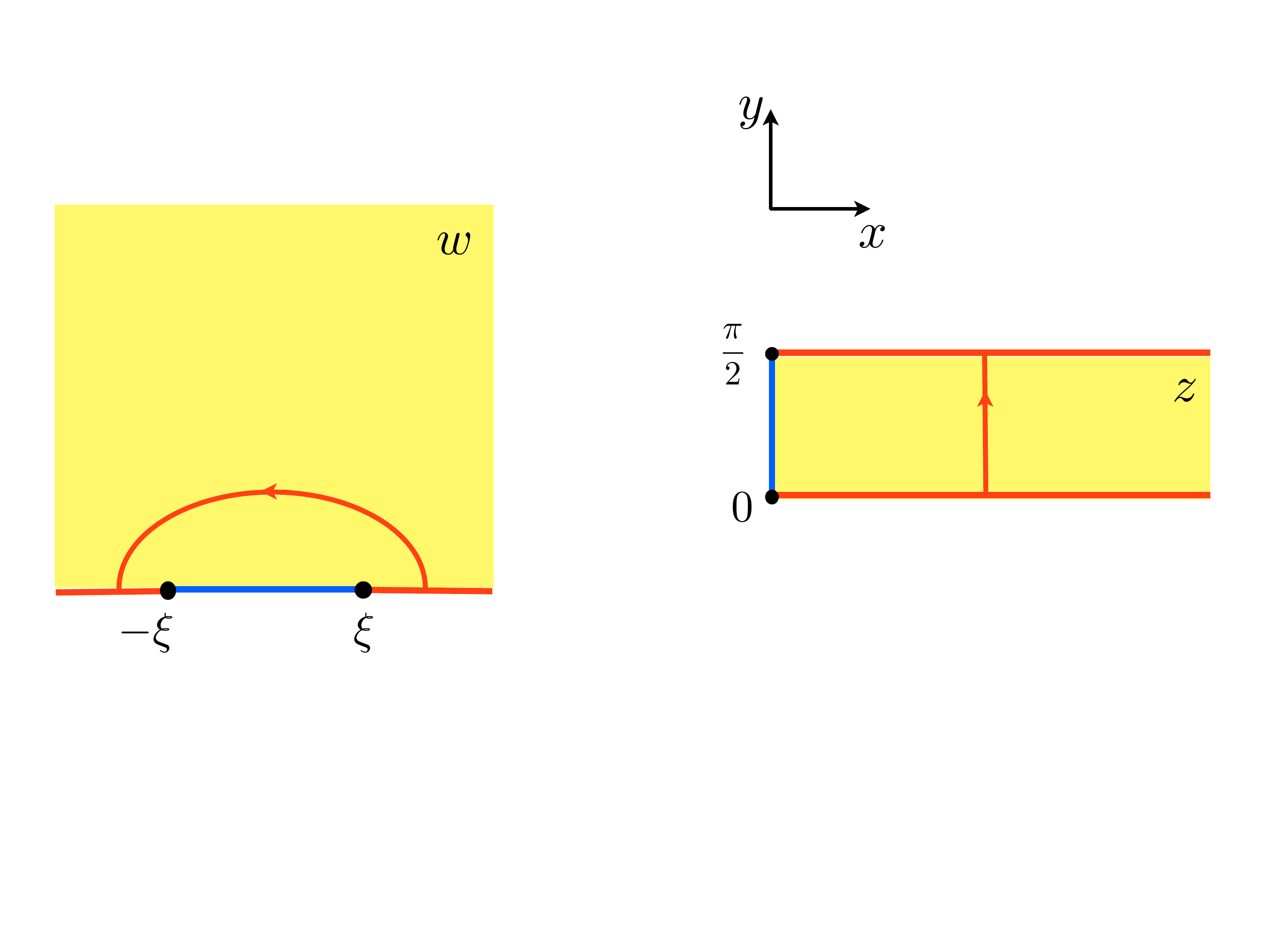}
\vskip -2.3 cm
\caption{\footnotesize The coordinate change $w\to z$  that maps the upper half plane to the semi-infinite strip,
as described in the text. The pole of
the holomorphic part of $h$ is at infinity.
The two flip singularities at $w= \pm\xi$ are mapped to the two corners of the
strip at $z=0, i\pi/2$.  The non-trivial 4-cycle has  as
basis the open curve marked in blue, following the color convention of  Figure \ref{fig:2}.}
\end{center}
\end{figure}


Consider next the deformed solution, obtained as $\g$ moves away from the special
value   $ \g = -1/2$.    From the results  of appendix \ref{sec:E}
one  can calculate  the deformed metric,  which approaches as $x \to \infty$
  the AdS$_7^\prime\times$S$^4$ metric given in
 \eqref{asymototicCases}.
 The nature of the deformation can be understood
more readily from the expressions for the flux potentials, eqs.\,\eqref{2e1} and \eqref{2e2}.
Using the results of  appendix  \ref{sec:E},  one finds in particular
\bea\label{flux6.5}
b_2 = {8\xi \,\nu_2\over c_2^3}    \,\g (2\g+1) \, {\cos(2y) \sinh^4\hskip -0.4mm x \over
(2\g \sinh^2\hskip -0.4mm x - 1) }  \ .
\eea
This vanishes  for $\g= -1/2$, consistently with the expectation that
the  only  flux  of the AdS$_ 7\times$S$ ^4$ background is the one threading $S_3^3$,  which is part of the
 4-sphere.\footnote{It can be checked that
also $b_1$ is constant when $\g=-1/2$.} For general $\g$,
the $db_2$ flux field is turned on, but it remains  cohomologically trivial. This is because  there
is only one boundary segment where $f_2=0$, so there is  no 4-cycle of   type  $\cC^{(2)}$.
 In \eqref{flux6.5} we have fixed
 the constant ambiguity  so that $b_2=0$ on this boundary segment, at $x=0$.
Thus $b_2$ is globally defined, and has trivial  cohomology.

\sm

The only non-trivial cycle  in the above geometry is actually the  4-cycle
$$\cC^{(3)} =
\{ y\in [0, \pi/2] \} \times S_3^3\  .
 $$
The corresponding magnetic   charge is  given by the change in  $b_3^c$
as $y$ ranges from  $y=0$ to $y=\pi/2$.  An easy computation, see appendix \ref{sec:E}, gives
 $b_3 ^c = 4\xi \g^{-1}\cos(2y)$, from which we obtain the M5-brane charge
 \bea
\mM^{(3)} \, = \,   { \nu _3 \over c_3^3}.
\, b_3 ^c \, \bigg | ^{y=\pi/2} _{y=0}\  =\   - { 8\xi   \over \g   }\,{ \nu _3 \over c_3^3}\ .
\eea
 Since $\xi$ and $\g$ are independent parameters, we may adjust the former while varying the latter so as
 to keep the M5 charge fixed. The net effect of the deformation at fixed M5 charge
is then  to turn  on  a non-zero M5$^\prime$-brane  flux.
  Since there is no corresponding charge,  this is a multipole field that should be attributed to dielectric
  M5$^\prime$ branes.

    \sm

  The $\g$-deformed background has also non-vanishing M2-brane flux, but there is no corresponding electric  charge.
  This follows from our  discussion in section \ref{sec:2h}, where we argued very generally that the $\g$ deformation
  may rescale the charges of a solution, but does not add new charges on top of the  ones that existed  before the
  deformation.  Alternatively, it can be seen that the deformed AdS$_7\times$S$^4$  geometry  does not have any
  non-trivial 7-cycles, because the 7-spheres around flip singularities can be contracted away to a point.
  The
  $ \cD^{(3)} =  \cC^{(3)} \times S_2^3$, in particular,
   can be written as the sum of two   trivial such cycles, so it is also trivial and it cannot support
  M2-brane charge.


\subsection{Self-dual strings and Young diagrams}
\label{stringsols}
\setcounter{footnote}{0}

The  AdS$_ 7\times$S$ ^4$ solution  was generalized in ref.\,\cite{D'Hoker:2008qm}  by the addition of an
   arbitrary even number of cap singularities  on the boundary of $\Sigma$.
 The functions $h$ and $G$ for these solutions read\footnote{We  have here  fixed the sign of $G$ at infinity,
so  these functions   give inequivalent solutions for all values of the parameter $\g$, with  $0<\g<-\infty$.   }
 \bea
 \label{stringGh}
  \underline{\rm    string\,\,solutions } :\quad    \quad     h = -   i   (w - \bar w), \quad
G =   - i \left( 1+   \sum_{j=1}^{2n+2}  (-)^j  \frac{w -  \xi_j }{|w - \xi_j |}   \right)\, ,
\eea
where $\xi_1 < \xi_2 \cdots < \xi_{2n+2}$. Since each term in the sum  obeys the linear  $G$-equation,
the same is true for their superpositions with real coefficients.
Choosing these coefficients as we did ensures that on
  the boundary
$G$ alternates between $-i$ and $+i$, in accordance with our regularity conditions.

Regularity  also requires that  $\vert G\vert <1$ in the $\Sigma$ interior,
which  has been shown in ref.\,\cite{D'Hoker:2008qm}.
A simple  proof follows from considering
 \bea
i G =     -1  +   \sum_{j=1}^{2n+2} (-)^j e^{i\theta_j}\ , \qquad {\rm where}\quad \theta_j = {\rm arg}(w-\xi_j)\ ,
\eea
as a function of the $2n+2$ variables $\xi_j$, or equivalently as a function of the $\theta_j$ in the
domain $0\leq \theta_1 \leq \theta_2 \cdots \leq \theta_{2n+2}\leq \pi$.
Clearly,  $G\bar G$ is bounded and real, so it is  enough to check the inequality at  the  extrema of this function
and   on the boundary of the domain  of the $\theta_j$.
On the boundary, at least  two of the angles are equal,
say $\theta_j=\theta_{j+1}$. This implies that $\xi_j = \xi_{j+1}$, so there are only $2n$ flips, and the proof
can proceed by induction on $n$.
 As for the extrema, these are
   given by the equations
\bea
e^{i\theta_j} \,\bar G + e^{-i \theta_j}\, G = 0\qquad {\rm for\ all}\ \ j\ ,
\eea
which imply that either $G=0$, in which case the inequality is satisfied, or that $e^{2i\theta_j}$ is the same for
all $j$. Since the angles must be different, the only remaining option is  $n=0$, $\theta_1=0$ and
$\theta_2=\pi$, which  happens on the boundary of $\Sigma$ where $G=\pm i$.  This completes the proof
  that  $\vert G\vert <1$ everywhere in the interior of $\Sigma$.

\sm

At $w\to\infty$,  the solution given by \eqref{stringGh} has (in the language of section \ref{sec:52})
 a "pole-no flip" singularity, so it approaches
 AdS$_7^\prime\times$S$^4$. Such solutions describe therefore conformal defects of the M5-brane theory.
 To analyze their nature, note  that the space-time manifold has now
$2n+1$ independent non-contractible  4-cycles,  as illustrated in Figure 5.
There are $n+1$ cycles of type  $\cC^{(3)}$, supporting M5-brane charges, and
$n$ cycles of type $\cC^{(2)}$ supporting M5$^\prime$-brane charges.
These cycles   are non-contractible  because the corresponding  open curves surround two
consecutive flip points.

\begin{figure}[htb]
\label{fig:4}
\begin{center}
\includegraphics[width= 5.1in]{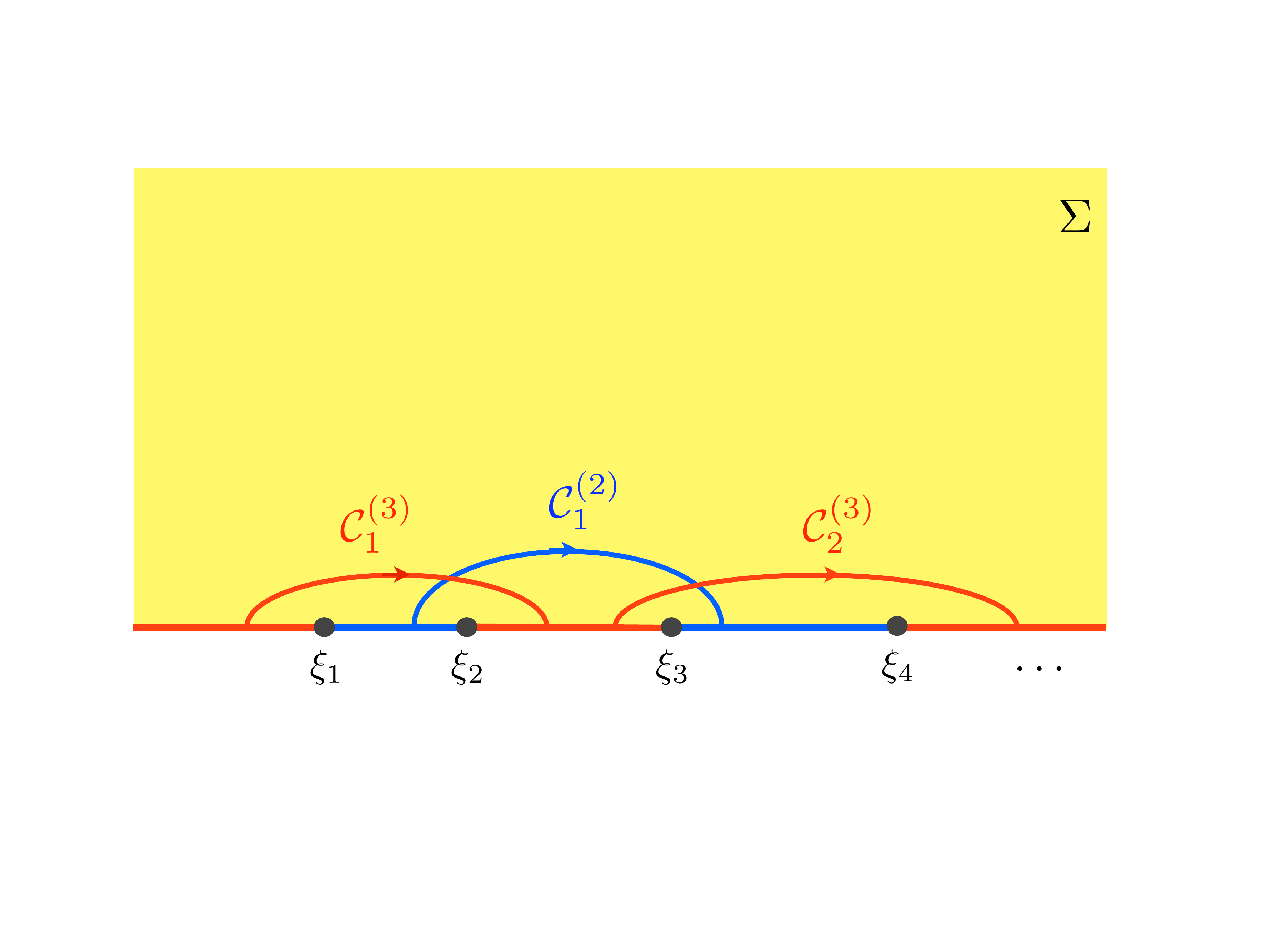}
\vskip -2.3 cm
\caption{\footnotesize At a "cap" singularity,      $G$ flips sign  from $\pm i$ to $\mp i$.
The asymptotic geometry  is AdS$_7^\prime\times$S$^4$  when the number of caps is even, and
AdS$_4\times$S$^7$ when there is an odd number of caps. For  $\g>0$ these cap singularities
should be replaced by curved-M5 regions, as we will discuss in  section \ref{sec:7}. The open curves
are the bases of independent non-contractible 4-cycles. Their number is one less   than the number of flips.
  }
\end{center}
\end{figure}


 In order to calculate  the  5-brane charges, we need
the real auxiliary function $\Phi$. This was
 defined by eq.\,\eqref{2e3} which, when
   $h = -iw + c.c.$,
   takes  the simple form $\p_w\Phi = -i \bar G$. Integrating this equation gives
\bea
\Phi(w,\bar w) \, = \,  2 \sum_{j=1}^{2n+2}  (-)^j  \vert w - \xi_j\vert + \Phi_0\ .
\eea
 Using also the dual harmonic function,  $\tilde h = -(w+\bar w)$,
 leads to   the following cohomological parts of the magnetic potentials, eq.\,\eqref{2e2},
 \bea\label{bsec72}
 &&b_2^c =  b_2^0 +  2\g   \sum_{j=1}^{2n+2}  (-)^j  \vert w - \xi_j\vert  + \g (w+\bar w)\ , \no \\
 && b_3^c =  b_3^0  -   {2\over \g}    \sum_{j=1}^{2n+2}  (-)^j  \vert w -    \xi_j\vert  +  {1\over\g}   (w+\bar w)\ .
 \eea
 The constant $\Phi_0$ has been absorbed   in the $b_j^0$. On the boundary, $w= x\in \mathbb{R}$, these
 potentials  are piece-wise linear functions of $x$, with the property that  $b_2^c$ is constant in the intervals
 $[\xi_1, \xi_2]$, $[\xi_3, \xi_4] \cdots , [\xi_{2n+1}, \xi_{2n+2}]$  where the sphere $S_2^3$ shrinks to a point,
whereas  $b_3^c$ is constant in the intervals
 $[-\infty, \xi_1]$, $[\xi_2, \xi_3] \cdots , [\xi_{2n+2}, \infty]$  at which  the sphere $S_3^3$ shrinks to a point.
 Outside these constant plateaux,   $b_2^c$ grows linearly with slope equal to $2\g$, and $b_3^c$ grows
 linearly with slope equal to $2/\g$.

  From the above potentials,  and our master formula  \eqref{2g4},
  one reads immediately the  invariant  M5-brane charges of the solution.
 They are given by
 \bea\label{M5selfdual}
\mM^{(2)}_a =  {4\nu_2\g\over c_2^3} (\xi_{2a+1} - \xi_{2a})\  \ \quad {\rm and}\quad\
\mM^{(3)}_b =  {4\nu_3\over \g c_3^3} (\xi_{2b} - \xi_{2b-1})\ ,
 \eea
 where $a=1, \cdots , n$ and $b= 1, \cdots , n+1$.
 The charges are proportional to the
 lengths of the intervals between successive flips, so they are free parameters.
 The charge of the asymptotic AdS$_7^\prime\times$S$^4$ region is $\sum_{b=1}^{n+1} \mM^{(3)}_b$,
 and its radius (at $\g=-1/2$) is
 \bea
   4L^3 c_1^3 = \sum_{j=1}^{2n+2} (-)^j  \xi_j  \ .
\eea


\begin{figure}[htb]
\label{fig:6}
\begin{center}
\includegraphics[width= 5.5 in]{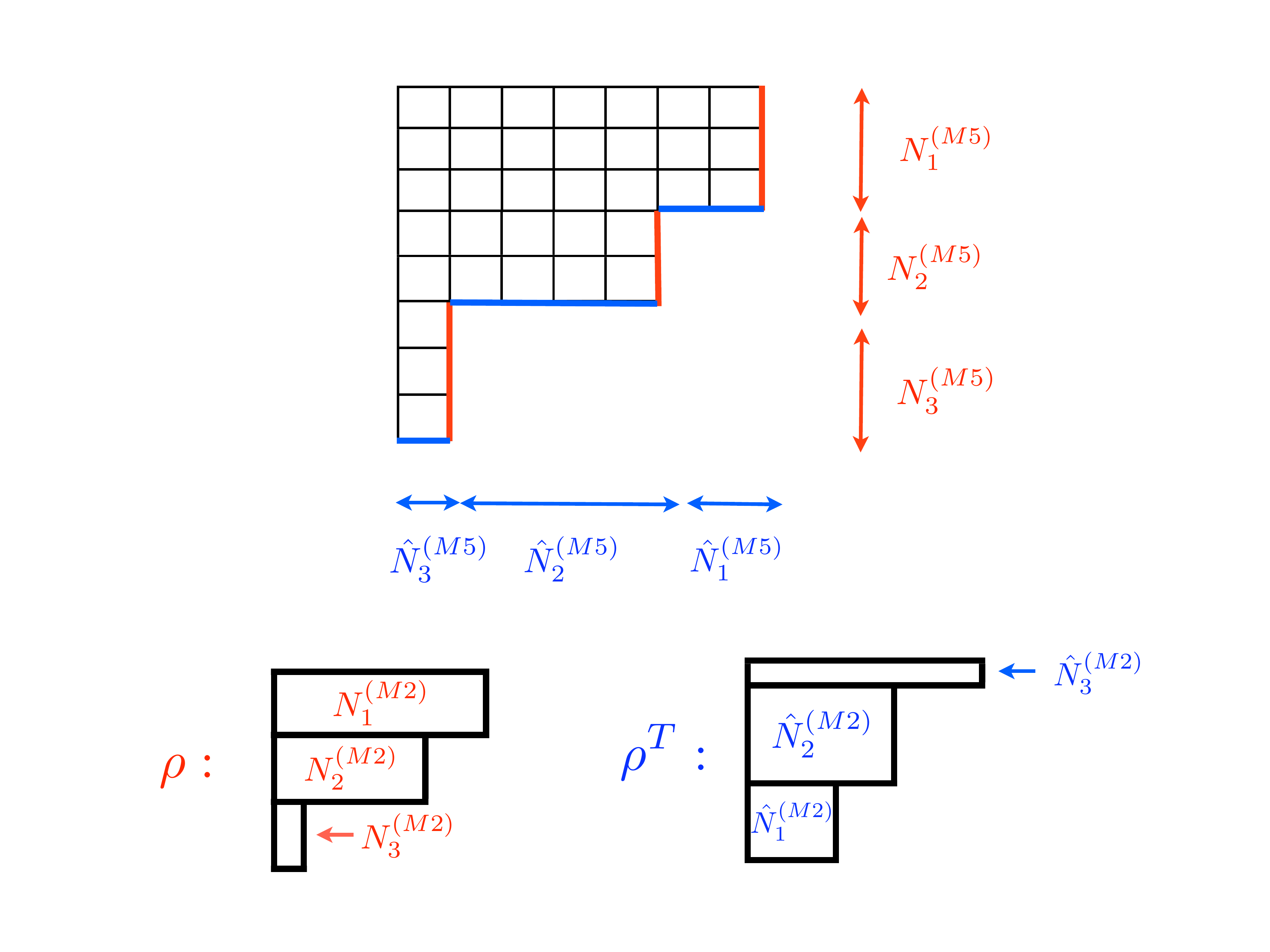}
 \caption{\footnotesize  The Young diagram associated to the 5-brane charges of the supergravity solution.
  The number of boxes in  the $b$th  M5-brane  stack can be identified with
 the membrane charge  $N_b^{(M2)}$, as shown in the lower left diagram. The transposed diagram on the right
 gives the numbers of boxes in M5$^\prime$ stacks.
  }
\end{center}
\end{figure}


In M theory, all electric and magnetic charges are quantized in units of the M2-brane  and M5-brane tensions,
according to the following rules,
\bea\label{Diracqnt}
2\pi^2 \mM = 2\kappa^2_{11} T_5^M \times {\rm (integer)}\ , \qquad
(2\pi^2)^2  \mE = 2\kappa^2_{11} T_2^M \times {\rm (integer)}\ ,
\eea
where in terms of the 11-dimensional Planck length $ (T_5^M)^{-1} = (2\pi)^5\ell_{11}^6$,
$ (T_2^M)^{-1} = (2\pi)^2\ell_{11}^3$ and $2 \kappa_{11}^2 = (2\pi)^8\ell_{11}^9$.
We will use the following notation for the integer 5-brane charges,
\bea
\mM^{(2)}_a \to \hat N_a^{(M5)}\ ,\qquad
 \mM^{(3)}_b\to N_b^{(M5)}\ , \qquad {\rm and}\ \ \
 N \equiv  \sum_{b=1}^{n+1} N_b^{(M5)}\ .
\eea
  These
  charges determine a Young diagram, and a corresponding representation of $SU(N)$,
in the way  illustrated by   Figure 6.  Note that $N$ is the number of M5 branes at the conformal boundary of the
solution.\footnote{For the purposes of this discussion, we may  choose the signs  $\nu_2$, $\nu_3$ so as
to make all  charges positive.}
 The M5 charges  and M5$^\prime$ charges,
 $$\{N_1^{(M5)}, N_1^{(M5)},  \cdots , N_n^{(M5)}\} \quad
 {\rm and}\quad \{ \hat N_1^{(M5)}, \hat N_1^{(M5)},  \cdots , \hat N_n^{(M5)}\}\  ,
 $$
 count, respectively,    lines of equal length and columns of equal height  in the Young diagram.
    The story  is familiar from the  study of holographic Wilson lines in the ${\cal N}=4$ super Yang-Mills
\cite{Gomis:2006sb,Yamaguchi:2006te}.  We conjecture by analogy that  the solution
\eqref{stringGh} describes surface operators in the representation of $SU(N)$ that corresponds to the
above Young diagram.
 Note that the total number of lines of the diagram,   $N- N_{n+1}^{(M5)}$, is less than $N$,
in agreement  with the fact that  all the columns   must contain less than $N$ boxes.
 \sm

   One  may wonder if the numbers of boxes in the diagram can be related to
    M2-brane charges.  This can be indeed  done
    by assigning an M2  charge to each 7-cycle of the form $S^3\times \cC$,
    where $\cC$ is one of the $2n+1$  independent non-contractible 4-cycles.
    We choose  a gauge such that  $b_3 =0$
    on the  boundary interval  $(-\infty, \xi_1]$, and $b_2=0$ on the
    boundary interval  $[\xi_{2n+1}, \xi_{2n+2}]$, and we deform the integration contours so that  $\cC^{(3)}_b$
 coincides  with the
 boundary interval $[\xi_{2a-1}, \xi_{2a}]$ along which the potential $b_2$ is constant, while
 $\cC^{(2)}_a$ coincides  with  the
 boundary interval  $[\xi_{2b}, \xi_{2b+1}]$ along which the potential $b_3$ is constant.
  It is simpler to   use directly the definitions \eqref{2f3} and \eqref{ambg}
     of the 7-form, rather than our reduced expressions for electric charges.
     Following the discussion of  section \ref{sec:3},
we choose  $\epsilon = +1$  for  the cycles with base $\cC^{(3)}_b$,  and $\epsilon = -1$ for  those with base $\cC^{(2)}_a$.
  A straightforward  calculation with the help of eq.\,\eqref{bsec72}  and the quantization conditions \eqref{Diracqnt}
  leads then to the following integer M2-brane charges,
   \bea\label{M2redh}
&& \hat N_a^{(M2)} =   \hat N_a^{(M5)} \left[   N_1^{(M5)} +   N_{2}^{(M5)} + \cdots    N_a^{(M5)} \right]\ ,
\no\\ \, &&\no\\
&& N_b^{(M2)} =   N_b^{(M5)} \left[  \hat N_n^{(M5)} +   \hat N_{n-1}^{(M5)} + \cdots  \hat N_b^{(M5)} \right]\ .
 \eea
One can indeed verify that   $\hat N_a^{(M2)}/ \hat N_a^{(M5)} \equiv \hat l_a$ and $N_b^{(M2)}/ N_b^{(M5)}\equiv l_b$ are, respectively,
 the number of boxes in a column of the $a$th stack, and
  the number of boxes in a row of the $b$th stack.
    Furthermore, with this choice of   gauge for $b_2$ and $b_3$,
the $(n+1)$th M5-brane stack has zero  M2-brane charge,  which is consistent with the fact that it
 does not enter in the determination of the
 Young diagram.

 We have thus succeeded in associating  the  boxes of the Young diagram with M2 branes. It should be noted, nevertheless,
that    this calculation is  somewhat of a red herring. Indeed, the geometry of the solution \eqref{stringGh}
   does not  have any non-contractible 7-cycles, because the 7-cycles  at the  "cap" singularities  are cohomologically trivial.
  Furthermore,   the  M2-brane charges
   \eqref{M2redh} carry no  new information, as is clear from eqs.\,\eqref{M2redh}.
  The  only topological invariants of the solution are    $N$,  and an irreducible representation of $SU(N)$,
   both of which are fully determined by the 5-brane charges.


\subsection{Semi-infinite M2 branes}
\label{sec: 73}

An intriguing set of new solutions can be obtained with little effort by a small modification of  the data
$(\g, h, G)$ given  in \eqref{stringGh}.  The function $G$ for these solutions has the same form
as \eqref{stringGh},  except for a small but significant
 difference:   the number of flip points or "caps"  is  odd. If $2n+1$ is the number of flips we have
  \bea\label{bnryGh}
  \underline{\rm     M2\  boundaries } :\quad    \quad     h = -   i   (w - \bar w), \quad
G =   - i   \sum_{j=1}^{2n+1} (-)^j  \frac{w -  \xi_j }{|w - \xi_j |}   \, .
\eea
Since at the $h$-pole on the boundary at infinity   $G$ flips  sign, the
 asymptotic geometry  is   (AdS$_4/Z_2)\times$S$^7$. The dual
 gauge theory is therefore a 3-dimensional gauge theory defined on the   world-volume of semi-infinite M2 branes.

Most of the analysis of these solutions proceeds as in the previous section. There are now
 $n$  independent M5-brane charges, and $n$ independent  M5$^\prime$-brane charges,
 related to the lengths of the intervals between successive flips as in
eqs.\,\eqref{M5selfdual}. We can use the corresponding integer charges  to construct a Young diagram as in Figure 6.
  We can also associate numbers of boxes to M2-brane charges, using a gauge in which  $b_3 =0$ on the
  boundary interval $(-\infty, \xi_1]$ and $b_2=0$ on the boundary interval $[\xi_{2n+1}, \infty)$.
  The reader can easily fill in the details of these calculations.
         \sm

      The only significant difference  between the solutions with even and odd number of flips
      has to do with the charges  at infinity.   In the string solutions \eqref{stringGh}, the asymptotic AdS$_7\times$S$^4$
      region  has   M5-brane charge equal to $N$, and   M2-brane charge equal to the total number, $M$,  of boxes in  the
      Young diagram. Note however that the  7-cycle at infinity has  the topology of $S^3\times S^4$,
      so  large gauge transformations can change the M2-brane charge by multiples of the M5 charge. More
      specifically,  a  constant  shift  of the magnetic potential $b_2$ will change
        $M$ by an integer multiple  of $N$. Thus the only invariant charges
      at infinity are $N$,  and the "$N$-ality"  of the Young diagram.

       In the solutions \eqref{bnryGh}, on the other hand, the 7-cycle at infinity has S$^7$ topology and
       the number $M$ of boxes in the Young diagram is an
       invariant asymptotic M2-brane charge.      There is now no  asymptotic M5-brane charge,
       and no dual $6d$ gauge theory with gauge group  $SU(N)$. The  dual gauge theory for these backgrounds is
       a  $3d$ gauge theory in half space, with the M2-brane  gauge group $SU(M)$.
       \sm

        It is plausible that the supergravity solutions \eqref{stringGh} and \eqref{bnryGh} are actually different descriptions of the
        same object, namely of a stack of M2-branes ending on M5-branes. The existence of such complimentary field theory
        descriptions is familiar from the study of monopoles, viewed   as D-strings ending on D3-branes
     \cite{Diaconescu:1996rk}.  Likewise, the M2$\perp$M5-brane system can be described as a string soliton of the
     M5-brane theory, or as a boundary of the M2-brane theory.  The unique feature of the M theory setup, however,  is
     that both M2 branes and M5 branes have regular near-horizon geometries, when they are allowed to back react.
  Thus the same physical system seems to admit two different superconformal limits, a  novel and intriguing
  phenomenon.

\section{Global   solutions with $\g>0$}
\setcounter{equation}{0}
\label{sec:7}

In this last section we turn our attention to global solutions with positive $\g$.
There is no limit, in this case,
to the number of singularities of the function $h$, so one can have more than one   higher-dimensional asymptotic region.
The basic example is the
 Janus solution  \cite{janus},  which  has two distinct  (AdS$_4/Z_2)\times$S$^7$
 asymptotic regions, with  conformal boundaries   joined along a $2d$ interface (see Figure 1 in the introduction).
There are no known    solutions, on the other hand,  that have  more than two asymptotic regions, and we don't expect to find
any such solutions
with holographic duals. If they  existed,  either the  conformal boundary would be disconnected, or  M2-branes
would  have to form non-trivial  junctions.


\subsection{Deformed  NS5 brane}
\label{sec:71}

Let us begin with a simple $\g>0$ solution,  given by  the following
  function $G$,
\bea\label{freelunch}
  G =  \pm\, ( i + \beta h )  \ , \qquad {\rm with}\quad   \beta \in \mathbb{R}\ .
  \eea
This satisfies the basic equation    $2 h \p_w G = (G+\bar G) \p_w h$, for any harmonic function $h$.
It also obeys the two regularity conditions,
$G=\pm i$ on the boundary of $\Sigma$,
and $\vert G\vert >1$ in the interior of $\Sigma$.
 Thus, \eqref{freelunch} seems to give a global solution for any admissible choice of the
 harmonic function $h$.

 As has been  explained, however,  in section  \ref{sec:4.4},  when
 $h$ has more than one singularity,  either  in the interior or on the boundary of $\Sigma$,
 there are interior points where $\p_w h=0$. Unless $G$ diverges at least as fast as $(\p_w h)^{-1}$
 at these same points, the space-time manifold would have
   conical singularities. Clearly the function in  \eqref{freelunch} cannot diverge faster than $h$, which is finite
   at the zeroes of  $\p_w h$,
 so we conclude that  $h$ has  at most one singularity.
 It thus  takes  one of the
 two canonical forms   \eqref{section6h}, so we have
   the following two  solutions,
 \bea\label{firstsolnt}
 \underline{\rm solution\ 1}:\quad   h=    y  \ , \quad  G = - i  + \beta   y \ ,  \qquad    {\rm with}   \ x+iy \in \mathbb{C}^+ \ ;
 \eea
  \bea\label{secondsolnt}
 \underline{\rm solution\ 2}:\quad
 h=  -a \log (z \bar z) \ , \quad  G = - i   +   \beta  h  \ ,  \qquad  {\rm with}   \   \vert z\vert \leq 1\ .
 \eea
Note that we  have used the freedom of conformal reparametrizations to  set $h=y$ in the first solution, and
to go to the more natural coordinate $z = r e^{i\theta}$ with $r\leq 1$ in the second solution. We have also fixed $G=- i$
on $\p\Sigma$, so inequivalent solutions are obtained for all values of $\g$ between $0$ and $\infty$.

\sm

Closer inspection shows that  \eqref{firstsolnt} and \eqref{secondsolnt}  are  almost equivalent, since they are related by
the  conformal change of variables
\bea
{1\over r} -i \theta = {1\over 2a} (y -ix)\ .
\eea
The only difference between the two  is that in the second solution  the coordinate $\theta$ is periodic. The extra
free parameter in \eqref{secondsolnt} accounts for  the length of the corresponding circle.
We may thus treat these two solutions as one and the same.

Inserting   \eqref{firstsolnt}   in \eqref{2d2} and \eqref{2d1} leads to  the supergravity metric\footnote{We
have here rescaled the coordinates $x$ and $y$ so as to absorb some irrelevant multiplicative
constants. }
\bea\label{NS5metric}
  ds^2 =    e^{4\phi/3}\,  \left[  { 1 \over 1+\g}\, ds^2_{AdS_3}  +   {1\over \g} \,  ds^2_{S^3_2}  +  dx^2+
   \left( dy^2 +  {y^2 \over 1 +y^2}\, ds^2_{S^3_3} \right)   \right]
\eea
 \bea\label{NS5metricd}
   {\rm with}\quad   \phi   =  {1\over 4}\, \log (1 + y^2) + \phi_0\ .
\eea
 The parameter $\beta$ in \eqref{firstsolnt}  enters only in the expression for $\phi_0$ which determines
the overall scale of the solution. The  metric \eqref{NS5metric}
 is conformally equivalent to the direct product of a  4-dimensional cigar
and   AdS$_3\times$S$^3\times\mathbb{R}$ space-time.
Since it approaches flat space-time at infinity, there is no holographic interpretation.
  The two auxiliary functions that enter   in the
expressions \eqref{2e2} for the magnetic potentials are
\bea
\tilde h = -x  \ , \qquad \Phi = x + {\beta\over 2} y^2\ .
\eea
If the coordinate $x$ is compact, $x=x+2\pi$, there is a  non-vanishing 5-brane charge
supported by the 4-cycle  $S_2^3 \times \{x\in [0, 2\pi]\}$. This  is the only non-trivial cycle of the above solution,
which describes a deformed M5$^\prime$ brane smeared along  the direction $x$.
\sm

  When $x$ is compactified on a circle,   the background given by \eqref{NS5metric} and \eqref{NS5metricd}
is  a solution of type-IIA string theory with  dilaton  $\phi$. The 10-dimensional   metric
in  string frame can be
extracted using  the reduction formula \cite{Witten:1995ex}
$$
ds^2_{11} = e^{4\phi/3} dx^2 + e^{-2\phi/3} ds^2_{10}\ ,
$$
with the result
\bea
ds^2_{10} = e^{2\phi} \,  \left[  { 1 \over 1+\g}\, ds^2_{AdS_3}  +   {1\over \g} \,  ds^2_{S^3_2}  +
   \left( dy^2 +  {y^2 \over 1 +y^2}\, ds^2_{S^3_3} \right)   \right]  \ .
\eea
This solution is reminiscent of the standard linear-dilaton background  of the flat NS5-brane \cite{Callan:1991at}.
The NS5-brane   worldvolume  wraps, however,  AdS$_3\times$S$^3$, the dilaton grows
only logarithmically at $y\to\infty$, and the space-time
caps off smoothly at $y=0$.
Furthermore, the solution has a non-vanishing Ramond-Ramond flux  background.
    We interpret this solution as describing a NS5 brane in the background of  dielectric
    D4 branes.


\subsection{Two-parameter Janus deformation }

The AdS$_4\times$S$^7$ solution of M theory is obtained for $\gamma = 1$. A one-parameter deformation
of this solution at constant $\g$, the Janus solution,
 was discovered in reference \cite{janus}. If we parametrize $\Sigma$ by the half plane, the functions $h$ and $G$  are
 given by
 \bea\label{janushG}
h =    i C ( w^{-1}  -  w) + c.c.\ , \quad
G =   i \, { \vert w\vert + {  \vert w\vert }^{-1} + \lambda (w - \bar w) \vert w\vert^{-1}
 \over
\bar w + { \bar w}^{-1}  }\ .
\eea
This solution describes a superconformal domain wall of the gauge theory on the M2 branes. The two asymptotic
(AdS$_4/Z_2)\times$S$^7$ regions emerge at the
two poles of  $h$, at $w=0, \infty$. At these poles the sign of $G$ flips, so that $G = +i$ for $w\in\mathbb{R}^+$
and $G = -i$ for $w\in\mathbb{R}^-$. Thus the geometry has a single non-contractible 7-cycle, and a conserved
charge proportional to the   number of M2 branes. The pure AdS$_4\times$S$^7$ solution corresponds to
 $\lambda =0$,  but more generally $\lambda$ can take  any arbitrary   real value.
\sm

Note that $\p_w h $ vanishes in the interior of $\Sigma$, at $w=i$, but the function $G$ diverges at this
point in such a way that the geometry remains smooth. This is  actually, at present,  the only   known
global solution  that has  no conical
singularity,  despite the presence of more than one higher-dimensional AdS   throats.

\sm
A more convenient parametrization of  \eqref{janushG}
  is in terms of the coordinate $w = e^{2z}$,  which covers the infinite strip
 $z= x+iy \in \mathbb{R} + i [0, \pi/2]$.
In terms of the strip coordinate  the functions  $h$ and $G$ read\,:
\bea\label{janus}
h =  -2iC \left[ \sinh(2z) - \sinh(2 \bar z) \right]\ , \quad
G = i \frac{\cosh(z + \bar z) + \lambda \sinh(z-\bar z)}{\cosh(2 \bar z)}\ .
\eea
Interestingly,  the $\lambda$-dependent
piece of this solution has the same functional form as the AdS$_7\times$S$^4$ solution \eqref{AdS7S4},  modulo
a  coordinate shift  $z\to z+ i\pi/4$.  This means that for $\lambda \gg 1$, the geometry will  approach in a local patch
the near-horizon geometry of the M5 branes. The global features of the two solutions are, however, very different --
in  particular, \eqref{janus} has no asymptotic AdS$_7\times$S$^4$ region and no 5-brane charge.
A possible interpretation is that  the M2 branes blow up into dielectric M5 branes in the middle region,
but it would  be interesting to understand   how this happens in detail.

\sm

 From the general analysis in  sections \ref{sec:2} and \ref{sec:3} we know that all solutions come in
 families parametrized by $\vert \g\vert$, and the same is true for the above Janus solution. We therefore have
 a  family of Janus solutions  parametrized by $-\infty < \lambda < \infty$, by $0<\g<\infty$,
 and by  the overall scale $C$.     The latter is related to the radius $L$ of the asymptotic AdS$_4\times$S$^7$
 solution, or equivalently to the M2-brane charge, by the equation
 \bea
 \qquad L^3 = \frac{C\sqrt{1 + \lambda^2}}{ (c_2 c_3)^{3/2}}  \ .
 \eea
This  two-parameter deformation of  the AdS$_4\times$S$^7$ background was discovered independently, as
a solution of gauged $4d$ supergravity, in ref.\,\cite{Bobev:2013yra}.
The map between the parameters of this reference and our solutions is as follows,\footnote{The choice
$a=0$ for any $\zeta_0$ gives AdS$_4\times$S$^7$, while
$\zeta_0=\pi/2$ implies $\gamma=1$ and yields the M-Janus solution parameterized by $a$
 (AdS$_4\times$S$^7$ corresponds now to $a=1$).
  Setting $a=1$ with arbitrary $\zeta_0$ is the $\g$-deformed solution found in  \cite{Estes:2012vm}.  }
\bea
\lambda=\frac{a\sin\zeta_0}{\sqrt{1-a^2}} \ , \qquad   \gamma=\frac{\sec\zeta_0+a}{\sec\zeta_0-a} \ , \qquad
C^2=\frac{\gamma^3}{8 g^6(1+\lambda^2)}\ .
\eea
Note  that the conformal transformation  $w\to -1/w$  of the upper-half plane
  maps   $\lambda\to -\lambda$ and $G\to -G$. When combined  with the symmetry $\mI$ of section \ref{sec:2i}, this
  allows us to identify as equivalent the solutions with parameters  $(\lambda, \g)$ and $(-\lambda, 1/\g)$.



\subsection{Strings and semi-infinite M2 branes}
\label{sec:73}

  Our last class of global solutions of the supergravity equations uses the
  local solution found in section \ref{sec:53}. Recall that this fourth type of  "Lego piece" arises  at
points on the boundary, $\p\Sigma$,  at which the function $G$ diverges
like $r^{-1}$, and the curvature grows unboundedly large, c.f. eq.\,\eqref{verycurvedasym}.
 These solutions carry non-vanishing M5 charge,
and are   analogous to the highly-curved NS5-brane and D5-brane
regions  of refs.\,\cite{Assel:2011xz,Assel:2012cj,Aharony:2011yc}. As shown in these references,
 such strongly-curved five-branes are necessary in order to properly account for the global symmetries
 of the dual superconformal gauge theories.  We conjecture that the local solutions of section  \ref{sec:53}
 are also admissible solutions of  M theory.

  \sm

  The  global solutions with this fourth Lego piece are given by the following data
\bea\label{curvedsols}
h = -i w + c.c.\ , \qquad
\pm\, G =  G_0\,  +\,  \sum_{a=1}^{n+1}  {\zeta_a \,{\rm Im}(w)\over (\bar w - x_a) \vert w - x_a\vert}\ ,
\eea
where $G_0 =  i$ (no flip),    or $G_0 = iw/\vert w\vert $ (flip).  All the parameters $\zeta_a$ and $x_a$ are real.
Each term in the above summand is of the form\   Im$(w)/( \bar w \vert w\vert )$, which solves our basic  equation
 \eqref{2c1}, and  the superposition of such terms  with real coefficients  is again
 a solution of the $G$ equation.
   Furthermore $G=\pm i$ on $\p\Sigma$, so the only remaining condition that
 we must check is   $\vert G\vert >1$ in the interior of $\Sigma$.
 \sm

 Let us first check this condition  in the  no flip case, namely  when  $G_0=i $. Setting $w=x+iy$ one finds after a
 little calculation
 \bea
 {\rm Im} (\pm G) \, =\,  1 + \sum _{a=1}^{n+1}  {\zeta_a\,  y^2 \over  \vert w - x_a\vert^3 }\ .
 \eea
This is $ \geq 1$   if all the $\zeta_a$ are positive, so  this condition suffices to prove  that   $\vert G\vert \geq 1$.
In the case of a flip, $G_0 = w/\vert w\vert  $,   one likewise finds
  \bea
   {\rm Re} (\pm G/G_0) =  1 + \sum _{a=1}^{n+1} {\zeta_a x_a\,  y^2 \over \vert w\vert  \vert w - x_a\vert^3 }\ .
  \eea
  This can be made greater or equal to one  if \   $\zeta_a x_a >0$\  for all $a$,    i.e. if  the $\zeta_a $
  are chosen to be  positive or negative
  according to whether the corresponding singularity    is  on the right or the left of the point of flip on the
  real axis. Since $\vert G_0\vert =1$, this condition   again  guarantees  that $\vert G\vert \geq 1$
  so that the solutions \eqref{curvedsols} are regular.\footnote{These conditions on the parameters $\zeta_a, x_a$ are sufficient,
  but we don't know if they are also necessary.}
\sm

 Since the  harmonic function $h$ in \eqref{curvedsols}  has a single pole,  at $w=\infty$,
  the conformal boundary is either that of  AdS$_7^\prime$ or of  AdS$_4/Z_2$.
   These solutions describe  therefore self-dual strings on the  world-volume of M5 branes,
   or semi-infinite M2 branes. They are the $\g>0$ counterparts of the regular solutions of section \ref{sec:6}.
\sm

   Let us   look more closely at   the solutions  with no  flip, which have  AdS$_7^\prime\times$S$^4$ asymptotics.
   Choose for definiteness  the minus  sign in front of $G$ in  \eqref{curvedsols}, so that $f_3=0$
    on the boundary [the entire boundary  is red,  in the color code that we have used previously].
   From eq.\,\eqref{624} one reads  the following M5-brane charges
   \bea
   \mM^{(3)}_a =  4 \nu_3\, \zeta_a  /(\g c_3^3)\, \ \to\  \, N^{(M5)}_a \ , \qquad N = \sum_{a=1}^{n+1}  N^{(M5)}_a\ ,
   \eea
   where we have denoted by $N^{(M5)}_a$ the integer charges, and $N$ is the total charge which
   is also the number of M5 branes  at   infinity.

   To calculate the M2-brane charges,  recall from
   section \ref{sec:53} that we only need to keep the term $-\tilde h \Phi + \epsilon \Phi^2/2$ in the expression
    \eqref{finalomegac} for $\Omega_1^c$.
   Combining the expression \eqref{dPhi} for $\Phi$, and   $\tilde h = -2(x-x_0)$, leads to
   the following potential on the real axis,
   \bea
   \Omega_1^c = 4(x-x_0) \sum_{a=1}^{n+1} \zeta_a\,  {\rm sign}(x-x_a) +
   2 \,  \Bigl[\, \sum_{a= 1}^{n+1} \zeta_a\,  {\rm sign}(x-x_a)
\Bigr]^2    + {\rm continuous}\ .
   \eea
   We need  not  write explicitly  the continuous piece because it  does not contribute to the electric charges.
   We have also  set  $\epsilon = +1$ because the solution has no M5$^\prime$ charges, so
   $b_2$ can be globally defined   (see the discussion in section \ref{sec:2f}).
  Now eq.\,\eqref{2h3} gives  the M2-brane charge of the $a$th singularity  in terms of the
  discontinuity  of the above function at $x_a$,
  \bea
  \mE^{(3)}_a  =   {8\nu_1 \sigma \over (c_2c_3)^3}\, \zeta_a\,
  \left[\,  (x_a -x_0)  -   \zeta_1-   \cdots  -  \zeta_{a-1} +  \zeta_{a+1} +  \cdots + \zeta_{n+1}
  \right]\ .
  \eea
  We denote by $N^{(M2)}_a$ the corresponding integer charge.
    Notice that this  is proportional to $N^{(M5)}_a$, so one can define the
  number  of M2 branes that emanate  from each M5 brane in the stack,   $N^{(M2)}_a \equiv  N^{(M5)}_a   \, l_a$\
  where
   \bea
  l_a =   N^{(M5)}_{n+1} + \cdots   + N^{(M5)}_{a+1} -  N^{(M5)}_{a-1} - \cdots - N^{(M5)}_{ 1}  + A\, (x_a-x_0) \ ,
 \eea
  and $A$ is a multiplicative constant. Note that these charges depend on the  free parameters $x_a$, namely on
  the positions of the singularities on the real axis. In    M theory, we expect   the $l_a$ to be  quantized.
  \sm

 It is convenient to order the charges $l_a$ as follows.  First define  auxiliary parameters
\bea
\chi_a \equiv   Ax_a - N^{(M5)}_{a}\  \ \Longrightarrow\ \  l_{a} - l_{a+1} = \chi_a - \chi_{a+1}\ .
 \eea
  Without loss of generality we can label the singularities so that $\chi_a$ is a decreasing function
  of $a$. By choosing $x_0$ appropriately, we can also set $l_{n+1}$ to zero.
  The data parametrizing  our solutions can then be repackaged as  follows,
  \bea
  N\ , \quad \{ N^{(M2)}_{1}, \cdots , N^{(M2)}_{n} \} \, , \qquad   l_1 \geq l_2\cdots \geq l_n > 0\ .
  \eea
   This data determines a Young diagram, and  an irreducible  representation of $SU(N)$, in the
  same way   as    in Figure 6. The only difference with the analysis of section \ref{stringsols}  is that, in the place  of the
  M5$^\prime$ charges, we are here given more naturally
 the  numbers  of boxes contained in a  row  of the $a$th stack. These are precisely the charges  $l_a$.
 Of course, these two parametrizations
 of the Young diagram are completely equivalent.
 \sm

    Thus the solutions \eqref{curvedsols} without  flip appear to be  the continuation to  positive $\g$,
      of the $\g<0$ self-dual string solutions
    given in  section \ref{stringsols}.    The transition from negative to positive
    $\g$ has  the effect of collapsing  the $a$th   interval  $[\xi_{2a-1}, \xi_{2a}]$  of the $\g<0$ solution,
    to the $a$th singularity of the solution  \eqref{curvedsols}.
    It should be very interesting to  understand the nature of this transition on the field theory side
   \sm

   The solutions \eqref{curvedsols} with  a flip, $G_0 = iw/\vert w\vert$, can be analyzed similarly.
    In this case  $f_3=0$ on the negative real axis,
    and $f_2=0$ on the positive real axis, so the  asymptotic region at infinity
     is (AdS$_4/Z_2)\times$S$^7$. These solutions correspond therefore to semi-infinite M2 branes.
     They are the counterparts, at positive $\g$, of the  solutions \eqref{bnryGh} of section  \ref{sec: 73}.

\acknowledgments

We thank   A. Amariti, N. Bobev, R. Feldman, N. Halmagyi, C. Hull, V. Niarchos, K. Pilch, K. Siampos, 
J. Troost, and  N. Warner for useful conversations, and correspondence. One of us (E.D.) gratefully
acknowledges the warm hospitality and the financial support of the Laboratoire de Physique Th\'eorique 
at the Ecole  Normale Sup\'erieure, where part of this work was carried out. The research  of E.D. is 
supported in part by National Science Foundation grant PHY-1313986. 

 \vskip 1cm

\appendix

\section{Redefining the reduced data ($\g, h, G$)}
\setcounter{equation}{0}
\label{sec:A}

In \cite{Estes:2012vm}, the reduced fields were expressed in terms of a M\"obius transform
$H$ of the field $\Gp$, and its associated composites $V_\pm$, defined by
\bea
\label{6a3}
H = { \Gp \over 1 - i c \Gp}\ ,
\hskip 0.5 in
V_\pm = 2 |H|^2 \pm i (H-\bar H)\ .
\eea
We use here a calligraphic letter, $\Gp$,  for the function $G$ of reference  \cite{Estes:2012vm},
and reserve  the symbol $G$ for the new function defined below.
The parameter $c$ is defined by $ c = (c_1 - c_2)/ c_3$, where $c_j$ are the three
real parameters that obey the constraint $c_1+c_2+c_3=0$. The relation of $c$  to the parameter
 $\g$ is   $c = - 2 \g -1$.\,\footnote{In  \cite{Estes:2012vm} the parameter $\g$ is denoted $c^\prime$, and
 our harmonic function $h$ is replaced by $\vert c_3\vert  \hat h$.}
 \sm

 In terms of $H$ and its composites the metric reads
 \bea
\label{6a4}
f_1 ^6 = + {  h ^2 \over c_1 ^6}  (1 - |H|^2) { V_+ \over V_-^2}\ ,
& \hskip 1in &
\rho^6 = - { |\p_w h |^6 \over c_3^6 \, h^4 } {1 - |H|^2 \over |1+ic H|^6} \,  V_+ V_-\ ,
\no \\
f_2 ^6 = - {  h^2 \over c_2 ^6}  (1 - |H|^2) { V_- \over V_+^2}\ ,
&&
f_3 ^6 = - {   h^2 \over c_3 ^6}   { V_+  V_- \over (1 - |H|^2)^2}\ .
\eea
 The necessary and sufficient conditions for positivity of  these metric factors
are given by the inequalities  $|H|^2 <1$, $V_+ >0$, and $V_-<0$. The condition $V_-<0$ is equivalent to
$H$ taking values in an open disk of radius 1/2 centered at $-i/2$,
\bea
\label{6a6}
\left | H + { i \over 2} \right | < \half\ .
\eea
It follows  from the definitions of $V_\pm$   that if   $V_-<0$,
  then the remaining conditions $|H|^2 <1$ and $V_+>0$ are satisfied automatically.
Thus  (\ref{6a6}) is the necessary and sufficient condition for positivity of all  metric factors.
\sm

In view of the above, it would seem that $H$ is the ideal variable in which to  formulate the BPS  problem.
 Indeed, its range is independent of the parameter $\g$, and
   metric factors are fairly simple combinations of $H$. Furthermore,  the regularity
condition  on the boundary $\p\Sigma$ is  $V_-=0$,  which implies the simple Dirichlet condition $H =0$ or $H=-i$.
\sm

  Unfortunately, the
 BPS equation (\ref{2c1}) expressed in terms of $H$
 will depend  explicitly on $\g$  and, more crucially,
it will become  non-linear  in  $H$.
  This is to  be contrasted to the function $\Gp$
 which obeys a simple linear equation, but whose range depends non-trivially on the parameter $\g$.
Another drawback of  the parametrization (\ref{6a4}) is that    the two spheres S$_2^3$ and S$_3^3$
do not enter  on   equal footing.


 Both of  these drawbacks can be  circumvented by a simple  shift and rescaling of
the function $\Gp$.  To see how, note  that M\"obius transformations preserve circles, so the
boundary of the disk  (\ref{6a6}) is also a circle in the complex $\Gp$ plane. Explicitly, the regularity
condition
can be written as follows
\bea
\left\vert  {\Gp (1-2\g)+i \over \Gp (1+2\g)-i} \right\vert\, <\, 1 \quad \Longrightarrow\quad
4\g \Gp\bar \Gp - i (\bar\Gp - \Gp) \, >\, 0\ .
\eea
Thus, the boundary of the allowed domain is a circle centered at $i/4\g$  with radius equal to $1/4\g$.
Furthermore, $\Gp$ must take values outside this circle, if $\g>0$, or inside this circle if $\g<0$.
 Defining the shifted and rescaled function
 \bea
G = -i + 4\g \Gp  \ ,
\eea
allows us to write the regularity conditions in the following convenient  form:
\bea
\label{6c5}
\g  ( |G|^2 -1) > 0 \qquad \Leftrightarrow \qquad
 \left \{   \begin{matrix}   |G|>1 & {\rm for} & \g >0\ ,  \\  |G|<1   & {\rm for} &  \g <0 \ .   \end{matrix}  \right .
\eea
 In words, $G$ takes values  in the interior or exterior of the unit disk, centered at the origin,
 for negative or positive, respectively $\g$.

Furthermore,  the  BPS equation (\ref{2c1}) is invariant under a real rescaling  and  imaginary shift
of the function $G$, so both $G$ and $\Gp$ obey the same equation.
The nice feature of the above redefinition is  that  the reduced BPS problem now depends only on the sign
of $\g$.  An immediate corollary   is that all supergravity solutions come in one-parameter families:
they can be continuously deformed by varying $\g$ without changing its sign.

\sm

Finally, the announced expressions   (\ref{2d2}) and (\ref{2d1}) for the space-time metric can be easily derived
from (\ref{6a3}) and (\ref{6a4}),
with the help of the   following relations:
\bea
W_+ = 4\g^2 V_+\ , \quad W_- = 4 (1- \vert H\vert^2)\ , \quad G\bar G - 1 = -4\g V_-\ .
\eea
 Verifying   these relations is a simple algebraic exercise that we leave to the reader.
Note that in contrast to (\ref{6a3}) and (\ref{6a4}), the expression
of the metric in terms of $G$ makes   manifest  the symmetry
under exchange of  the two spheres and inversion of the parameter $\g$.


\section{Magnetic   potentials  in terms of $(\g, h, G)$ }
\setcounter{equation}{0}
\label{sec:B}

In this Appendix, we shall express the  4-form field strength, and the corresponding potentials $b_i$,
  in terms of the canonical $G$ defined in appendix \ref{sec:A}.
   The starting point is the expression for the field strength  obtained in \cite{Estes:2012vm}
    in terms of the variable $c=-2 \g-1$, the real
harmonic function $h$ (after conversion from $\hat h$ by setting $h = c_3 \hat h$), and the
complex-valued function $H$ given in (\ref{6a3}).
After some slight
rearrangements,  the one-forms that multiply the canonical volume forms of the (peudo)spheres read:
 \bea
\label{A2}
\p_w b_1 & = &
2 i \nu_1 \frac{h}{c_1^3} \, { H(1-i H) \p_w\bar H  \over  V_- ^2}
\no \\&&
+2 i \nu_1 \frac{h}{c_1^3} \, { \p_w H  \over (-i+c H)(H+\bar H) V_- ^2}
\Bigg [ \bar H(-i+c H)(H-3 \bar H+4H \bar H^2)
\no \\ && \qquad
- H\bar H(H-3 \bar H+4H \bar H^2)
 -i  c \Big ( (H-\bar H)^2-H\bar H^3+3H^2\bar H^2 \Big )  \Bigg]\ ,
\eea
\bea
 \p_w b_2 & = &
-2 i \nu_2 \frac{h}{c_2^3} \, { H (1+i H) \p_w\bar H  \over  V_+ ^2}
\no\\&&
-2 i \nu_2 \frac{h}{c_2^3} \, { \p_w H  \over (-i+c H)(H+\bar H) V_+ ^2}
\Bigg[\bar H(-i+c H)(H-3 \bar H+4H \bar H^2)
\no \\ && \qquad
+ H\bar H(H-3 \bar H+4H \bar H^2)
+ i \, c \Big((H-\bar H)^2-H\bar H^3+3H^2\bar H^2 \Big)  \Bigg]\ ,
\eea
\bea
\p_w b_3 & =&
\nu _3 \frac{h}{c_3^3} \, { (1+H^2) \p_w \bar H  \over (1 - H \bar H)^2}
\no\\&&
+\nu _3 \frac{h}{c_3^3} \, {  \p_w H \over (-i+c H)(H+\bar H)(1 - H \bar H)^2}
\Bigg[  8 i H^2 \bar H^3 + 3 c H \bar H^3 - i \bar H^3 + 3 c H^2 \bar H^2
\no\\ && \qquad \qquad
 - 15 i H \bar H^2 - 2 c \bar H^2 + 2 i H^2 \bar H - c H \bar H
+ 5 i \bar H + c H^2 -3 i H  \Bigg]\ ,
 \eea
where the prefactors $\nu_i$ are signs taking the values $ \pm 1$.
The BPS equations do not completely determine these signs,
but they constrain their product,
\bea
\label{A5}
\nu_1 \nu_2 \nu_3 = -\sigma \ .
\eea

The above expressions for the  $\p_w b_i$  are rather complicated.
We will now show, however, that  the  potentials $b_i$ are  simpler.
This is fortunate since the  brane charges,  which interest us ultimately,
are easier to compute  from the potentials than from the fluxes.

\sm

 To organize the calculation of $b_i$, we  separate a piece that vanishes on the boundary,
 $b_i^s$,  from a
  cohomological piece, $b_i^c$,  that will contribute to brane charges,
 \bea
\label{A6}
b _i = { \nu _i \over c_i ^3} \Big ( b_i ^s + b_i ^c  \Big )\ .
\eea
This separation is not unique, but a natural choice is to  integrate terms in $\p_w b_i$ that are proportional
to $\p_w \bar H$ and collect them in the contribution   $b_i^s$, and the remaining terms proportional to
$\p_wh$ and  $\p_w  H$
into $b_i^c$.  The former   integrate  nicely  into the following  local functions  of $h,H$:
\bea
\label{A7}
b_1 ^s & = & - { h (H + \bar H) \over V_-} ~
= ~ - { h (G + \bar G) \over 1-G \bar G}\ ,
\no \\
b_2 ^s & = & - { h (H + \bar H) \over V_+} ~
= ~ - \g \, { h (G + \bar G) \over W_+}\ ,
\no \\
b_3 ^s & = & - { h (H + \bar H) \over H \bar H -1} ~
 = ~ + { 1 \over \g}  \, { h (G + \bar G) \over W_-}\ .
\eea
Here, the second  equality on each line gives the expression in terms of the canonical $G$, which is
related to $H$ by the equations  $H = \Gp / (1 - i c \Gp)$ and  $\Gp = (G+i)/(4 \g)$.
\sm

The remaining contributions to  $\p_w b_i$ contain no terms proportional to $\p_w \bar H$,
but only terms proportional to  $\p_w H$. Since $H$ is holomorphic in $G$, these terms
are all proportional to $\p_w G$, which by virtue of the reduced field equation (\ref{2c1}),
is in turn proportional to  $\p_w h$. Some straightforward algebra leads to the expressions
 \bea
\label{A8}
&&\p_w b^c _1 =  \Big [ (2 + \gamma + \gamma^{-1} ) \bar G + i (\gamma - \gamma^{-1} ) \Big ] \p _w h\ ,
\no \\
\p_w b^c _2&=&\gamma (  \bar G + i  ) \p _w h\ , \qquad
 \p_w b^c _3 =    - { 1 \over \gamma} ( \bar G - i  ) \p _w h\ .
\eea
Note that all  three one-forms  $\p_w b_i^c$ are   linear superpositions
of the two basic combinations $i \p_w h$ and $\bar G \p_w h$, with real coefficients.
\sm

The integration of  the combination $i \p_w h$ is   subtle. One might be tempted to
integrate it to a contribution for $b_i^c$ proportional to $ih$, but this is not acceptable since
 $b_i^c$ must be
real.    Instead, we use the harmonicity of $h$
to recast $i \p_w h$ in terms of the real harmonic function $\tilde h$ dual to $h$,
which may be defined by the relation,
\bea
\label{A9}
i \p _w h = - \p_w \tilde h\ .
\eea
The resulting contribution to $b_i^c$ will now be real.

The integration of the combination $\,\bar G \p_w h\,$   is   more involved. Note first
 that   $\p_{\bar w}  ( \bar G \p_w h )$ is real in view of the reduced field
equations (\ref{2c1}). As a result, we have
\bea
\label{A10}
\p_{\bar w} \Big ( \bar G \p_w h \Big ) - \p_w \Big ( G \p_{\bar w} h \Big )=0\ .
\eea
Equivalently, the differential form
\bea
\label{A11}
dw\,  \bar G \p_w h + d \bar w\,  G \p_{\bar w} h\
\eea
is closed, and can be written as the total differential of a local real function $\Phi$, as defined in (\ref{2e3}).
This function was first encountered  in equations (8.6) and (8.23)
of \cite{D'Hoker:2008wc}, though in the slightly different notation,  $\Phi = h \Psi$.

\sm

We have thus succeeded in  integrating  the three expressions (\ref{A8}) into real flux potentials,  $b_i^c$,
which can be written in terms of the two real function  $\tilde h$ and $\Phi$,
\bea
\label{A13}
&& b^c _1 =  b^0_1 +  (2 + \gamma + \gamma^{-1}) \Phi - (\gamma - \gamma^{-1}) \tilde h\ ,
\no \\
b^c _2 &=&b^0_2  +\gamma (\Phi - \tilde h)\ , \qquad
 b^c _3 =   b^0_3 - \frac{1}{\gamma} (\Phi + \tilde h)\ .
\eea
The constant residual gauge variations  $b_i^0$ may be of course added freely to $b_i$.

\section{Electric   potentials  in terms of $(\g, h, G)$}
\setcounter{equation}{0}
\label{sec:C}

In this Appendix, we shall compute the conserved   7-form flux $d\Omega$ in
terms of the reduced data ($\g,h, G$)  on $\Sigma$. Recall that $d \Omega$ was defined by
\bea
d \Omega = * F + \half C \wedge F\ .
\eea
To work out the  components $d \Omega _i$ of $d \Omega$  in
(\ref{2f2}), we shall need a careful evaluation of various Poincar\'e duals.
A useful formula is
\bea
F _{(p)} \wedge * G_{(p)} = { 1 \over p!} F_{a_1 \cdots a_p } G^{a_1 \cdots a_p } \, e^{0123456789\natural }\ ,
\eea
where $\natural$ stands for the 11-th dimension, indices are raised and lowered with the Minkowski metric
$(- + \cdots +)$,  and
\bea
F_{(p)} = { 1 \over p!} F_{a_1 \cdots a_p } \, e^{a_1 \cdots a_p}\ ,
\qquad
G_{(p)} = { 1 \over p!} G_{a_1 \cdots a_p } \, e^{a_1 \cdots a_p}\ .
\eea
We shall also use the $\ep$-frame anti-symmetric symbol on $\Sigma$,
normalized to
\bea
\ep ^{9 \natural } =\ep ^9 {}_\natural = \ep _{9 \natural} = 1\ ,
\qquad  \ep _z {}^z = - \ep ^z {}_z =i\ .
\eea
As a result, we find
\bea
{}*_\Sigma e^a =  + \ep ^a {}_b e^b\ ,
\eea
where ${}*_\Sigma$ denotes the Poincar\'e dual  taken in the 2-dimensional space appropriate for $\Sigma$,
and
\bea
{}* e^{012a} =  - \ep ^a{}_b \, e^{345678b}\ ,
\quad
{}* e^{345a} =  - \ep ^a{}_b \, e^{012678b}\ ,
\quad
{}* e^{678a} =  + \ep ^a{}_b \, e^{012345b}\ ,
\eea
where  duals are here taken  in the eleven-dimensional space-time.

\medskip

Having set our conventions, we next  decompose
  the fields $C$ and $F$ onto the unit-volume forms of  S$^3$  and AdS$_3$
with the help of the reduction of  (\ref{2b1}).
One  readily obtains  the  reduced expression
for $d\Omega$, given in (\ref{2f2}), in terms of component one-forms $d\Omega _i$ on $\Sigma$
defined by (\ref{2f3}).  The  explicit expressions for the $dw$ components of  the one-forms
are
\bea
\p_w \Omega _1  & = &  i {f_2^3 f_3^3 \over f_1^3} \p_w b_1
+ \half \left ( b_2 \p_w b_3 - b_3 \p_w b_2 \right ) + \p_w\eta_1 \ ,
\no \\
\p_w \Omega _2  & = &  i {f_1^3 f_3^3 \over f_2^3} \p_w b_2
- \half \left ( b_3 \p_w b_1 - b_1 \p_w b_3 \right ) + \p_w\eta_2\ ,
\no \\
\p_w \Omega _3  & = & i {f_1^3 f_2^3 \over f_3^3} \p_w b_3
- \half \left ( b_1 \p_w b_2 - b_2  \p_w b_1 \right ) + \p_w\eta_3\ ,
\eea
where we have included   the total-derivative ambiguity, see equation \eqref{2f3}.

 Next, we   factor out the dependence on $c_i$ and $\nu_i$, just as we had
done for the components of the magnetic   potentials $b_i$ in (\ref{2e1}).
Let us define
 $ b_i = { \nu _i  } \hat b_i / c_i^3
$\ ,
where the $\nu_i $  are the signs introduced earlier which are subject to the constraint
$\nu_1 \nu_2 \nu_3= - \sigma$. We  also rescale the $\eta_i$ accordingly, and
make use of relation (\ref{6a5}), namely $c_1 c_2 c_3 f_1 f_2 f_3 = \sigma h$ where $\sigma = \pm 1$.
In terms of these data, we may scale out of $\Omega$ the following dependence on $c_i, \nu_i$ and $\sigma$,
\bea
\Omega _i = { \nu_i \sigma c_i^3 \over c_1^3 c_2^3 c_3^3} \, \hat \Omega _i\ ,
\qquad
{\rm so\  that}
\ \ \
 d \hat \Omega _i =  - { h^3 \over  c_i^6 f_i^6} \left ( {}* _\Sigma d \hat b_i \right )
+  \half   \ep_i{}^{jk}  \hat b_j d \hat b_k + d\hat\eta_i \ .
\eea
The most important flux component will be $i=1$, since its contribution to $d\Omega$ is dual to the
compact cycle $S^3 \times S^3$ warped over a curve in $\Sigma$. For this component we have
\bea
\p_w ( \Omega _1^s +  \Omega _1^c )
= i  h  {(G \bar G -1)^2 \over W_+ W_-} \p_w \hat b_1
- \half  \hat b_2 \p_w \hat b_3 + \half \hat b_3 \p_w \hat b_2  +    \p_w\hat\eta_1\ ,
\eea
where we have again written the one-form as the sum of a cohomological piece, and one that does not
contribute to the electric charges, $\hat \Omega_1 \equiv  \Omega _1^s +  \Omega _1^c $.
 The components $\p_w \Omega _2$ and $\p_w \Omega_3$ may be similarly expressed,
but we shall not need them. Furthermore, for our purposes $\hat\eta_1$  is
 proportional to $\hat b_2 \hat b_3$,    which  has been computed earlier and
 can be added back when needed.  We therefore set it   to zero in the rest  of this appendix,
 and add it back only at the end of section \ref{sec:2f}.

\sm

 From here the analysis parallels the one for the magnetic  potentials in Appendix \ref{sec:B}.
  We first single out the $\p_w \bar G$ terms in
$\p_w \hat\Omega_1$ and integrate them   to obtain $\Omega _1 ^s$.
One may choose  the integration constant so that $\Omega_1^s$ is real.  The result is,
\bea
\Omega _1 ^s  & = &
{ h \over 2 W_+} \left ( \g h (G \bar G-1 ) + ( \Phi + \tilde h) (G + \bar G) \right )
\no \\ &&
- { h \over 2 W_-} \left ( { 1 \over \g} h (G \bar G-1) + (\Phi - \tilde h ) (G + \bar G) \right )
\eea
where the function $\Phi$ was defined in (\ref{2e3}).
 The remaining part   of $\p_w \Omega _1$ is then   given by
\bea
\label{defOmeg1}
\p_w \Omega_1^c  =  - \Big ( \bar G (\tilde h - i h) + i \Phi \Big ) \p_w h\ .
\eea
 Using the facts that $h$ is harmonic, and that $\tilde h - i h$ is holomorphic, it
is straightforward to compute the $\p_{\bar w}$ derivative of the above equation,
and check that it is real,
\bea
\p_{\bar w} \p_w \Omega^c _1
= - \half (G + \bar G)  { \tilde h \over h}  \p_w h \p _{\bar w} h - {i \over 2}  (G- \bar G) \p_w h \p_{\bar w} h\ .
\eea
 From this one can show that
  $\p_w \Omega _1 ^c$ may indeed be integrated, as expected.

To be more specific, we begin by eliminating $\bar G$ in \eqref{defOmeg1} in  favor of
the real function $\Phi$. After straightforward manipulations, we are led to  solve the following equation:
\bea
\p_w \Omega ^c_1 = -(\tilde h - i h) \p_w \Phi  - i \Phi \p_w h\ ,
\eea
or equivalently, using $\p_w h =  i\p_w \tilde h$,
\bea\label{equivOmega}
\p_w \Big ( \Omega ^c_1 + \tilde h \Phi \Big ) =   i h \p_w \Phi - 2 i \Phi \p_w h\ .
\eea
We shall now show that, whenever $h$ and $G$ satisfy (\ref{2c1}),    there
exists a real smooth function $\Lambda$, which can be computed explicitly in terms of $(h,G)$, and
 such that
\bea
\Omega _1 ^c = - \tilde h \Phi +   \Lambda\ .
\eea

\sm

 In order to prove this final step, it is convenient to make the usual  holomorphic  change of coordinate
 so that  $w=  - \tilde h +  i h \equiv x  + iy   $.
 Equation (\ref{equivOmega}) then takes the form
\bea
( \p_x -  i \p_y )  \Big ( \Omega ^c_1 - x  \Phi \Big ) =   i y ( \p_x -  i \p_y)  \Phi - 2 \Phi\ .
\eea
  Decomposed  into real and imaginary parts this reads
\bea
 \p_x    ( \Omega ^c_1 - x \Phi   ) =    y \p_y  \Phi - 2 \Phi\ ,
\qquad {\rm and}\ \  \ \
 \p_y    ( \Omega ^c_1 - x \Phi   ) =   \p_x   ( -y \Phi   )\ .
\eea
The second equation may be thought of as a conservation equation, and thus, locally,
there exists a real function $\Lambda$ such that
\bea
\label{6e5}
- y \Phi =  \p_y   \Lambda
\qquad {\rm and} \ \ \  \Omega ^c_1 - x \Phi =  \p_x   \Lambda  \ .
 \eea
Inserting these expressions in  the first equation gives
an  equation for $\Lambda$ alone,
\bea\label{Lambdaeq}
\left ( \p_x^2  + \p_y ^2  +{1 \over y} \p_y  -{ 4 \over y^2} \right ) (y^{-2}\Lambda)\,  =\, 0\ .
\eea



\subsection{Solving for the function $\Lambda$}

Equation (\ref{Lambdaeq}) resembles \eqref{eqfor Psi},
 the only difference between the two is in  the  coefficient  of the $1/y^2$ term.
Both equations may be solved by Fourrier transform, and we will
then show that the solutions can be matched with the first equation in (\ref{6e5}).
\sm

The Fourier transforms in $x$ are given by,
\bea
\Phi (x,y) & = &   {  y}\, \Psi (x,y) =  \int _0 ^\infty { dk \over 2\pi} \Big ( \Phi _k(y) \, e^{-ikx} + \Phi _k^* (y) \, e^{ikx} \Big )\ ,
\no \\
\Lambda (x,y) & = & {  y^2 }\,  \hat \Psi (x,y) =
\int _0 ^\infty { dk \over 2\pi} \Big ( \Lambda _k(y) \, e^{-ikx} + \Lambda_k^* (y) \, e^{ikx} \Big )\ ,
\eea
 where $\Psi_k(y)$ and $ \Lambda_k(y)$ satisfy the modified Bessel equation for indices $\nu=1,2$ respectively,
\bea
\left ( -k^2  + \p_y^2 + {1 \over y} \p_y  -{ 1 \over y^2} \right ) \Psi_k(y)=0\ ,
\no \\
\left ( -k^2   + \p_y^2 + {1 \over y} \p_y  -{ 4 \over y^2} \right )  \hat \Psi_k(y)=0\ .
\eea
The general solution of these equations gives
\bea
\Phi _k(y) & = &  -  \psi _1 (k)\, y  I_1(ky) + \psi _2 (k)\,  y K_1(ky)\ ,
\no \\
\Lambda_k(y) & = &  \hat\psi _1 (k) \, y^2  I_2(ky) + \hat\psi _2 (k)\, y^2  K_2(ky)\ .
\eea
Next, we use the first equation of (\ref{6e5}) to relate $\Phi $ and $\Lambda$. It is
manifest that the absence of $x$-dependence and derivatives in $x$ implies that
the first relation in (\ref{6e5}) descends to the Fourier coefficients,  and we have
\bea
- y \Phi _k (y) = \p_y   \Lambda _k (y)  \ .
\eea
The standard recursion relations of modified Bessel functions,
 \bea
 { d \over  dz}  \left ( z^2 I_2 (z) \right ) =  z^2 I_1 (z)\ , \qquad
  { d \over  dz}   \left ( z^2 K_2 (z) \right ) =  - z^2 K_1 (z)\ ,
\eea
allows us then to match  the Fourrier coefficients,
\bea
\hat\psi _i (k) = { \psi _i (k) \over k} \hskip 1in i=1,2\ .
\eea

The second relation in (\ref{6e5}) leads now to  a complete expression for $\Omega_1^c$,
\bea
\Omega _1^c (x,y)
& = & \int _0 ^\infty {dk \over 2\pi} e^{-ikx}    \psi _1 (k)  \left ( - xy  I_1 (kr) - i y^2 I_2(ky)  \right ) + {\rm c.c.}
\no \\ &&
+  \int _0 ^\infty {dk \over 2\pi} e^{-ikx}  \psi _2(k) \left ( xy K_1(kr) - i y^2 K_2(ky) ) \right ) + {\rm c.c.}
\eea
 Thus $\Omega_1^c$ and the function $\Lambda$ can be computed from $\Phi$,
  at least in a local patch, as advertised.

 \if
Using the integral formulas for modified Bessel functions,
\bea
I_1(r) = - {1 \over \pi} \int _{-1} ^1 { t dt \over \sqrt{1-t^2} } \, e^{-tkr}
& \hskip 0.8in &
K_1(r) = \int _1 ^\infty { t dt \over \sqrt{t^2-1} } \, e^{-tkr}
\no \\
I_2(z) =  {1 \over \pi} \int _{-1} ^1 { (2t^2-1) dt \over \sqrt{1-t^2} } \, e^{-tkr}
& \hskip 0.8in &
K_2(z) = \int _1 ^\infty { (2t^2-1) dt \over \sqrt{t^2-1} } \, e^{-tkr}
\eea
and using the old definitions of $C$-functions,
\bea
C_i (tr+is) = \int _0 ^\infty { dk \over 2 \pi} \, \psi _i (k) \, e^{-k (tr+is)} \hskip 1in i=1,2
\eea
we can re-express the result just in terms of those,
\bea
\Omega_1^c & = & \int _{-1} ^1 { dt \over \sqrt{1-t^2}} \Big ( srt -i r^2(2t^2-1) \Big ) C_1 (tr+is) + {\rm c.c.}
\no \\ &&
+ \int _1 ^\infty {  dt \over \sqrt{t^2-1} } \Big ( xrt -i r^2(2t^2-1) \Big ) C_2 (tr+is) + {\rm c.c.}
\eea
\fi


\section{Solving the BPS equations with  $h$  constant}
\setcounter{equation}{0}
\label{sec:D}


  In this appendix we adapt  the analysis carried out in \cite{D'Hoker:2008wc, Estes:2012vm}
  to the special case of constant harmonic function $h$. Reference \cite{Estes:2012vm} introduced  the
  functions  $\oa = \rho \alpha \bar\alpha^3/\kappa $ and $\ob = \rho \beta \bar\beta^3/\kappa $,
   where $\rho$ is the scale factor of the metric on $\Sigma$,
  $\alpha$ and $\beta$ are  the  components  of the Killing spinor defined in  (\ref{zeta}), and $\kappa = \p_w h$.
   In terms of these auxiliary functions, the BPS
  equations reduce  to the pair of differential equations
  \bea\label{systemolambda}
\p_w\oab-\p_w(\bar\l^4\oab)&=&{\kappa\over 2}(c_1-c_2)
 \left({\rho\over\vert\kappa\vert}\right)^{3\over 2}\oa^{1\over 4}\oab^{1\over 4}\, {\bar\l^2+\l^2\over(\l\bar\l)^{1\over 2}} \ , \\
{3\over 4}\p_w\ln(\l^2\oa)&=& {3\over 2} \p_w\ln\left({\rho\over\vert\kappa\vert}\right)
+{1\over 8}\p_w\ln\oab\,{\bar\l^2+9\l^2 \over \bar\l^2+\l^2}+{1\over 8}\p_w\ln(\bar\l^4\oab)\,{9\bar\l^2+\l^2 \over \bar\l^2+\l^2}
\ , \no
\eea
   and to the algebraic constraint
  \bea\label{constraintll}
\bigg(1-(2\gamma+1) \lambda\bar\lambda\bigg)-\lambda^4\left(1-{2\gamma+1\over \lambda \bar\lambda}\right) = {1 \over
\oa} \ ,
\eea
where  $\lambda$ is  the ratio   $(\ob/\oa)^{1/4}$.
  \sm

  When $\kappa = \p_w h = 0$,  the  above definitions  of $\oa$,  $\ob$ are singular.
  To proceed, we go through the same steps  as  \cite{Estes:2012vm}  using instead the functions
  $\om_\a=\rho \a\bar\a^3 $ and $\ob = \rho \beta \bar\beta^3$, which are not scaled by $\kappa$ but only by $\rho$.
  After some simple algebra, one arrives at  the same equations (\ref{systemolambda}) but with $\kappa$ set
  equal to 1, and at the same constraint (\ref{constraintll}) but with  the right-hand-side  set to zero.
  \sm

   The constraint equation can in this case be explicitly solved, with the result
   \bea
\label{lambdavalue}
\lambda=|\lambda |\exp{i \lambda_\phi}, &&\qquad  \lambda_\phi={n \pi/4} \qquad {\rm where} \qquad n \in\{0, 1, 2 ... 7\} \ ,
 \cr \,  && \cr
 \qquad&& |\lambda |\in\{1,(2\gamma+1 \pm 2 \sqrt{\gamma(1+\gamma)})^{1/2}\}\ .
\eea
The case $|\lambda |=1$ implies $|\a|=|\b|$,  which makes the factor $f_2=0$  everywhere.
Likewise, the cases $\l_{\phi} = 0, \pi/2 $ imply that $f_3 \sim (\bar\alpha\beta -  \alpha\bar \beta)\sim \sin(2\l_{\phi}) = 0$
everywhere. Thus, these solutions are  singular, and  though
it is conceivable  that they  can be reinterpreted as a decompactification limit,
  we do not pursue this possibility here.  We focus therefore on $\l_{\phi} \in  \{\pi/4, 7\pi/4\}$, and
   $|\lambda| \neq 1$ which requires $\gamma>0$.

\sm

Using the fact that $\l$ is  constant, one can write the  second equation in (\ref{systemolambda})
  as a total derivative  whose integral is an antiholomorphic function.
  Without loss of generality, we may write the antiholomorphic function as the derivative of a real harmonic function $X$.
  Using also $\l_{\phi} \in  \{\pi/4, 7\pi/4\}$ finally gives
\bea
{\rho^{3/2}\, \bar\om_\a^{5/4}   \om_\a^{-3/4}}=(\p_{\bar w} X)^2\ .
\eea
  The power on the right-hand side of this equality has been chosen so that $X$ is a scalar on $\Sigma$. Since
   the conformal factor $\rho$ is real, we can furthermore write
    \bea
\rho^{3 }={|\p_{\bar w} X|^4   |\om_\a|^{-1 }}.
\eea
 Substituting into the first equation in \ref{systemolambda} leads to
\bea\label{omegaeq}
\p_w\bar\om_\a=\p_w X \p_{\bar w}X \,  {(c_1-c_2) |\l|\over 1-\bar\l^4} \cos(2\l_{\phi}) = 0 \ ,
\eea
where we have used   the fact that  $\cos(2\l_{\phi}) =0$.
\sm

The above equation tells us that   $\omega_\a$ is   holomorphic, so
 without loss of generality  we  write   $\omega_{\a}=\p_w Y$ for  some  real harmonic function $Y$.
From the reality condition of $\rho$ we deduce that $\p_w Y= C \p_w X $,
where $C$ is a complex constant.
Putting everything together, one arrives at the  following  expressions for
the  metric factors of the three (pseudo)spheres:
\bea
\label{5a1}
  f_1^3={(1+|\l|^2)^3\over c_1^3}|C|^2\ ,  \qquad
  f_2^3=-{(1-|\l|^2)^3\over c_2^3}|C|^2\ , \qquad
  f_3^3=  \tilde\sigma\, {8|\l|^3\over c_3^3} |C|^2 \ ,
  \eea
where   $\tilde\sigma  = -\sin(2\l_{\phi}) = \pm 1$. In addition, the Weyl factor of the metric on
 $\Sigma$ reads
 \bea
\label{5a1a}
   \rho^3={1\over |C|}|\p_w X|^3\ .
\eea
Note that the metric factors $f_i$ are constant, and that the metric on $\Sigma$ is conformally-flat, i.e. it
can be made flat by a conformal change of coordinates. This proves that the fibration is trivial,
and the geometry is locally AdS$_3\times $S$^3\times $S$^3\times $E$_2$  as advertized.
The  relations (\ref{CFT2relns}) among the (pseudo)sphere radii follow  from
the identity $4\g \vert\lambda\vert^2 =  (1 - \vert\lambda\vert^2)^2$,
and  the   relations   $c_1 + c_2 +c_3 =0$ and $c_2= \g c_3$.

\sm

 We may also compute the  gauge potentials and charges of the solution. Choosing a local coordinate such that
 $X = x$ with $w = x + i y$, and picking a definite sign for  $b_2$, one finds
\bea
b_1= - {2  (1+|\l|^2)^2\over c_1^2 } |C| \, x , \qquad
b_{2}= {2  (1-|\l|^2)^2\over c_2^2 } |C| \, x , \qquad
b_{3}= \tilde\sigma  \, {8  |\l|^2 \over c_3^2 } |C| \, x .
\eea
Note that the corresponding one-forms all point in the same direction $x$ in $\Sigma$.
If this direction is compact, there exist two 4-cycles that support M5-brane charge: ${\cal C}_x \times S^3_2$
and  ${\cal C}_x \times S^3_3$, where $C_x$ is the circle parametrized by $x$. The
definition (\ref{2g3}) then gives
\bea
\mM^{(2)} =   {2  (1-|\l|^2)^2\over c_2^2 } |C| \ell_x\ , \qquad \quad
\mM^{(3)} =   \, \tilde\sigma\,   {8  |\l|^2 \over c_3^2 } |C| \ell_x\ ,
\eea
where $\ell_x$ is the circumference of the $x$-circle. Note that
$  \mM^{(3)} =  \tilde\sigma \g\,  \mM^{(2)}$, so the parameter $\g$ is the ratio of the two M5-brane charges of the solution.

\sm

\if
Finally, if the second coordinate of $\Sigma$  is also compact, one can compute the M2-brane charge associated to the
7-cycle ${\cal C}_y \times S^3_2 \times S^3_3$, where ${\cal C}_y$ is the $y$-circle of circumference $\ell_y$.
The definiton
(\ref{2f3}) leads to
 \bea
\mE^{(1)}
= \int_{{\cal C}_y} - \frac{f_1^3 f_2^3 f_3^3}{f_1^6} (*_{\Sigma} d b_1)
=   - 16\, \nu_3     \frac{c_1}{c_2^3 c_3^3}\, \frac{|\lambda|^3(1-|\lambda|^2)^3}{1+|\lambda|^2} |C|^3 \ell_y\  .
\eea
\fi

The M2-brane charge  corresponding  to the 7-cycle ${\cal C}_x \times S^3_2 \times S^3_3$
is formally the integral of $b_2 db_3$,  or of $-b_3 db_2$. Neither of these one-forms  is however well-defined on the
$x$-circle. The ambiguity can be attributed to the Hanany-Witten effect \cite{Hanany:1996ie}
and to  the smearing of the M5-branes.
To see why, let us compactify    $x^1$ in the configuration of Table 1, thereby converting it
to a type-IIA configuration of orthogonal  D4-branes intersecting fundamental strings. As the D4-branes cross each other,
they create or destroy fundamental strings -- a phenomenon dual to anomaly inflow \cite{Bachas:1997ui}.
In principle this leads to an integer ambiguity of the fundamental-string charge, but if the D4-branes are
actually smeared, the ambiguity becomes continuous.


\section{Detailed formulae for the deformed AdS$_7\times$S$ ^4$ solution}
\setcounter{equation}{0}
\label{sec:E}


As a check of our formulae, we here compute explicitly some of the intermediate quantities for the solution
 \eqref{AdS7S4} of the BPS equations. The harmonic function,
\bea\label{appE1}
h =  2 \xi \sinh(2x)\,\sin(2y) \ ,
\eea
vanishes on the real and on the  imaginary-$z$ axes, and is everywhere positive in the interior of the strip
$0<y<\pi/2$,  as required. Furthermore,  some straightforward algebra gives:
 \bea
 (G\bar G -1)  =  X^{-1} \left[ \sin(2y)\right]^2 (1 - \cosh(2x))\ , \quad
  \vert G + i\vert^2 = X^{-1} \left[ \sin(2y)\right]^2 \ , \no
\eea
 \bea
   \vert G - i\vert^2 = X^{-1} \left\{ [  \sinh(2x) ]^2 +   2 [ \sin(2y) ]^2 (1-  \cosh(2x))  \right\}\ , \no
  \eea
where
\bea\label{appE2}
4 X = \left\vert {\sinh(2z)}\right\vert^2 =   \left[  \sinh(2x) \right]^2 +  \left[  \sin(2y)\right]^2 \ .
\eea
Note that $(G\bar G -1)$ is negative in the interior of the strip, consistently with the fact that these data
give regular solutions for negative values of the parameter $\g$. From the above formulae we may
compute
   \bea
  W_+  = X^{-1} \left[ \sin(2y)\right]^2 (1 +  \g  -\g \cosh(2x))\ ,  \no
\eea
 \bea\label{appE3}
    W_- =  X^{-1} \left\{ [ \sinh(2x) ]^2 +   (2 + {1\over  \g}) \left[  \sin(2y)\right]^2 (1-  \cosh(2x))
\right\}\ .
\eea
 Note that $W_\pm$  simplify greatly if $\g = -1/2$, in which case
  the expressions  \eqref{2d2} for the metric factors read:  $f_1^2 = 4L^2 \cosh^2x$,
 $f_2^2 = 4L^2 \sinh^2x$ and $f_3^2 = 4L^2 [\sin(2y)]^2$. These are precisely the factors
 of the  AdS$_7\times$S$^4$ metric, eq.\,\eqref{AdS7S4metric}.
  For general (negative) values of $\g$, there is  no separation of  $x$ and $y$,
  and the AdS$_7\times$S$^4$ geometry gets deformed in a way that can be
easily computed from the above expressions.

 \sm

 To calculate the fluxes and the charges, one needs the auxiliary functions
 \bea\label{appE4}
 \tilde h = - 2 \xi \,\cos(2y)\,\cosh(2x) \ ,  \qquad
 \Phi = 2\xi  \cos(2y) ( 2 \sinh^2 \hskip -0.4mm x -1)\ .
 \eea
The reader may easily check that these solve the two defining equations,
$\p_z \tilde h  = - i\p_z h$ and $\p_z\Phi = \bar G \p_z h$ or, more conveniently,
$ G =  i \bar \p_{\bar z} \Phi / \bar\p_{\bar z} \tilde h$.  One other useful formula is
\bea\label{appE5}
X (G + \bar G) =  \sinh(2x)\,\sin(2y)\,\cos(2y)\ .
\eea
 Note that the expressions in \eqref{appE1}, \eqref{appE2}, \eqref{appE4} and \eqref{appE5}
 are the same for all values of $\g$.    Inserting
\eqref{appE1} to \eqref{appE5}       in the expressions   \eqref{2e2} for the flux fields
 gives
  \bea
 b_2^s\  =\   -\g {h(G+\bar G)\over W_+}\  =\ 8\xi\g\,\cos(2y)\,{\sinh^2\hskip -0.4mm x\, \cosh^2\hskip -0.4mm x\over
 (2\g \sinh^2\hskip -0.4mm x - 1) } \ , \no
\eea
\bea
b_2^c\  =\  b_2^0 + \g (\Phi -\tilde h) \ =\ b_2^0 + 8\xi\g\,\cos(2y)\,\sinh^2\hskip -0.4mm x \ ,  \no
\eea
   \bea
 b_3^s\  =\    {1\over \g}  {h(G+\bar G)\over W_-}\  =\ {2\xi\over \g}\,\cos(2y)\,
 { [\sinh(2x)\,\sin(2y)]^2 \over
 \left\{ [ \sinh(2x) ]^2 +   (2 + {1\over  \g}) \left[  \sin(2y)\right]^2 (1-  \cosh(2x))
\right\} } \ , \no
\eea
\bea\label{E.6b3}
b_3^c\  =\  b_3^0 - {1\over \g}  (\Phi + \tilde h) \ =\ b_3^0 + {4\xi\over \g} \,\cos(2y)\,  .
\eea
For $\g = -1/2$ these magnetic  gauge potentials  simplify to $b_2^s + b_2^c = b_2^0$, which can be chosen equal to zero,
and
\bea
b_3^s + b_3^c &=& b_3^0 +  {2\xi\over \g} \,\cos(2y) \left( \left[ \sin(2y)\right]^2+2 \right) \no \\
&&\Longrightarrow\,  d(b_3^s + b_3^c) = - {12\xi\over \g}\, \left[ \sin(2y)\right]^3 dy\ ,
\eea
which is precisely the volume element on the four-sphere. This is as expected for the AdS$_7\times$S$^4$
solution, which has only one type of M5-brane flux. For $\g\not=-1/2$, the second 5-brane flux is also
turned on, as is evident from the above expressions.


\end{document}